\RequirePackage{fix-cm}
\documentclass[twocolumn]{svjour3}          
\smartqed  
\usepackage{graphicx}

\usepackage[T1]{fontenc} 
\usepackage{amsmath}
\usepackage{amssymb}
\usepackage{amsfonts}       
\usepackage{xspace}
\usepackage{multirow}
\usepackage[toc]{appendix}
\usepackage{url}

\providecommand{\PH}{\ensuremath{\mathrm{H}}\xspace} 
\providecommand{\PZ}{\ensuremath{\mathrm{Z}}\xspace} 
\providecommand{\PW}{\ensuremath{\mathrm{W}}\xspace} 
\providecommand{\Pg}{\ensuremath{\mathrm{g}}\xspace} 
\providecommand{\PQu}{\ensuremath{\mathrm{u}}\xspace} 
\providecommand{\PQd}{\ensuremath{\mathrm{d}}\xspace} 
\providecommand{\PQs}{\ensuremath{\mathrm{s}}\xspace} 
\providecommand{\PQc}{\ensuremath{\mathrm{c}}\xspace} 
\providecommand{\PQb}{\ensuremath{\mathrm{b}}\xspace} 
\providecommand{\PQt}{\ensuremath{\mathrm{t}}\xspace} 
\providecommand{\PQq}{\ensuremath{\mathrm{q}}\xspace} 
\providecommand{\PAQq}{\ensuremath{\overline{\mathrm{q}}}\xspace} 
\newcommand{\qqbar}{\ensuremath{\PQq\PAQq}\xspace}

\newcommand{\pt}{\ensuremath{p_{\mathrm{T}}}\xspace}
\newcommand{\kt}{\ensuremath{k_{\mathrm{T}}}\xspace}

\newcommand{\TeV}{\ensuremath{\,\text{Te\hspace{-.08em}V}}\xspace}

\newcommand{\ptvecmiss}{\ensuremath{{\vec p}_{\mathrm{T}}^{\kern1pt\text{miss}}}\xspace}
\usepackage{cite}
\begin{document}

\title{JEDI-net: a jet identification algorithm based on interaction networks}


\author{
Eric A. Moreno \and
Olmo Cerri \and
Javier M. Duarte \and
Harvey B. Newman \and
Thong Q. Nguyen \and
Avikar Periwal \and
Maurizio Pierini \and
Aidana Serikova \and
Maria Spiropulu \and
Jean-Roch Vlimant
}

\institute{Eric A. Moreno  
  Olmo Cerri \\
  Harvey B. Newman 
  Thong Q. Nguyen \\
  Avikar Periwal 
  Aidana Serikova \\ 
  Maria Spiropulu 
  Jean-Roch Vlimant \at California Institute of Technology, Pasadena, CA 91125, United States
              \and
              Javier M. Duarte \at Fermi National Accelerator Laboratory (FNAL), Batavia, IL 60510, United States
              \at University of California San Diego, La Jolla, CA 92093, United States
              \and
              Maurizio Pierini \at European Center for Nuclear Research (CERN), CH-1211 Geneva, Switzerland
}


\maketitle

\begin{abstract}
We investigate the performance of a jet identification algorithm based
on interaction networks (JEDI-net) to identify all-hadronic decays of
high-momentum heavy particles produced at the LHC and distinguish them
from ordinary jets originating from the hadronization of quarks and
gluons.  The jet dynamics are described as a set of one-to-one
interactions between the jet constituents. Based on a representation
learned from these interactions, the jet is associated to one of the
considered categories.  Unlike other architectures, the JEDI-net
models achieve their performance without special handling of the
sparse input jet representation, extensive pre-processing, particle
ordering, or specific assumptions regarding the underlying detector
geometry. The presented models give better results with less model
parameters, offering interesting prospects for LHC applications.
\end{abstract}


\section{Introduction}
\label{sec:intro}

Jets are collimated cascades of particles produced at particle accelerators. 
Quarks and gluons originating from hadron collisions, such as the proton-proton collisions at the CERN Large Hadron Collider (LHC), generate a cascade of other particles (mainly other quarks or gluons) that then arrange themselves into hadrons. 
The stable and unstable hadrons' decay products are observed by large particle detectors, reconstructed by algorithms that combine the information from different detector components, and then clustered into jets, using physics-motivated sequential recombination algorithms such as those described in Ref.~\cite{CA,Catani:1993hr,Cacciari:2008gp}. 
Jet identification, or \emph {tagging}, algorithms are designed to identify the nature of the particle that initiated a given cascade, inferring it from the collective features of the particles generated in the cascade. 

\begin{figure*}[htpb!]
\centering
\includegraphics[width=\textwidth]{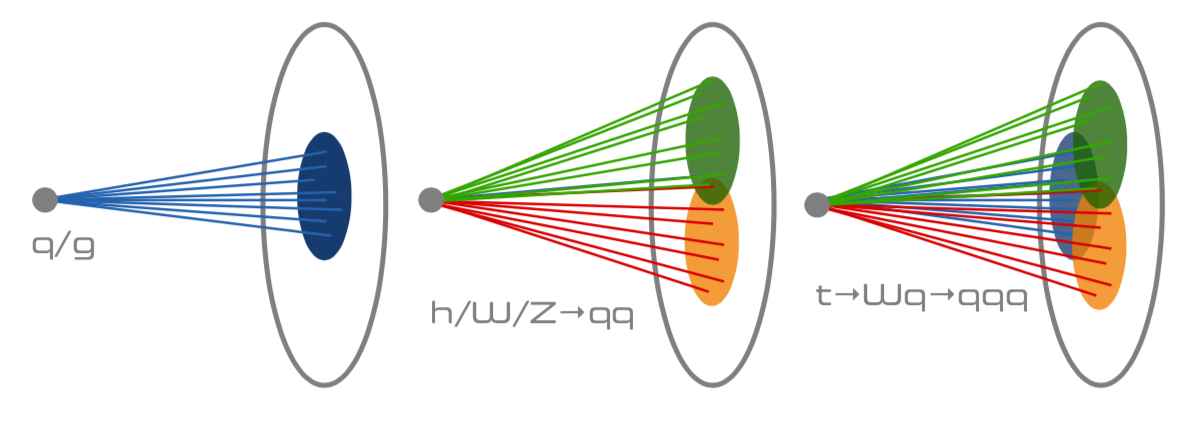}
\caption{Pictorial representations of the different jet categories considered in this paper. Left: jets originating from quarks or gluons produce one cluster of particles, approximately cone-shaped, developing along the flight direction of the quark or gluon that started the cascade. Center: when produced with large momentum, a heavy boson decaying to  quarks would result in a single jet, made of 2 particle clusters (usually referred to as prongs). Right: a high-momentum $\PQt \to \PW \PQb \to \qqbar^{\prime}\PQb$ decay chain results in a jet composed of three prongs. \label{fig:boostedJetCartoon}}
\end{figure*}

Traditionally, jet tagging was meant to distinguish three classes of jets: light flavor quarks $\PQq=\PQu,\PQd,\PQs,\PQc$, gluons $\Pg$, or bottom quarks ($\PQb$). 
At the LHC, due to the large collision energy, new jet topologies emerge. 
When heavy particles, e.g. $\PW$, $\PZ$, or Higgs ($\PH$) bosons or the top quark, are produced with large momentum and decay to all-quark final states, the resulting jets are contained in a small solid angle.
A single jet emerges from the overlap of two (for bosons) or three (for the top quark) jets, as illustrated in Fig.~\ref{fig:boostedJetCartoon}.
These jets are characterized by a large invariant mass (computed from the sum of the four-momenta of their constituents) and they differ from ordinary quark and gluon jets, due to their peculiar momentum flow around the jet axis. 

Several techniques have been proposed to identify these jets by using physics-motivated quantities, collectively referred to as ``jet substructure'' variables. 
A review of the different techniques can be found in Ref.~\cite{Larkoski:2017jix}. 
As discussed in the review, approaches based on deep learning (DL) have been extensively investigated (see also Sec.~\ref{sec:related}), processing sets of physics-motivated quantities with dense layers or raw data representations (e.g. jet images or particle feature lists) with more complex architectures (e.g. convolutional or recurrent networks).

In this work, we compare the typical performance of some of these approaches to what is achievable with a novel jet identification algorithm based on an interaction network (JEDI-net). 
Interaction networks~\cite{interactionnetwork} (INs) were designed to decompose complex systems into distinct objects and relations, and reason about their interactions and dynamics.
One of the first uses of INs was to predict the evolution of physical systems under the influence of internal and external forces, for example, to emulate the effect of gravitational interactions in $n$-body systems. 
The $n$-body system is represented as a set of objects subject to one-on-one interactions. 
The $n$ bodies are embedded in a graph and these one-on-one interaction functions, expressed as trainable neural networks, are used to predict the post-interaction status of the $n$-body system. 
We study whether this type of network generalizes to a novel context in high energy physics.
In particular, we represent a jet as a set of particles, each of which is represented by its momentum and embedded as a vertex in a fully-connected graph. 
We use neural networks to learn a representation of each one-on-one particle \emph{interaction}~\footnote{Here, we refer to the abstract message-passing interaction represented by the edges of the graph and not the physical interactions due to quantum chromodynamics, which occur before the jet constituents emerge from the hadronization process.} in the jet, which we then use to define jet-related high-level features (HLFs).
Based on these features, a classifier associates each jet to one of the five categories shown in Fig.~\ref{fig:boostedJetCartoon}. 

For comparison, we consider other classifiers based on different architectures: a dense neural network (DNN) \cite{MLP} receiving a set of jet-substructure quantities, a convolutional neural network (CNN)~\cite{fukushima,Cun1990,Waibel1990} receiving an image representation of the transverse momentum ($\pt$) flow in the jet~\footnote{We use a Cartesian coordinate system with the $z$ axis oriented along the beam axis, the $x$ axis on the horizontal plane, and the $y$ axis oriented upward. The $x$ and $y$ axes define the transverse plane, while the $z$ axis identifies the longitudinal direction. The azimuthal angle $\phi$ is computed from the $x$ axis. The polar angle $\theta$ is used to compute the pseudorapidity $\eta = -\log(\tan(\theta/2))$. We use natural units such that $c=\hbar=1$ and we express energy in units of electronVolt (eV) and its prefix multipliers.}, and a recurrent neural network (RNN) with gated recurrent units~\cite{GRU} (GRUs), which process a list of particle features. 
These models can achieve state-of-the-art performance although they require additional ingredients: the DNN model requires processing the constituent particles to pre-compute HLFs, the GRU model assumes an ordering criterion for the input particle feature list, and the CNN model requires representing the jet as a rectangular, regular, pixelated image. 
Any of these aspects can be handled in a reasonable way (e.g. one can use a jet clustering metric to order the particles), sometimes sacrificing some detector performance (e.g., with coarser image pixels than realistic tracking angular resolution, in the case of many models based on CNN). It is then worth exploring alternative solutions that could reach state-of-the-art performance without making these assumptions.
In particular, it is interesting to consider architectures that directly takes as input jet constituents and are invariant for their permutation.
This motivated the study of jet taggers based on recursive~\cite{RecursiveJets}, graph networks~\cite{Egan:2017ojy,Cheng:2017rdo}, and energy flow networks~\cite{Komiske:2018cqr}.
In this context, we aim to investigate the potential of INs. 

This paper is structured as follows: we provide a list of related works in Sec.~\ref{sec:related}. 
In Sec.~\ref{sec:dataset}, we describe the utilized data set. 
The structure of the JEDI-net model is discussed in Sec.~\ref{sec:models} together with the alternative architectures considered for comparison.
Results are shown in Sec.~\ref{sec:results}.
Sections~\ref{sec:learned}~and~\ref{sec:resources} discuss what the JEDI-net learns when processing the graph and quantify the amount of resources needed by the tagger, respectively.
We conclude with a discussion and outlook for this work in Sec.~\ref{sec:conclusions}. Appendix~\ref{appendix:otherModelOpt} describes the design and optimization of the alternative models.

\section{Related work}
\label{sec:related}

Jet tagging is one of the most popular LHC-related tasks to which DL solutions have been applied. 
Several classification algorithms have been studied in the context of jet tagging at the LHC~\cite{deOliveira:2015xxd,Guest:2016iqz,Macaluso:2018tck,Datta:2017lxt,Butter:2017cot,Kasieczka:2017nvn,Komiske:2016rsd,Schwartzman:2016jqu} using DNNs, CNNs, or physics-inspired architectures. 
Recurrent and recursive layers have been used to construct jet classifiers starting from a list of reconstructed particle momenta~\cite{RecursiveJets,Egan:2017ojy,Cheng:2017rdo}. 
Recently, these different approaches, applied to the specific case of top quark jet identification, have been compared in Ref.~\cite{Kasieczka:2019dbj}. 
While many of these studies focus on data analysis, work is underway to apply these algorithms in the early stages of LHC real-time event processing, i.e. the trigger system.
For example, Ref.~\cite{hls4ml} focuses on converting these models into firmware for field programmable gate arrays (FPGAs) optimized for low latency (less than 1~$\mu$s).
If successful, such a program could allow for a more resource-efficient and effective event selection for future LHC runs. 

Graph neural networks have also been considered as jet tagging algorithms~\cite{graphjettagging,Qu:2019gqs} as a way to circumvent the sparsity of image-based representations of jets.
These approaches demonstrate remarkable categorization performance. 
Motivated by the early results of Ref.~\cite{graphjettagging}, graph networks have been also applied to other high energy physics tasks, such as event topology classification~\cite{Abdughani:2018wrw,Choma2018GraphNN}, particle tracking in a collider detector~\cite{GraphTracking}, pileup subtraction at the LHC~\cite{Martinez:2018fwc}, and particle reconstruction in irregular calorimeters~\cite{Qasim:2019otl}.

\section{Data set description}
\label{sec:dataset}

This study is based on a data set consisting of simulated jets with an energy of $\pt\approx 1$ TeV, originating from light quarks $\PQq$, gluons $\Pg$, $\PW$ and $\PZ$ bosons, and top quarks produced in $\sqrt{s} = 13\TeV$ proton-proton collisions.
The data set was created using the configuration and parametric description of an LHC detector described in Ref.~\cite{hls4ml,Coleman:2017fiq}, and is available on the Zenodo platform~\cite{data1,data2,data3,data4}.

\begin{figure*}[htp!]
\centering
\includegraphics[width=0.3\textwidth]{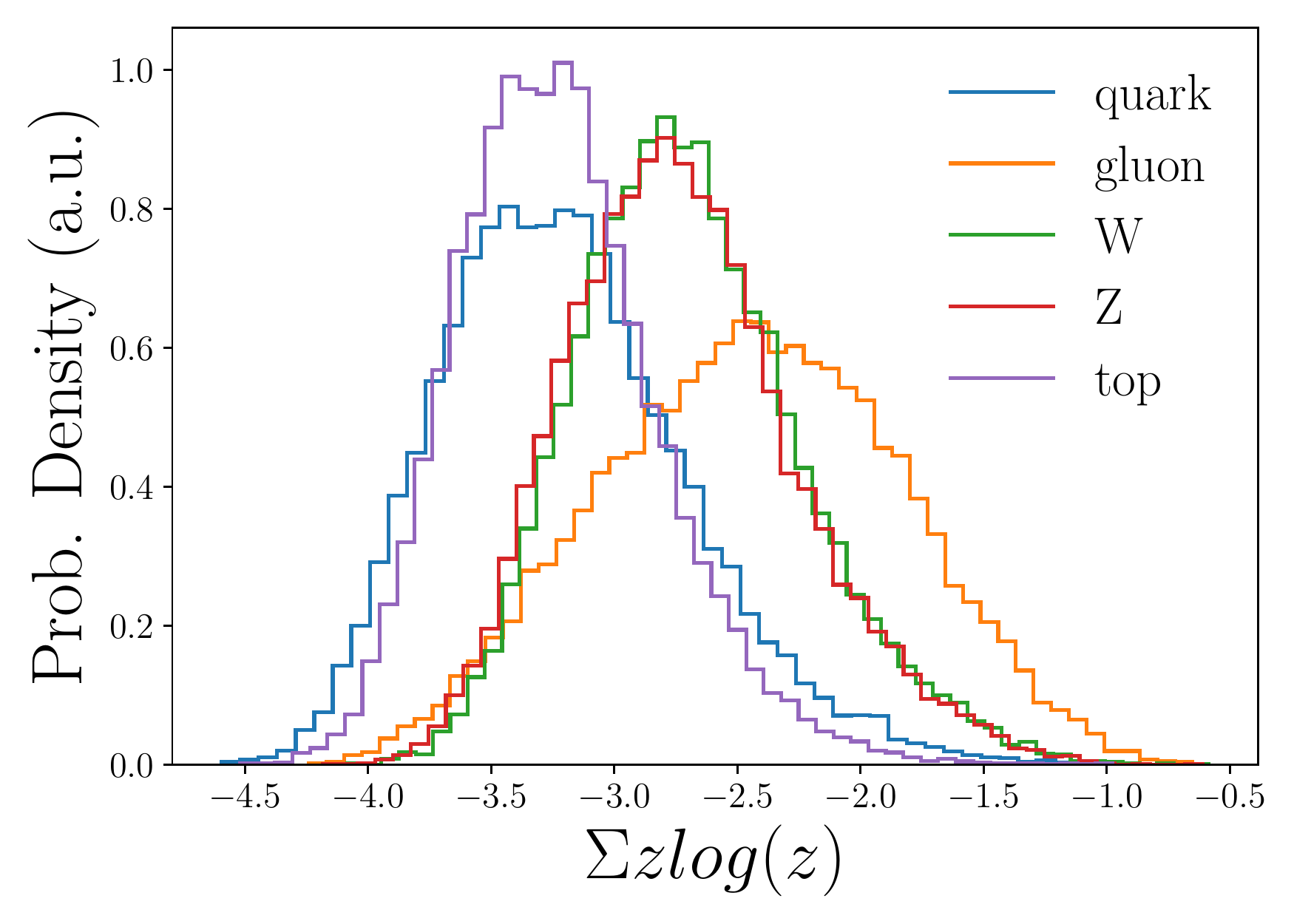}
\includegraphics[width=0.3\textwidth]{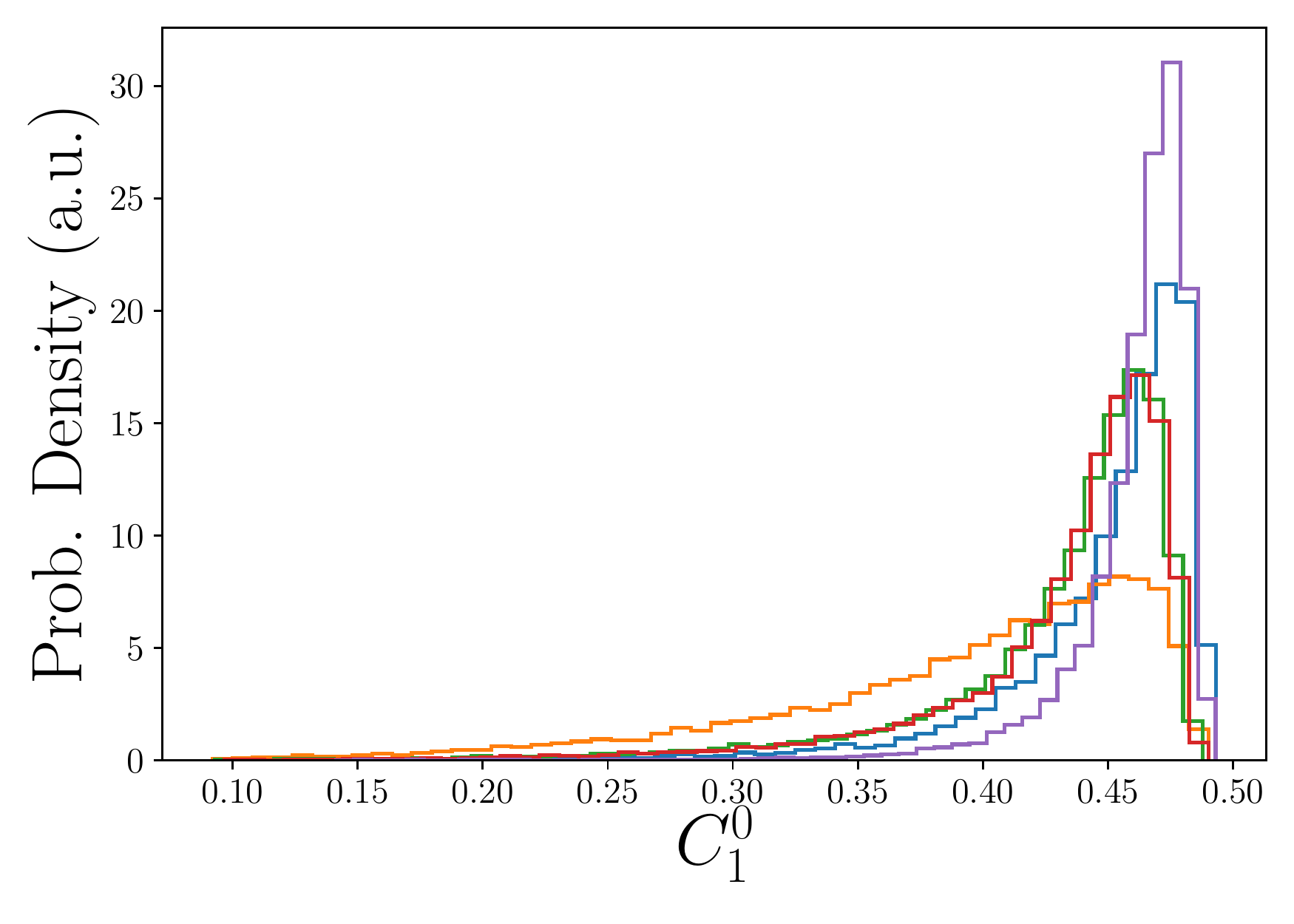}
\includegraphics[width=0.3\textwidth]{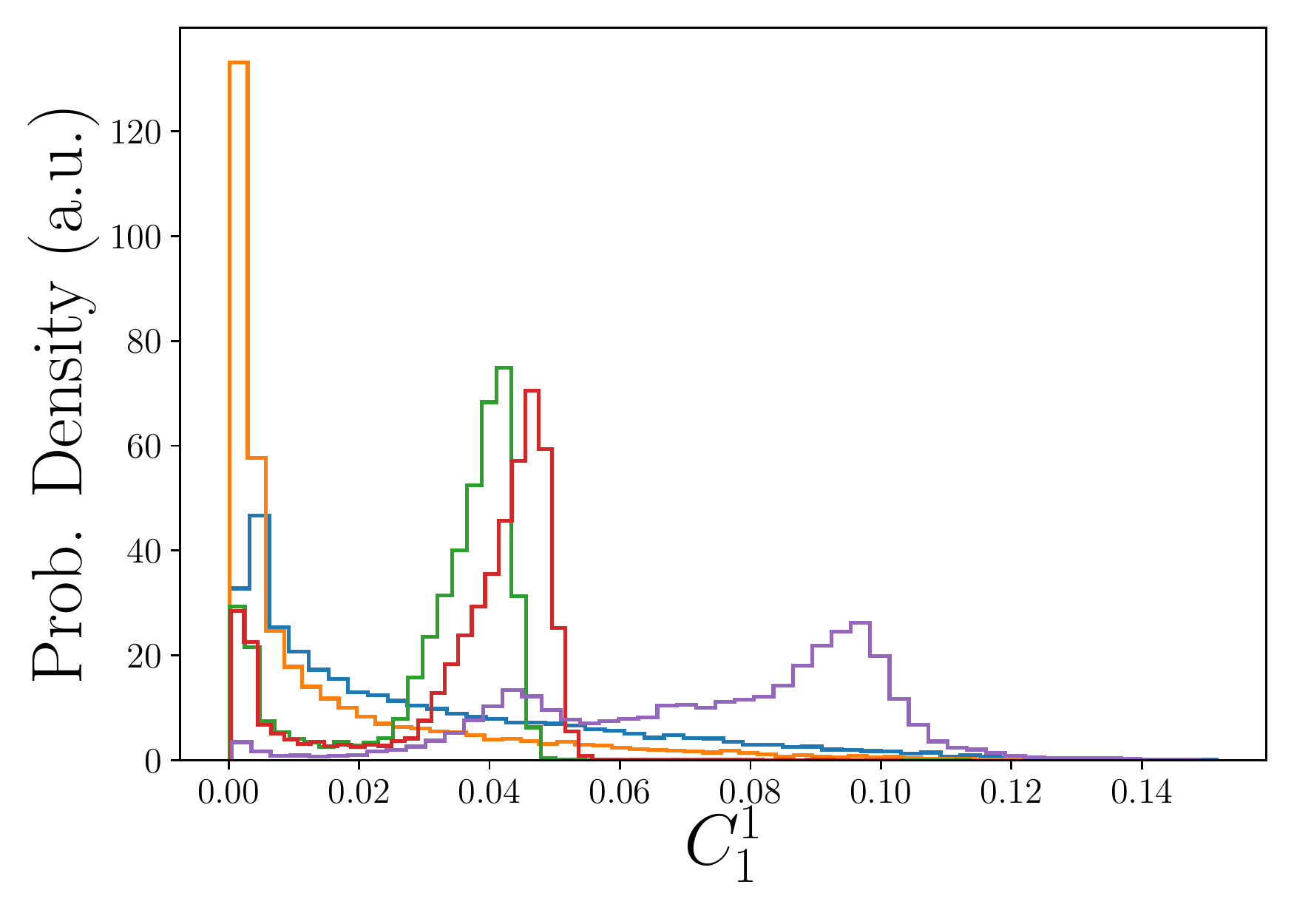}
\\
\includegraphics[width=0.3\textwidth]{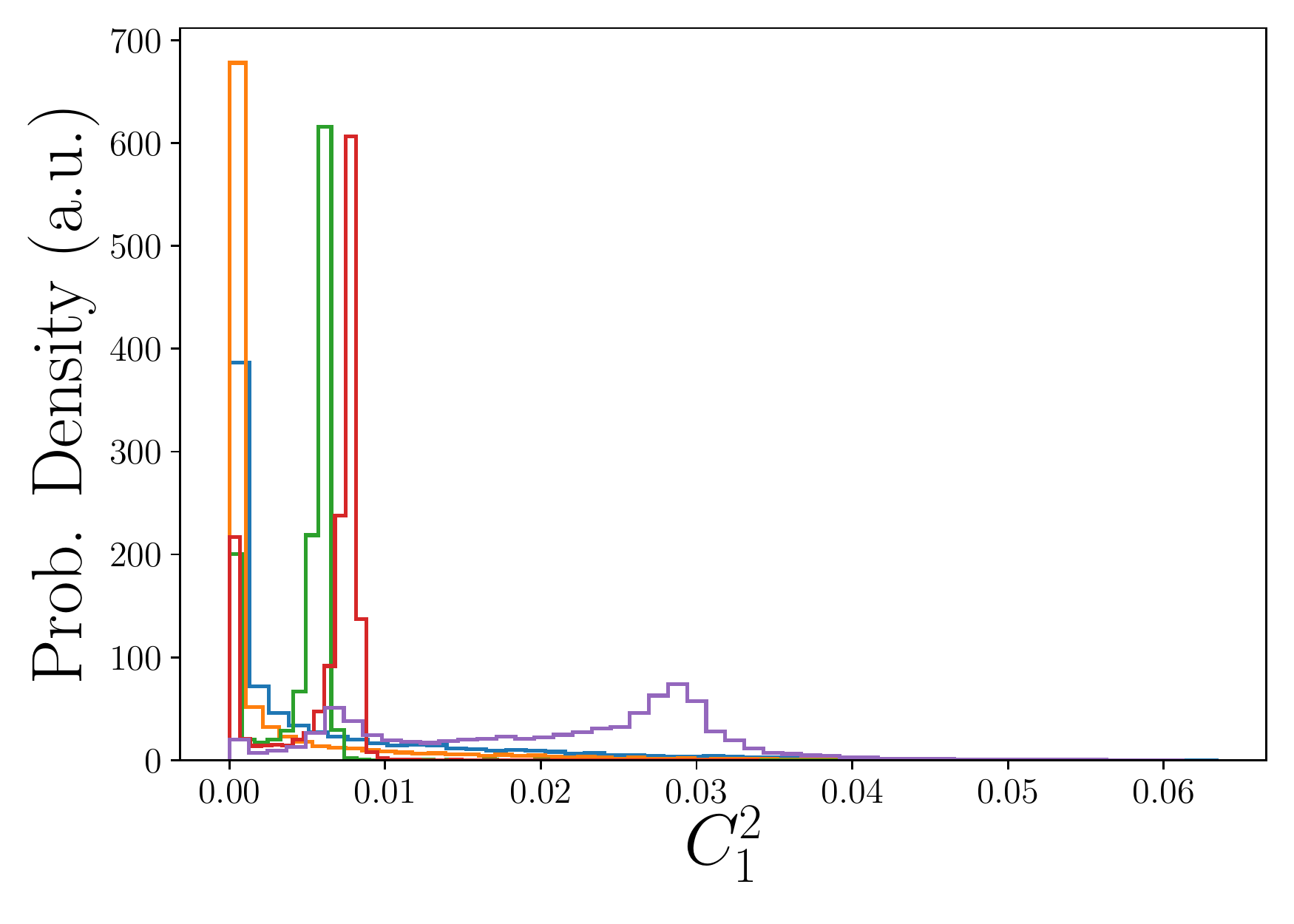}
\includegraphics[width=0.3\textwidth]{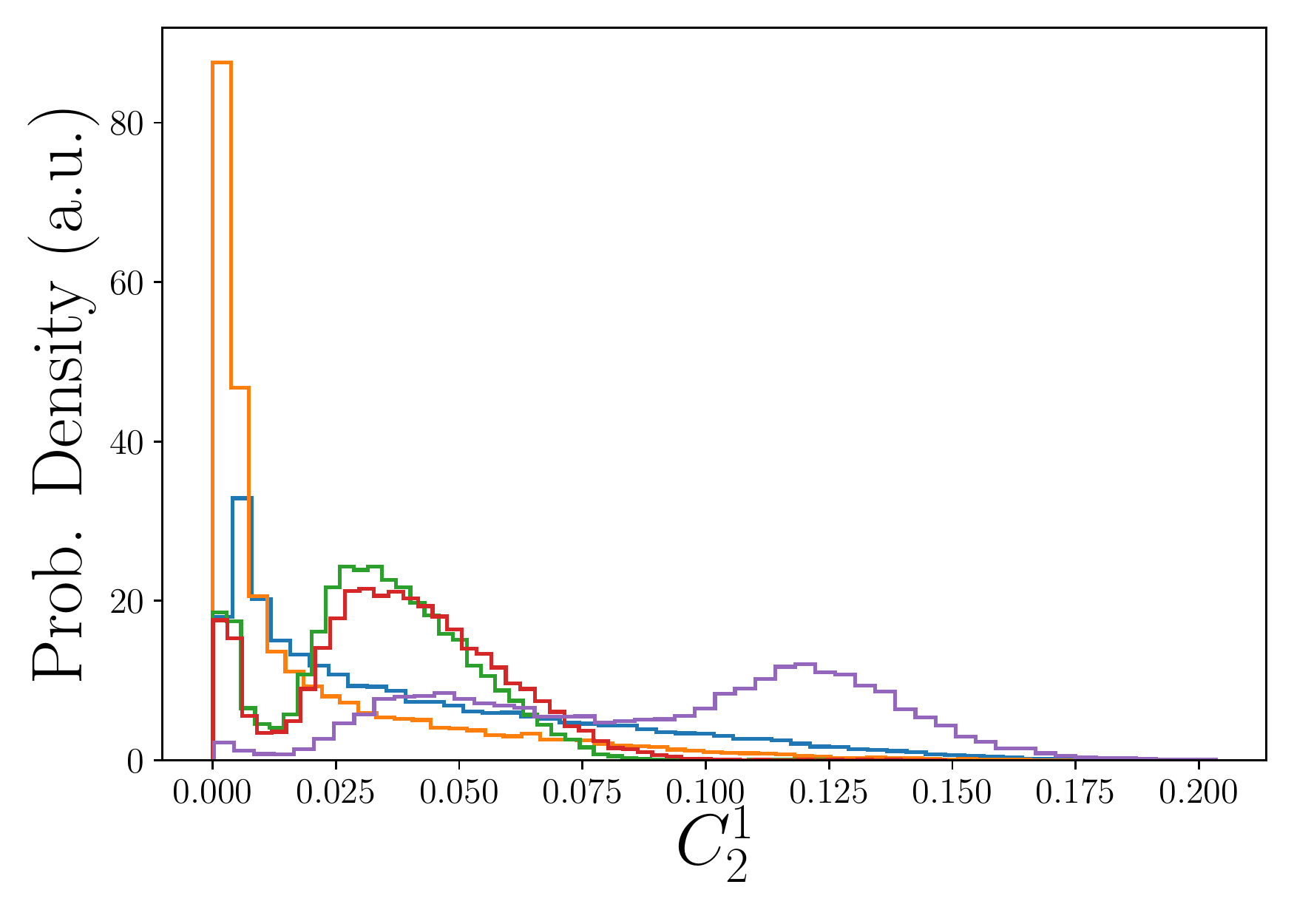}
\includegraphics[width=0.3\textwidth]{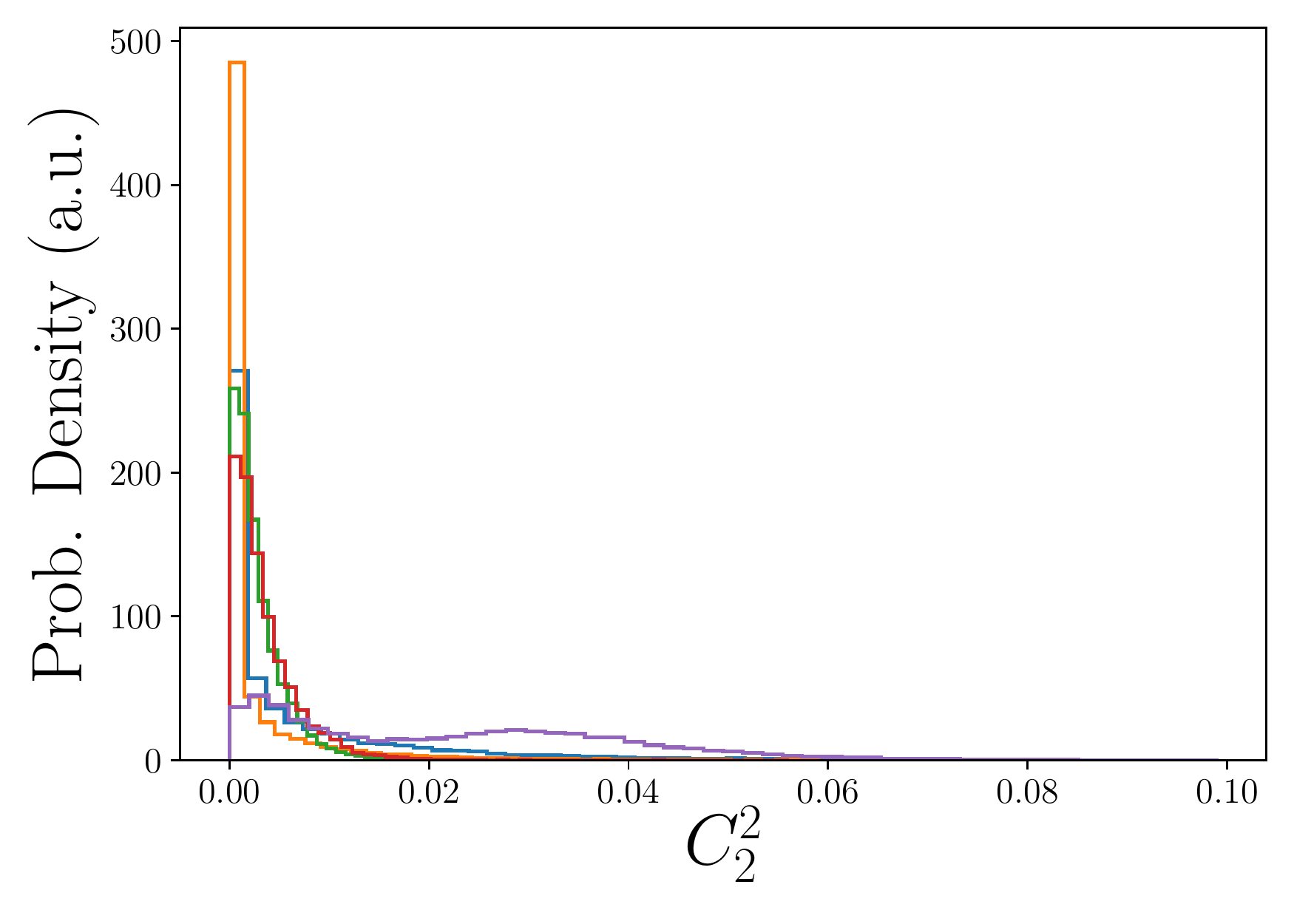}
\\
\includegraphics[width=0.3\textwidth]{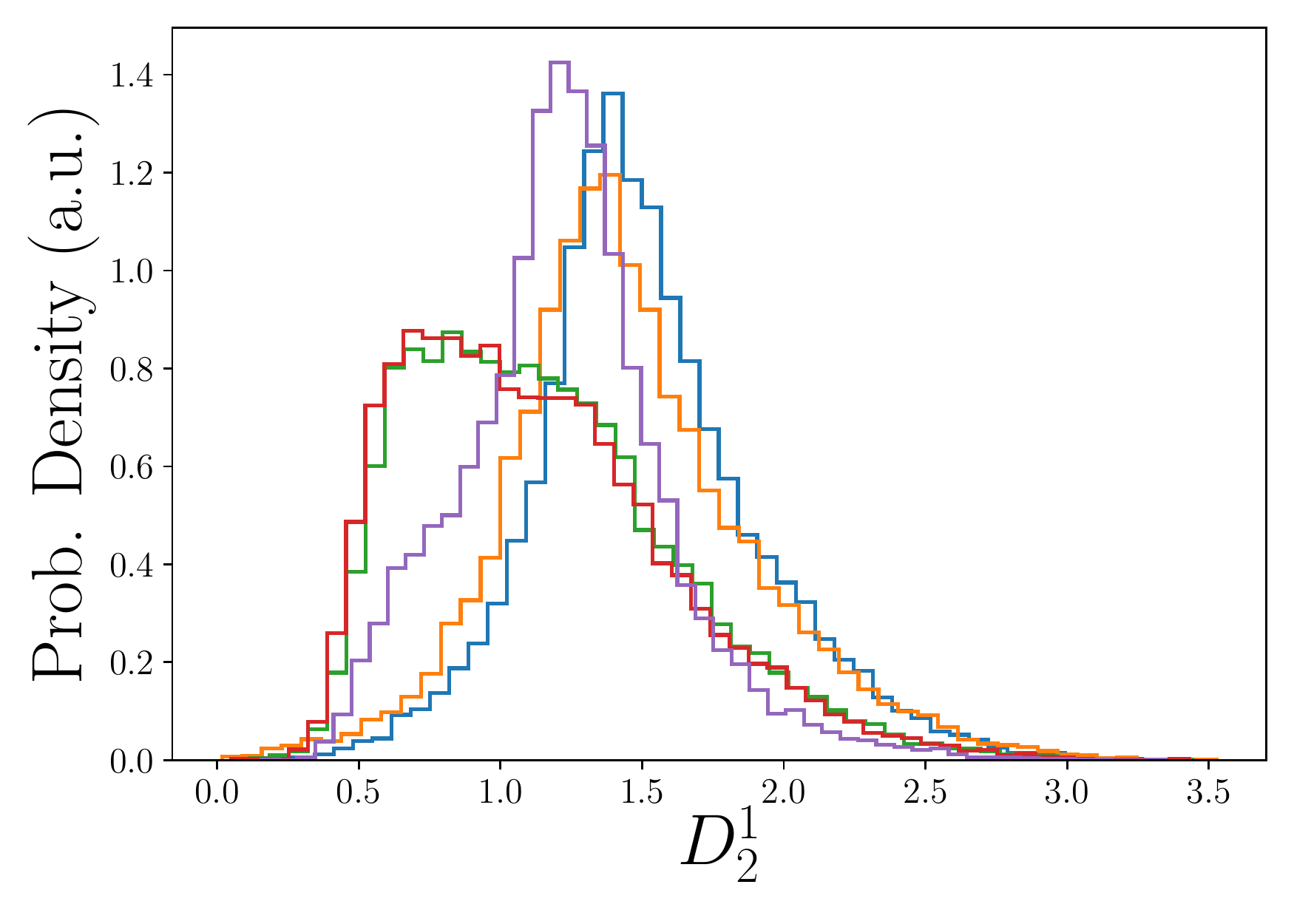}
\includegraphics[width=0.3\textwidth]{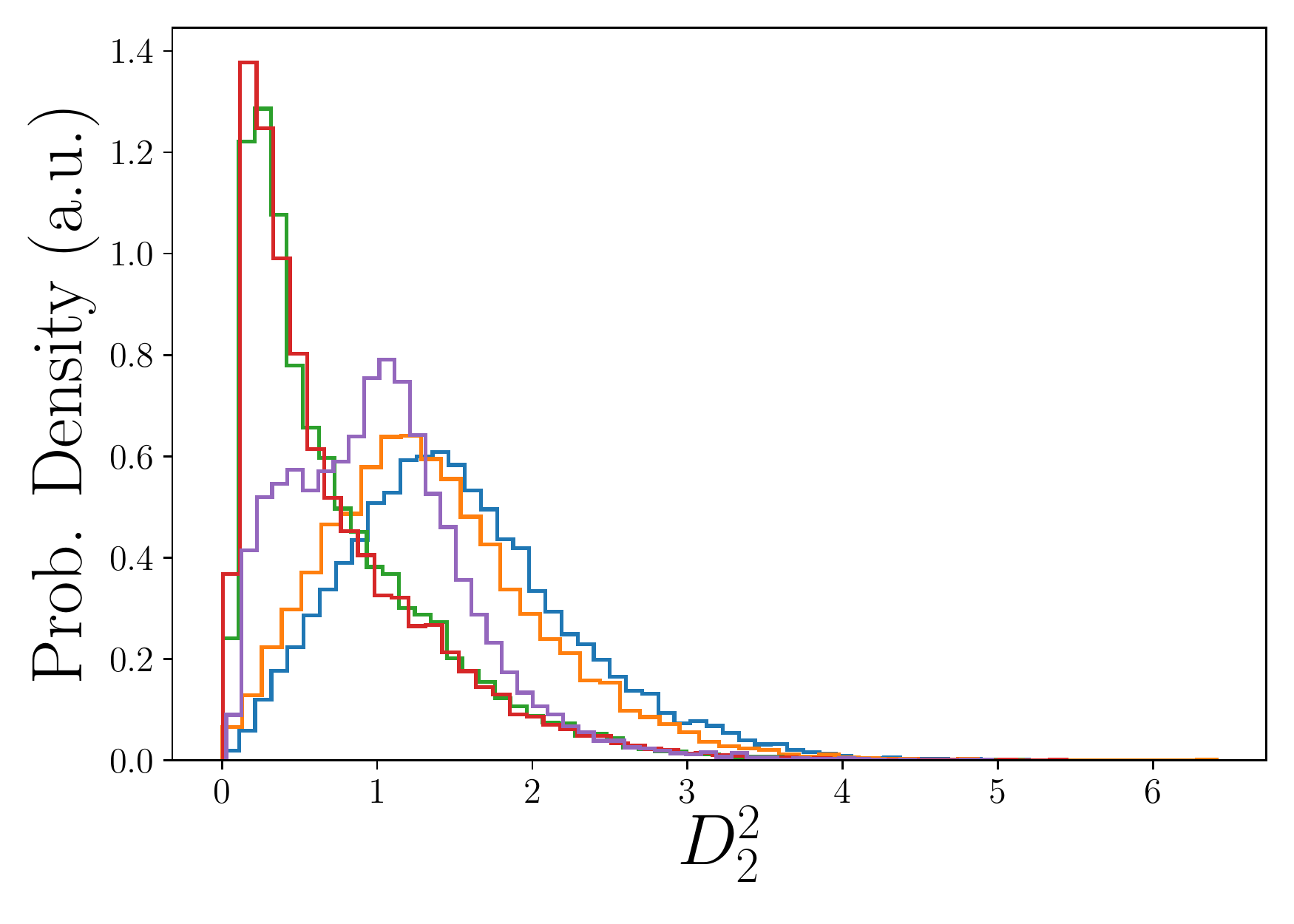}
\includegraphics[width=0.3\textwidth]{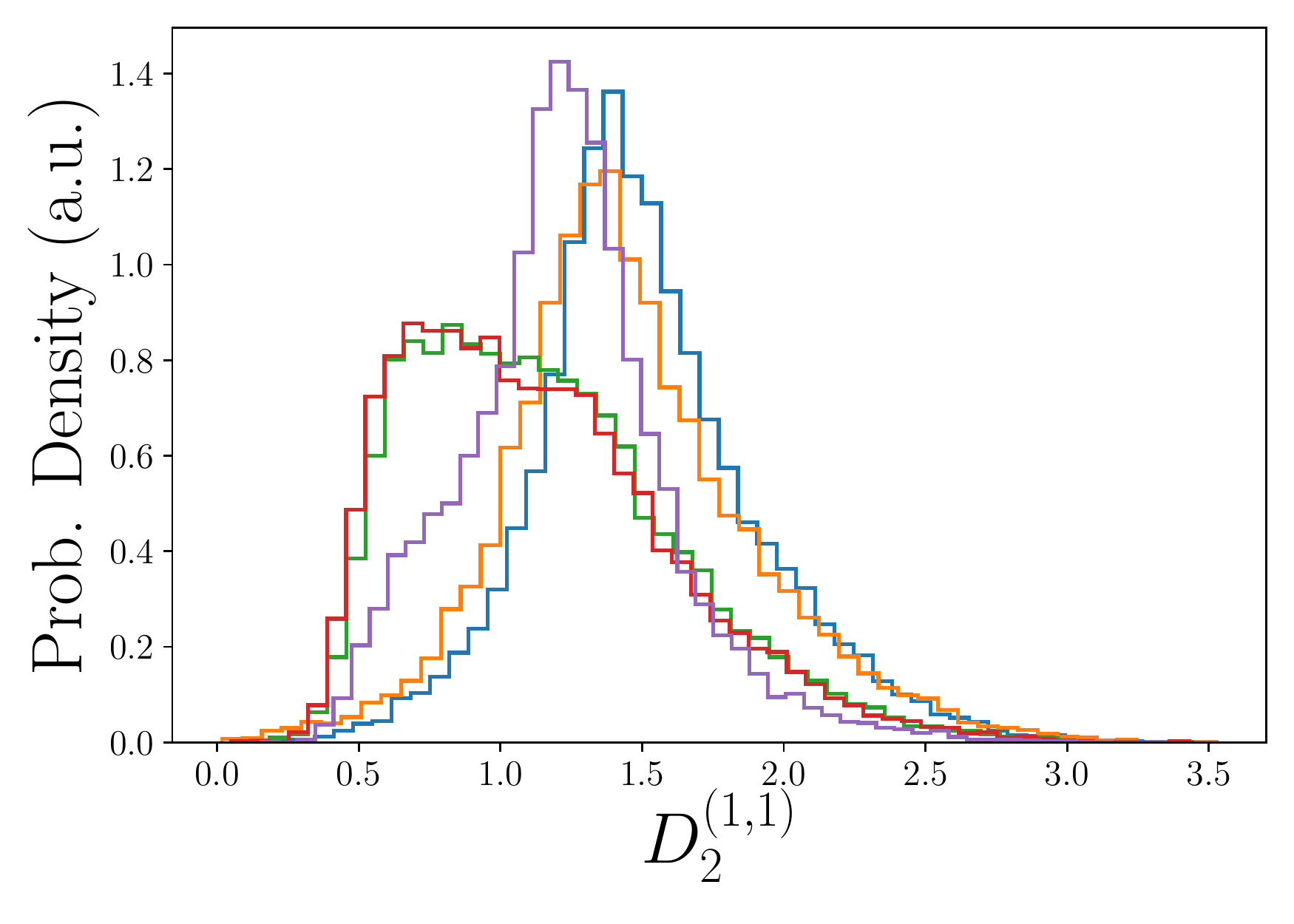}
\\
\includegraphics[width=0.3\textwidth]{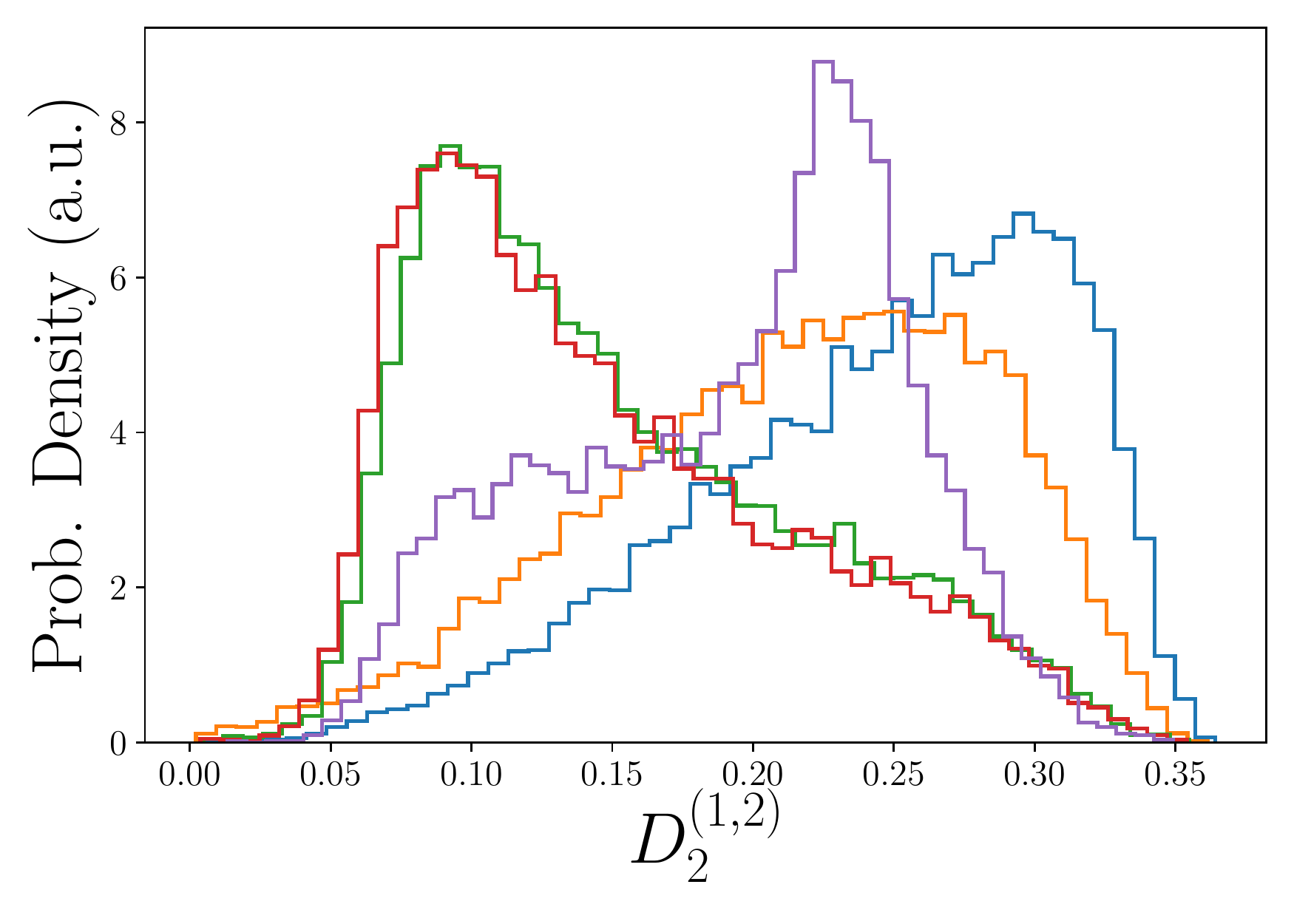}
\includegraphics[width=0.3\textwidth]{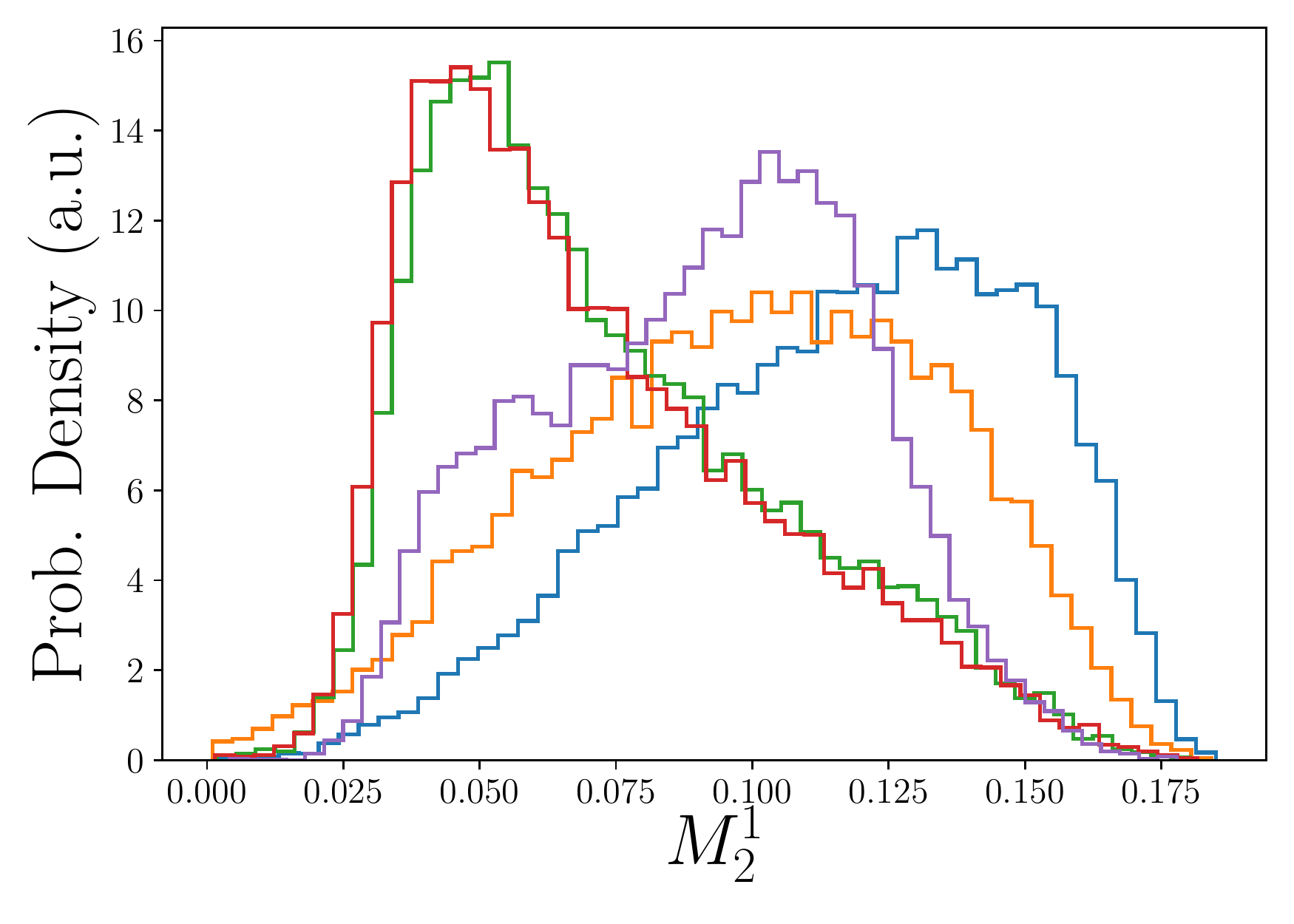}
\includegraphics[width=0.3\textwidth]{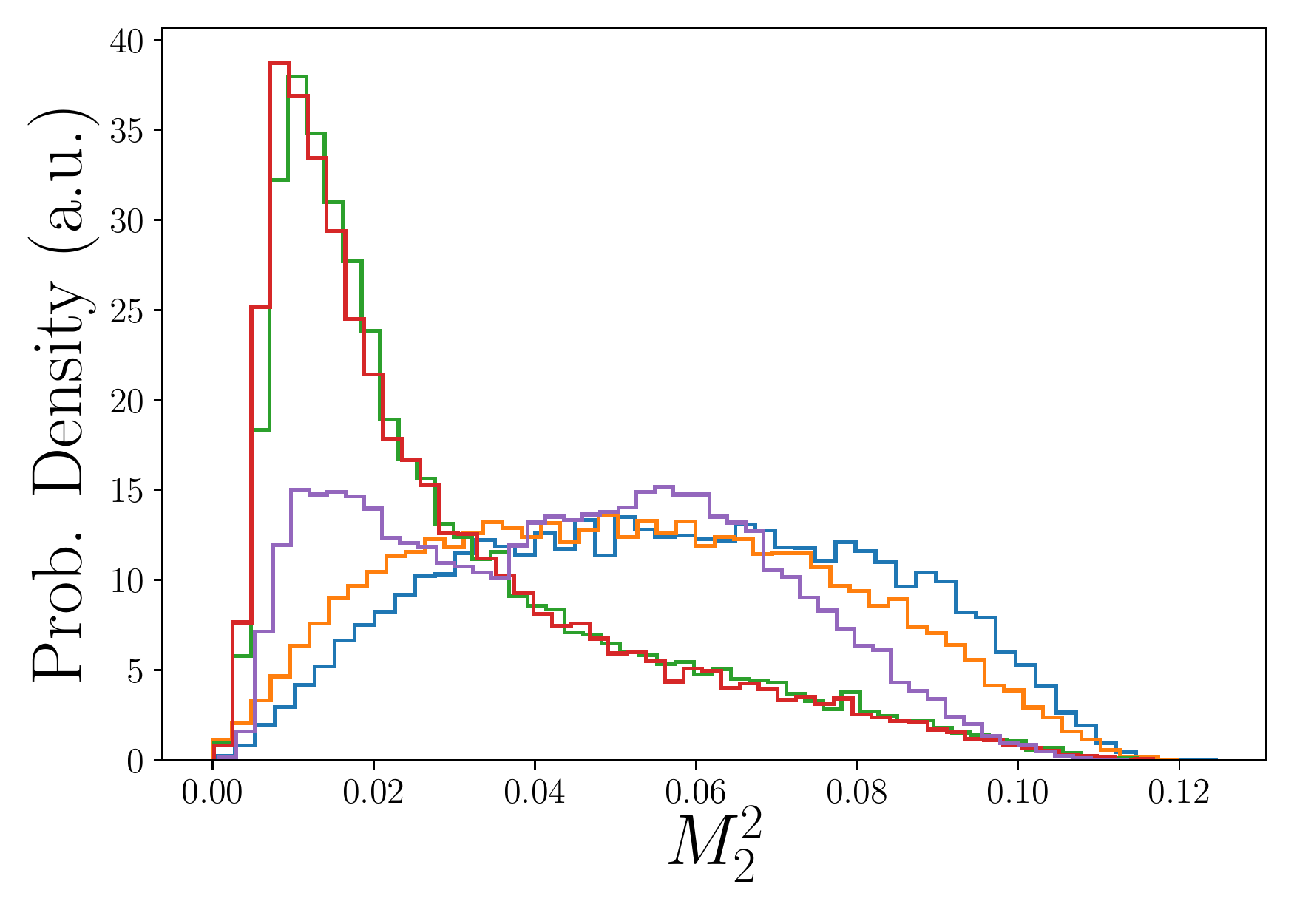}
\\
\includegraphics[width=0.3\textwidth]{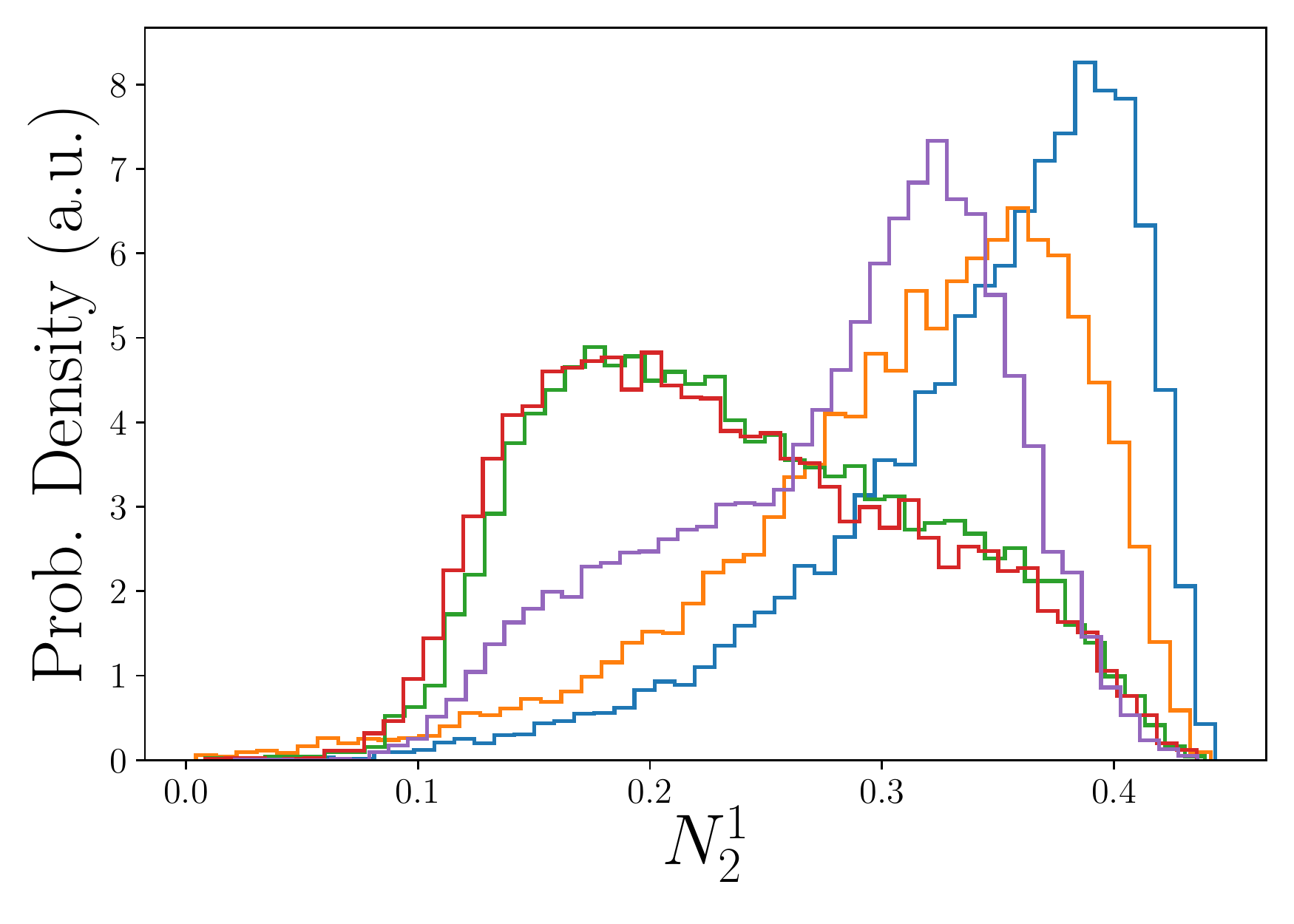}
\includegraphics[width=0.3\textwidth]{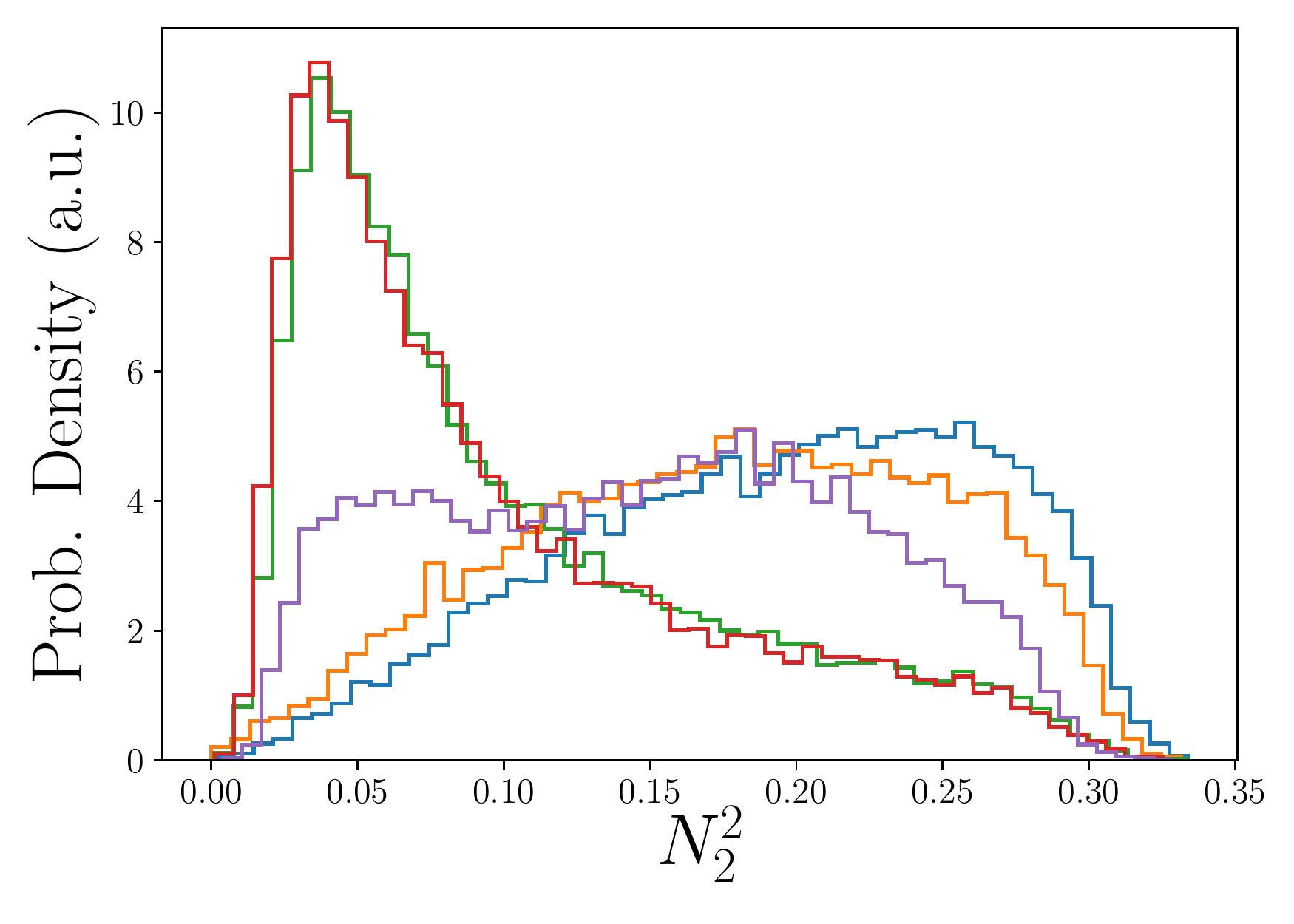}
\\
\includegraphics[width=0.3\textwidth]{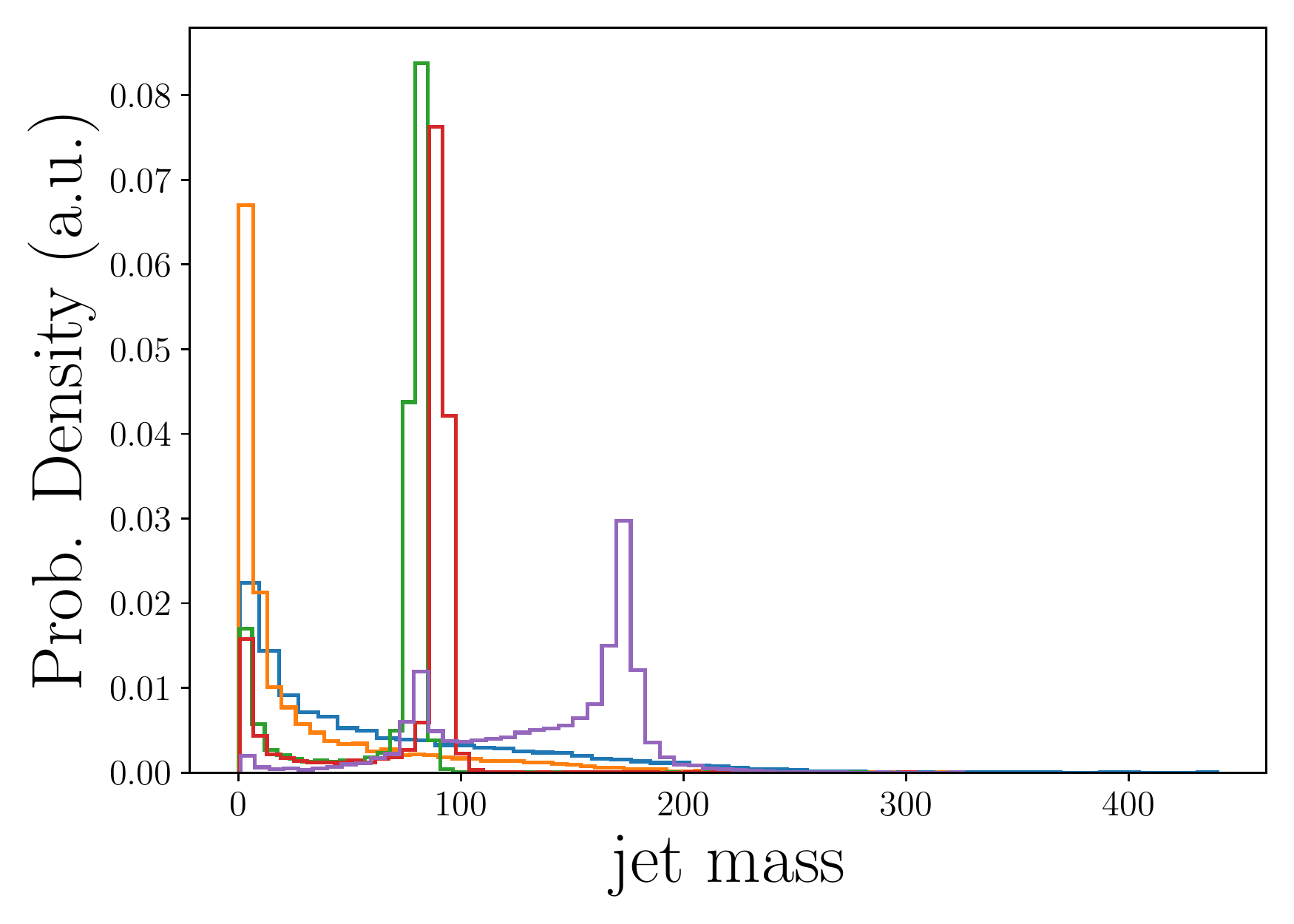}
\includegraphics[width=0.3\textwidth]{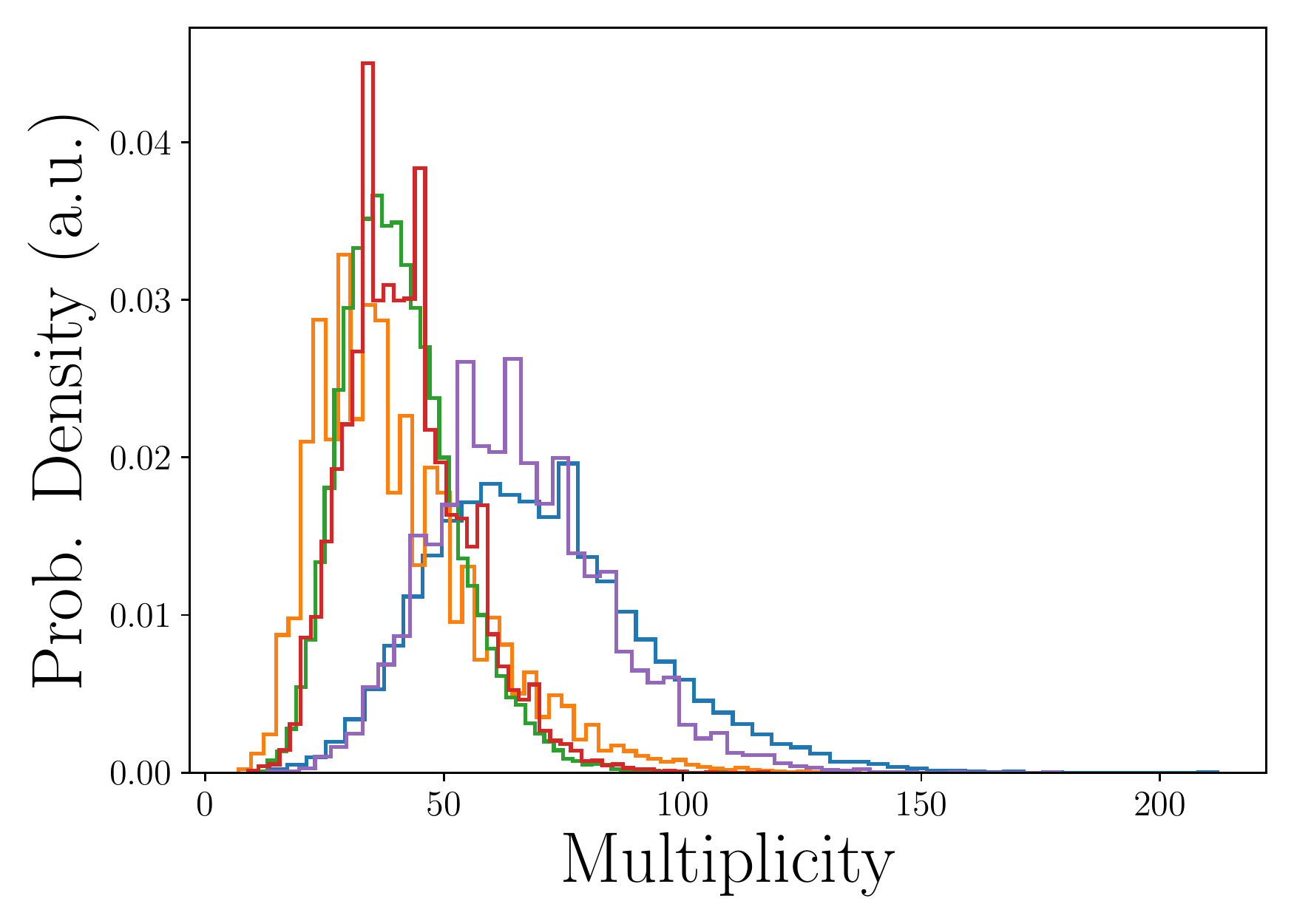}
\caption{Distributions of the 16 high-level features used in this
  study, described in Ref.~\cite{hls4ml}.\label{fig:HLS}}
\end{figure*}

\begin{figure*}[htp!]
\centering
\includegraphics[width=0.32\textwidth]{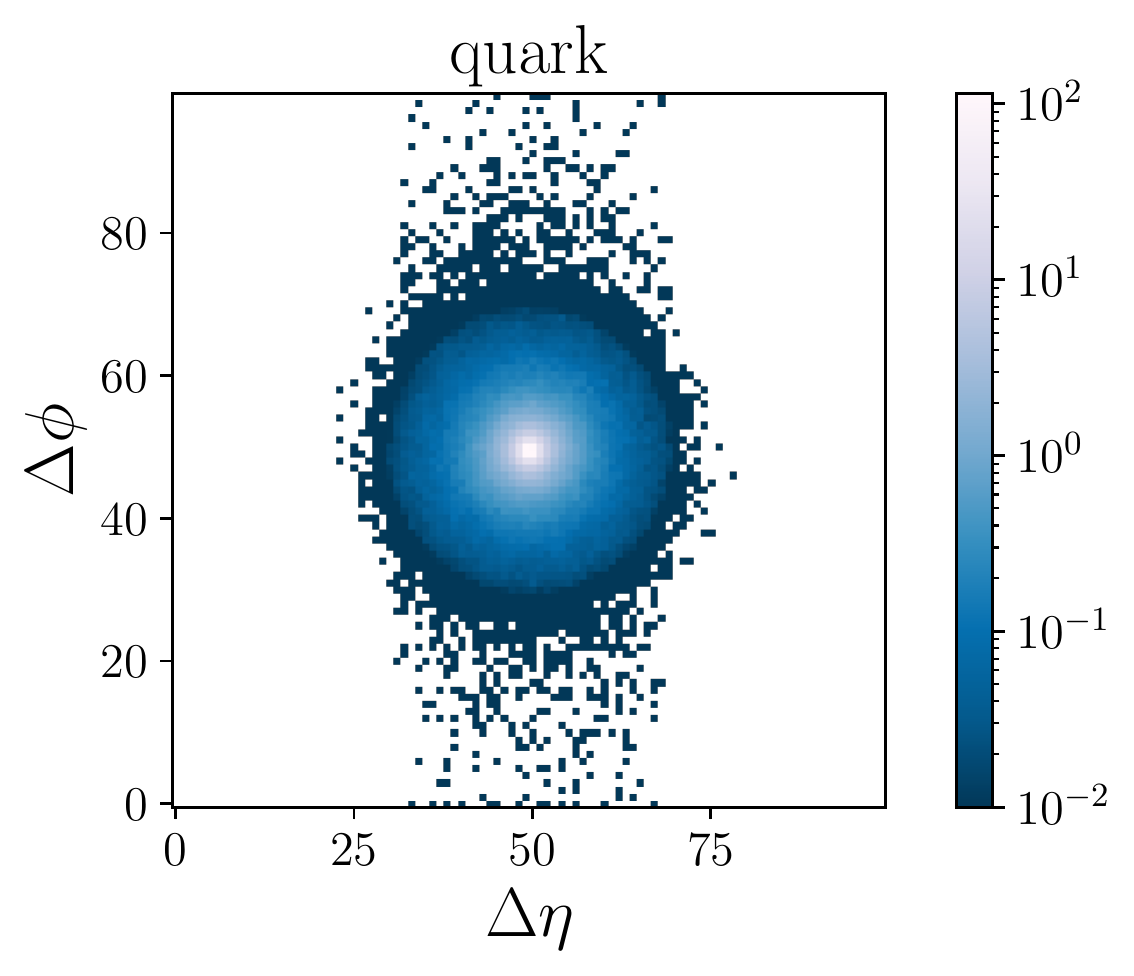}
\includegraphics[width=0.32\textwidth]{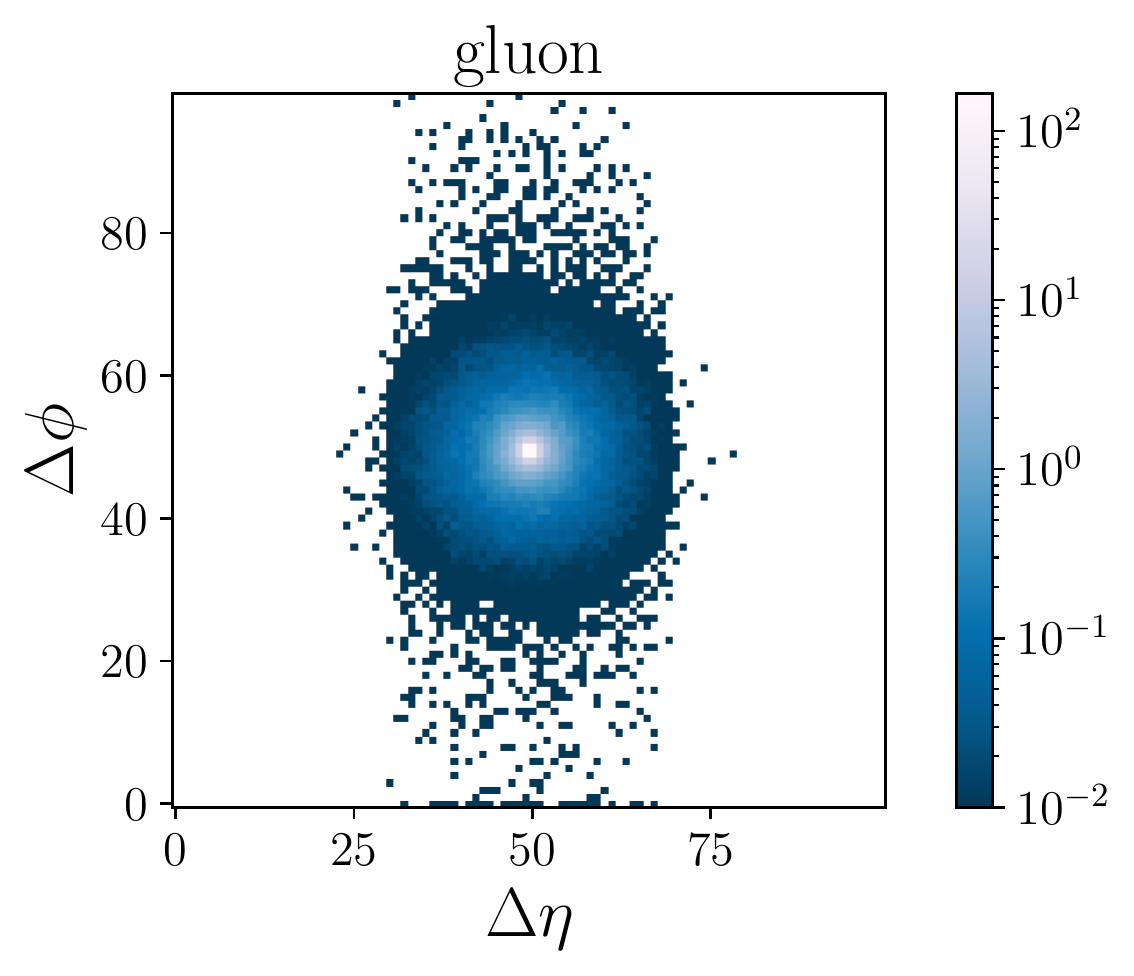} 
\includegraphics[width=0.32\textwidth]{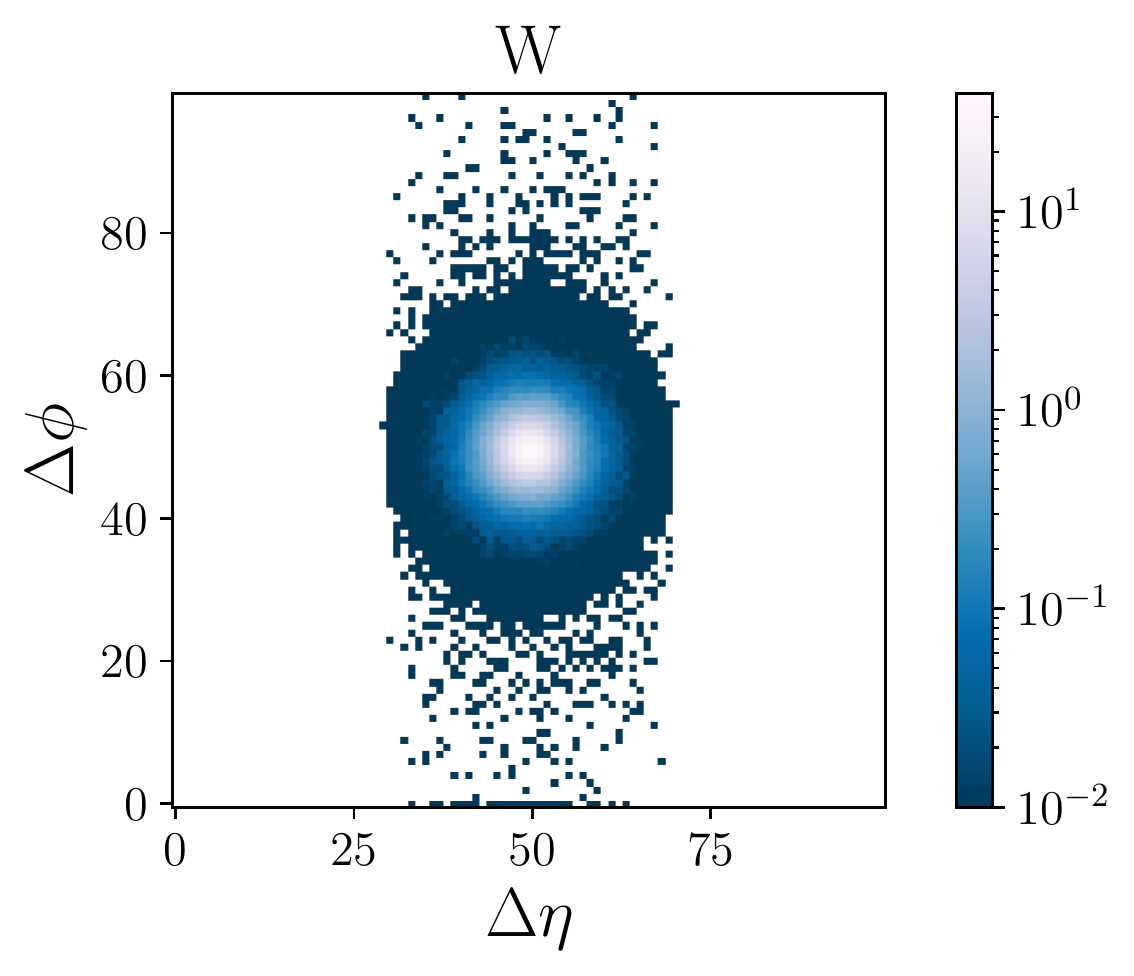} \\
\includegraphics[width=0.32\textwidth]{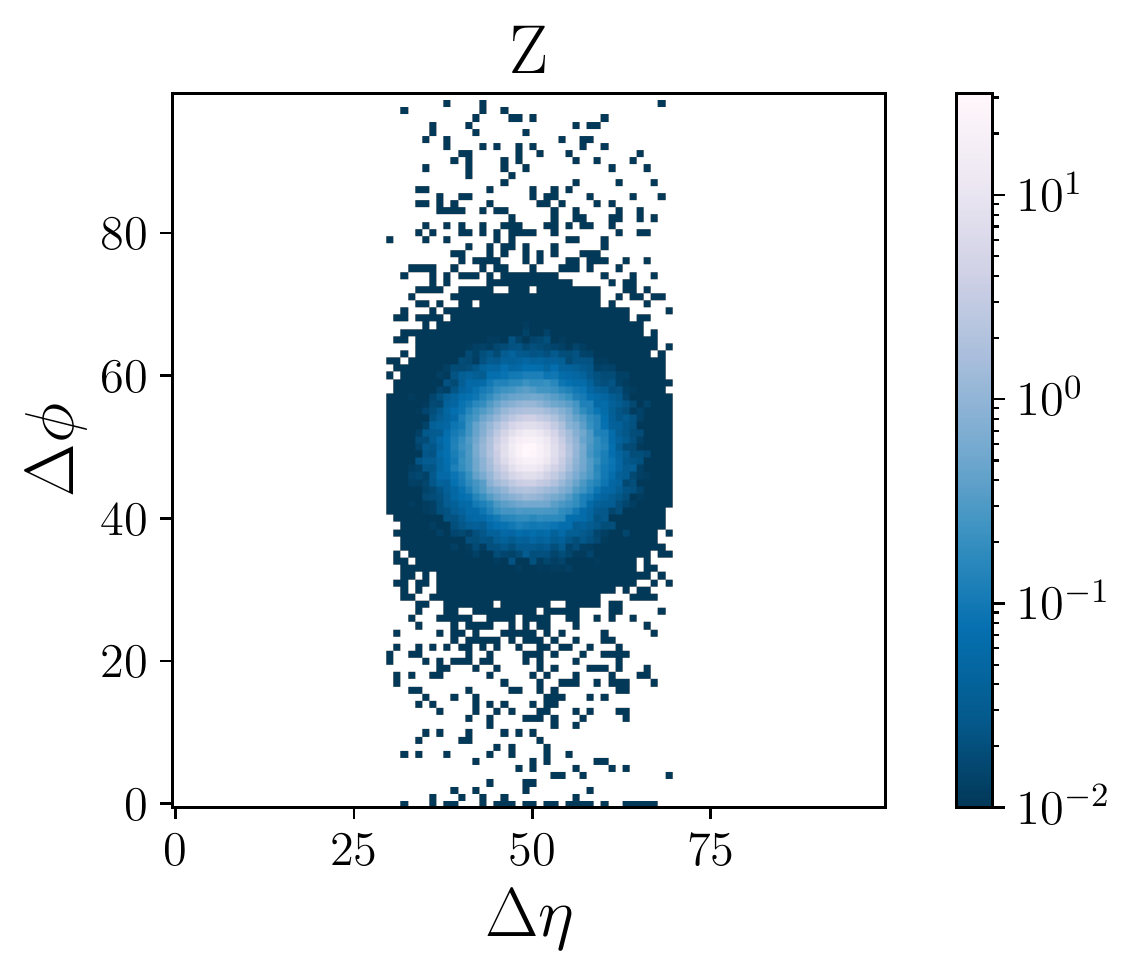} 
\includegraphics[width=0.32\textwidth]{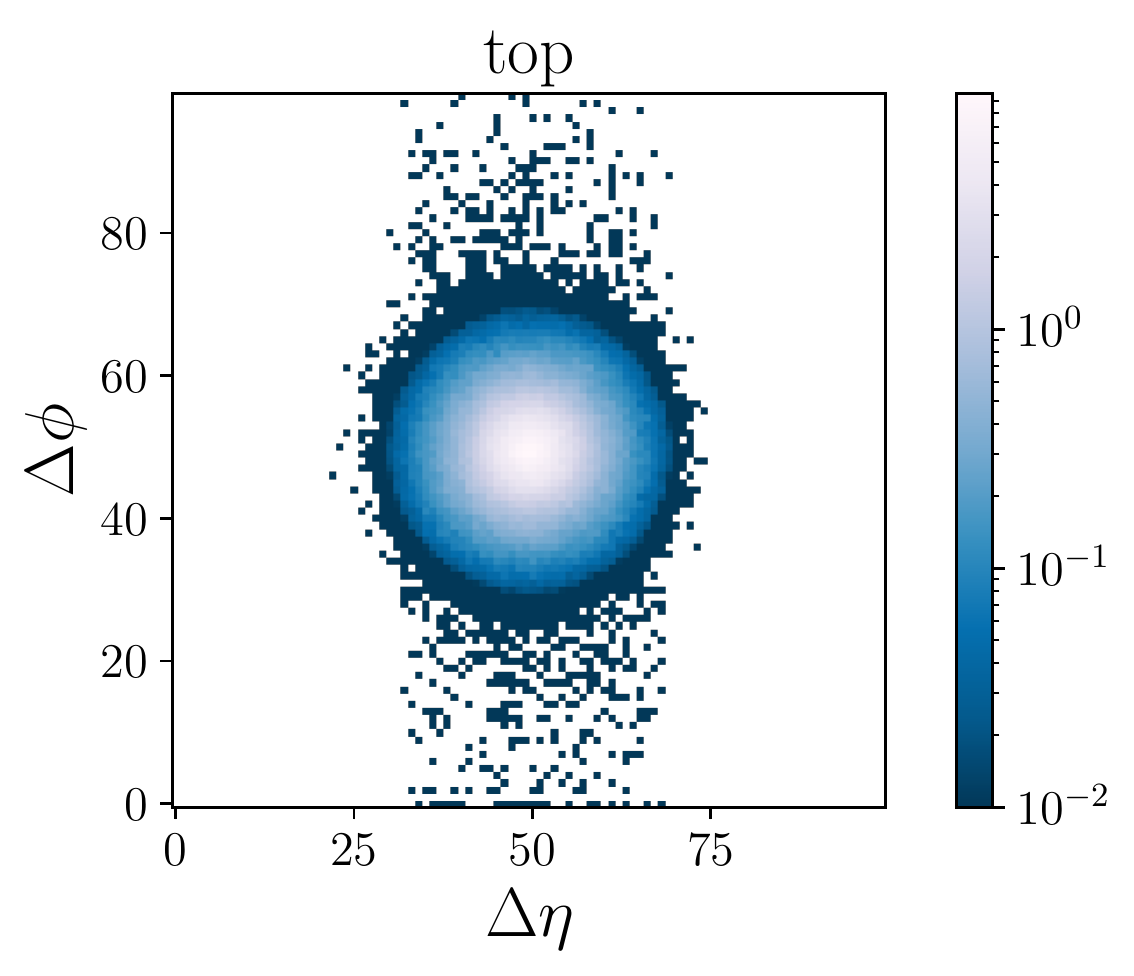}
\caption{Average $100\times100$ images for the five jet classes
  considered in this study: $\PQq$ (top left), $\Pg$ (top center),
  $\PW$ (top right), $\PZ$ (bottom left), and top jets (bottom
  right). The temperature map represents the amount of $\pt$ collected
  in each cell of the image, measured in GeV and computed from the
  scalar sum of the $\pt$ of the particles pointing to each
  cell.\label{fig:jetImage}}
\end{figure*}

\begin{figure*}[ht!]
\centering
\includegraphics[width=0.32\textwidth]{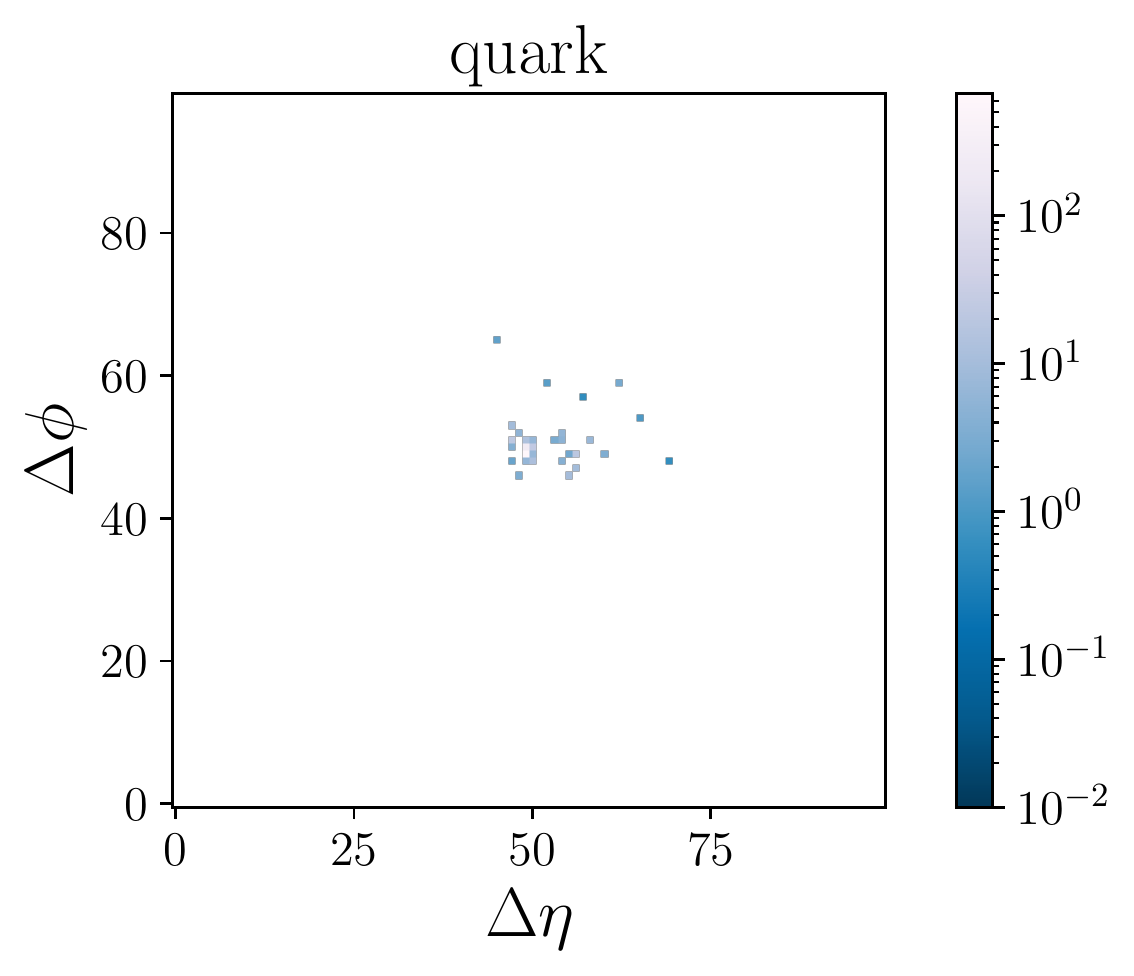} 
\includegraphics[width=0.32\textwidth]{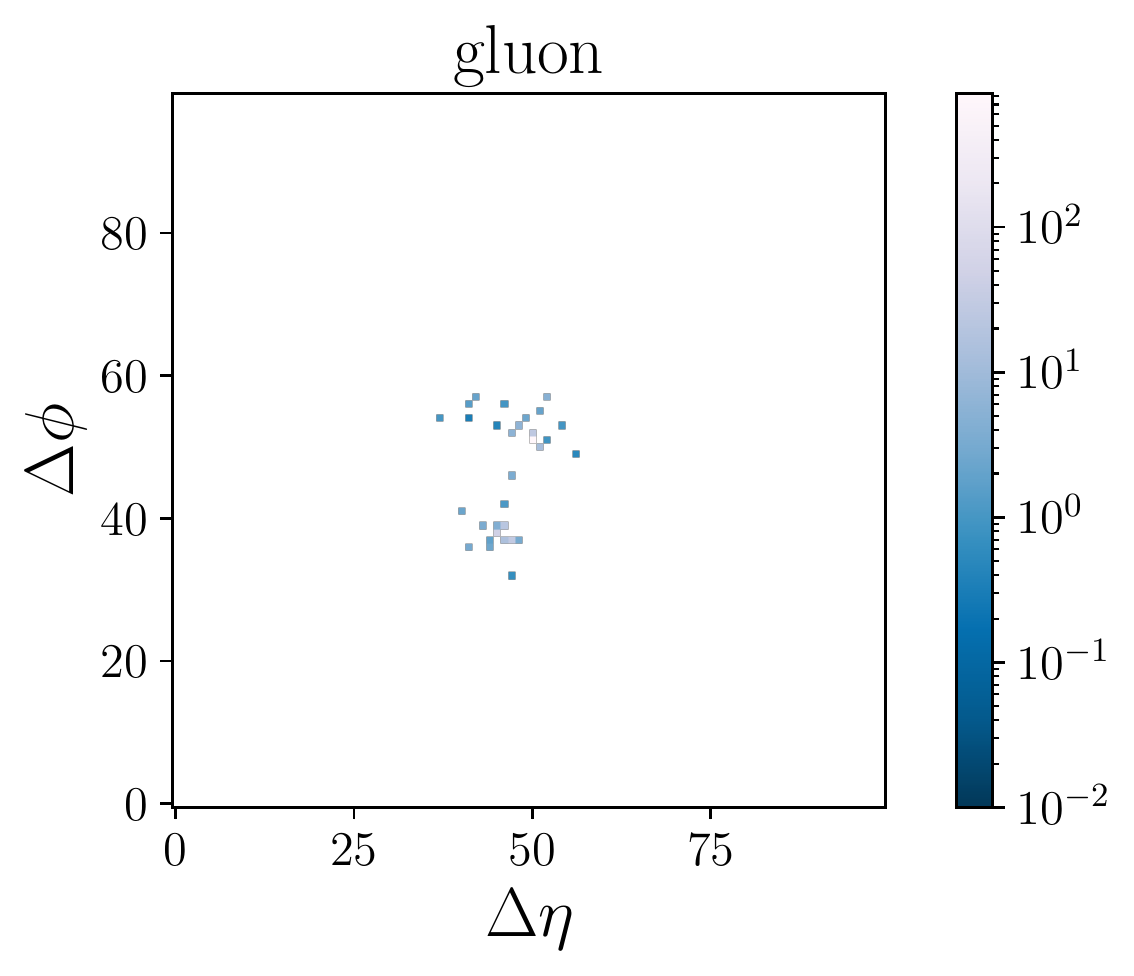} 
\includegraphics[width=0.32\textwidth]{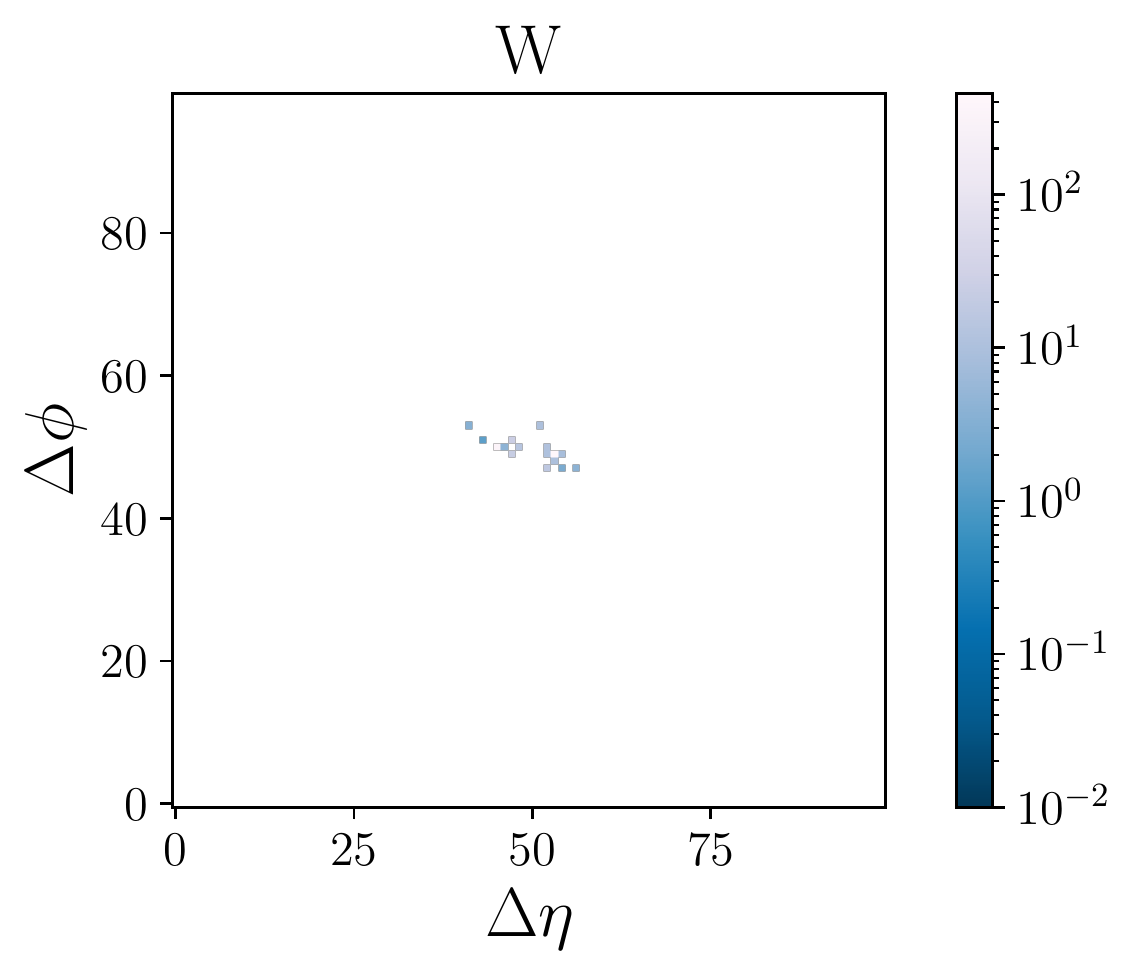} \\
\includegraphics[width=0.32\textwidth]{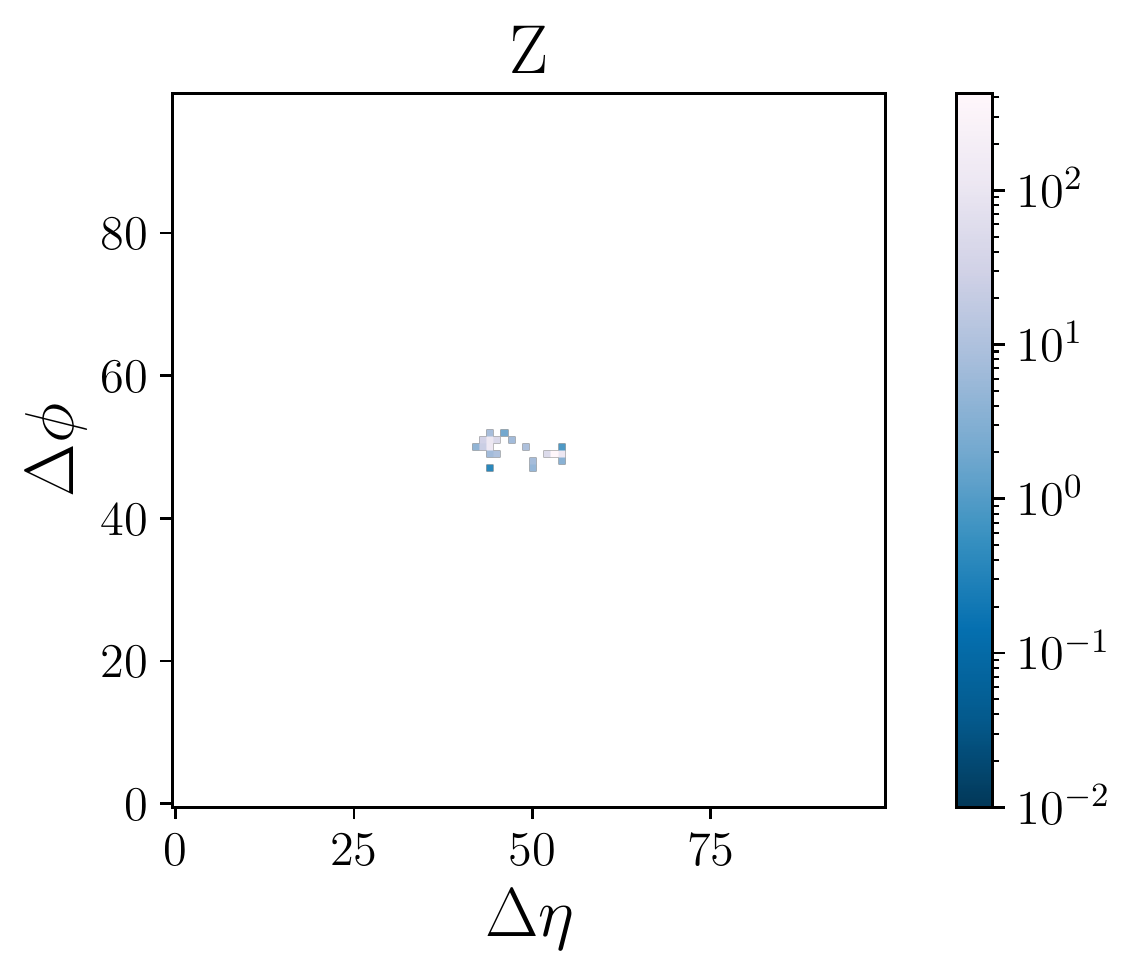} 
\includegraphics[width=0.32\textwidth]{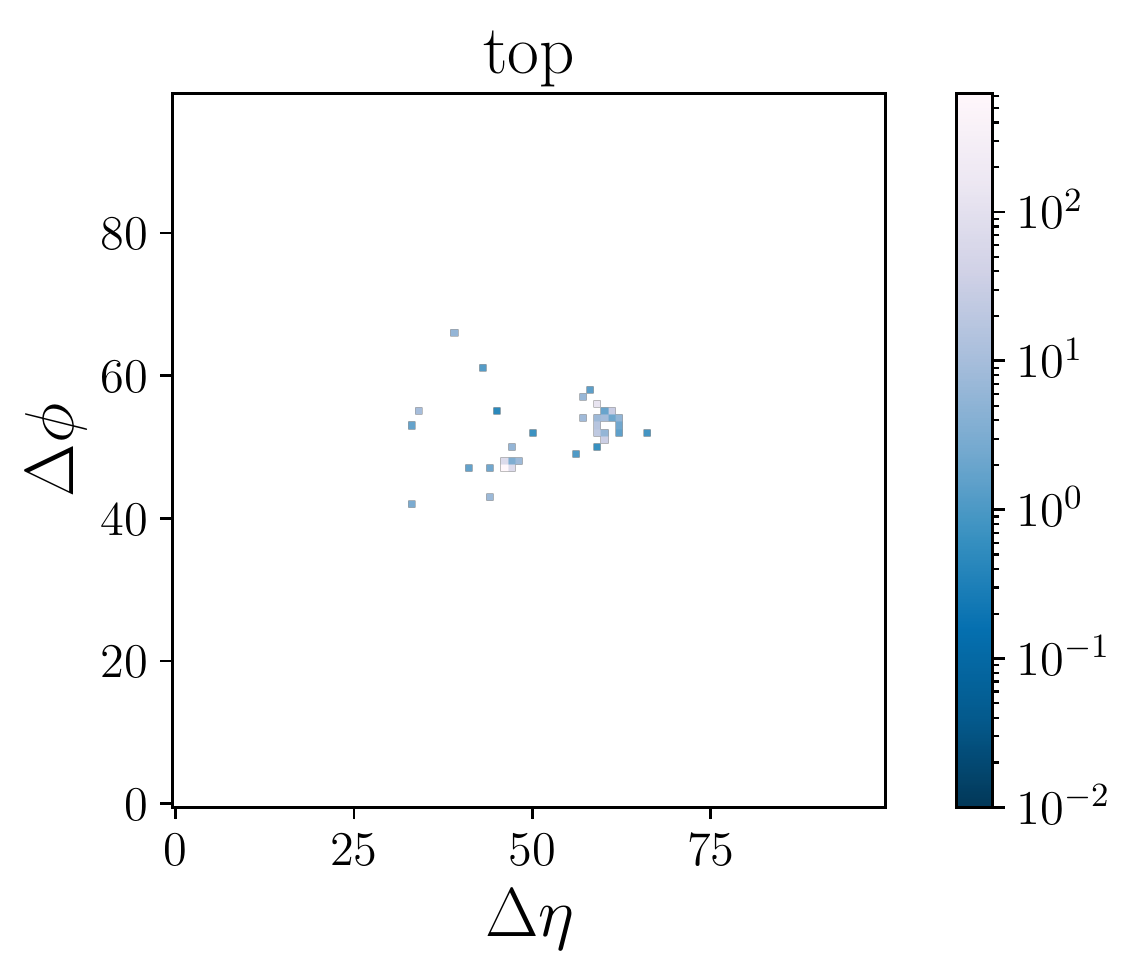} 
\caption{Example of $100\times100$ images for the five jet classes
  considered in this study: $\PQq$ (top-left), $\Pg$ (top-right),
  $\PW$ (center-left), $\PZ$ (center-right), and top jets (bottom). The
  temperature map represents the amount of $\pt$ collected in each
  cell of the image, measured in GeV and computed from the scalar sum
  of the $\pt$ of the particles pointing to each
  cell.\label{fig:OnejetImage}}
\end{figure*}

\begin{figure*}[htp!]
\centering
\includegraphics[width=0.3\textwidth]{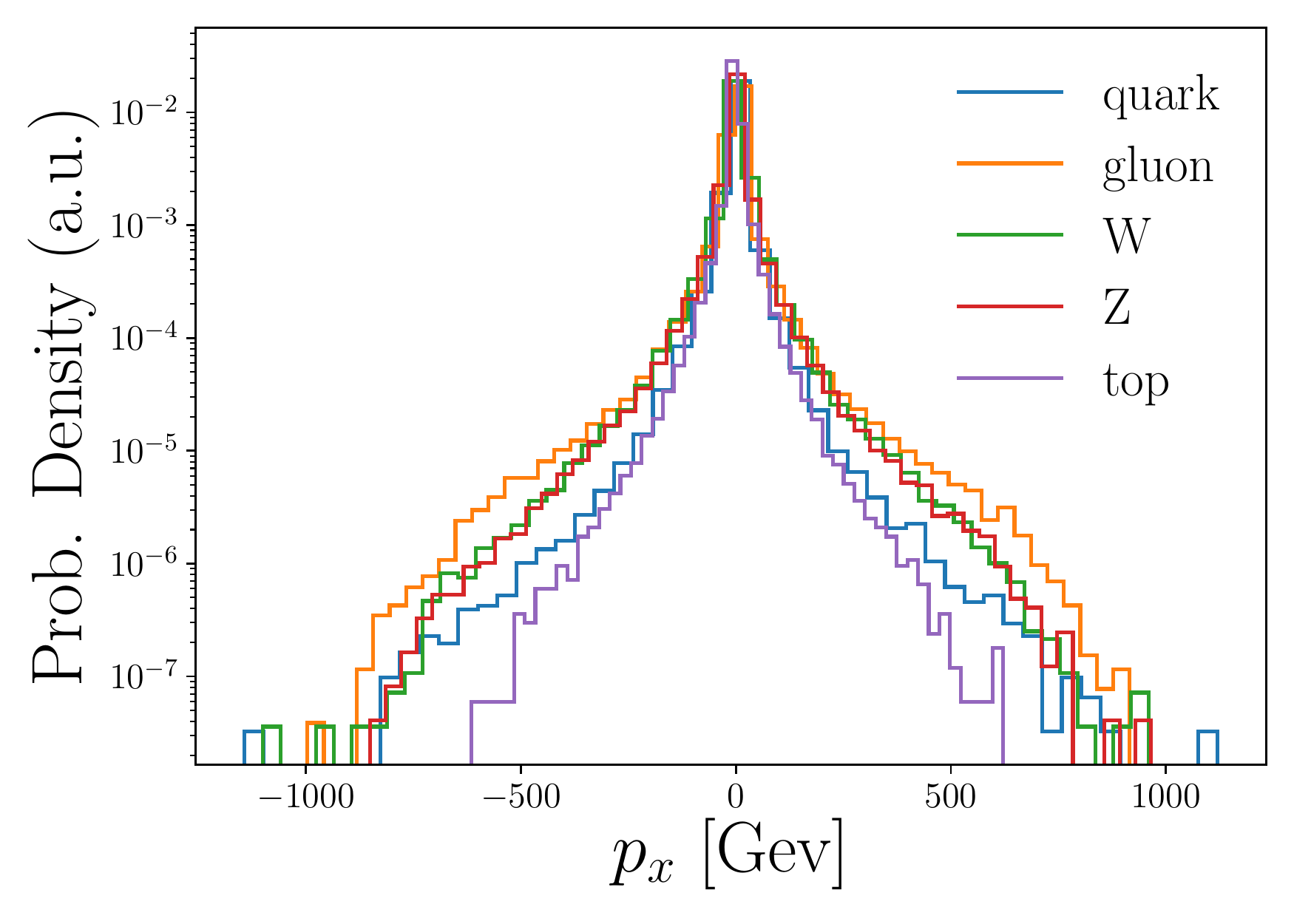} 
\includegraphics[width=0.3\textwidth]{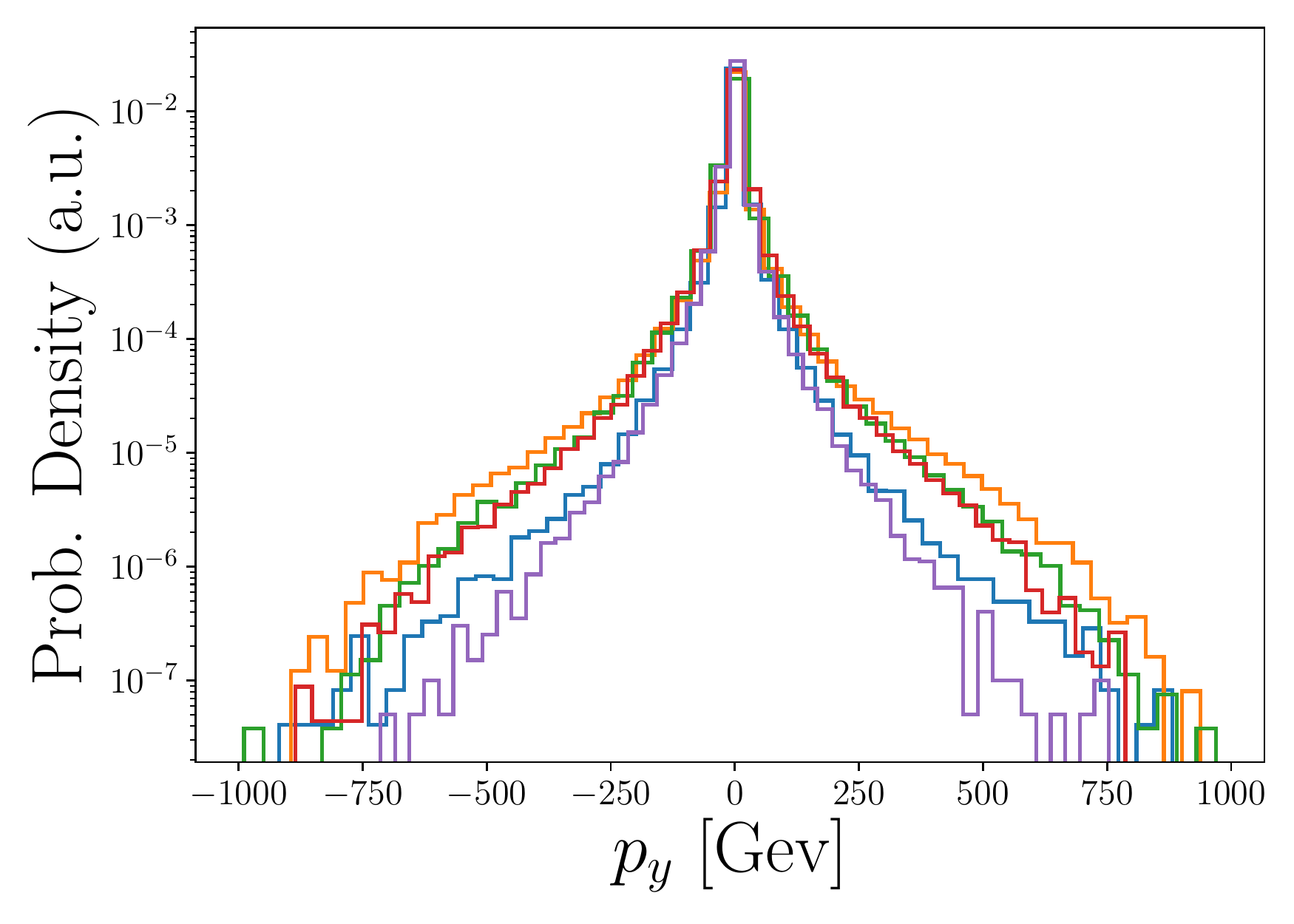} 
\includegraphics[width=0.3\textwidth]{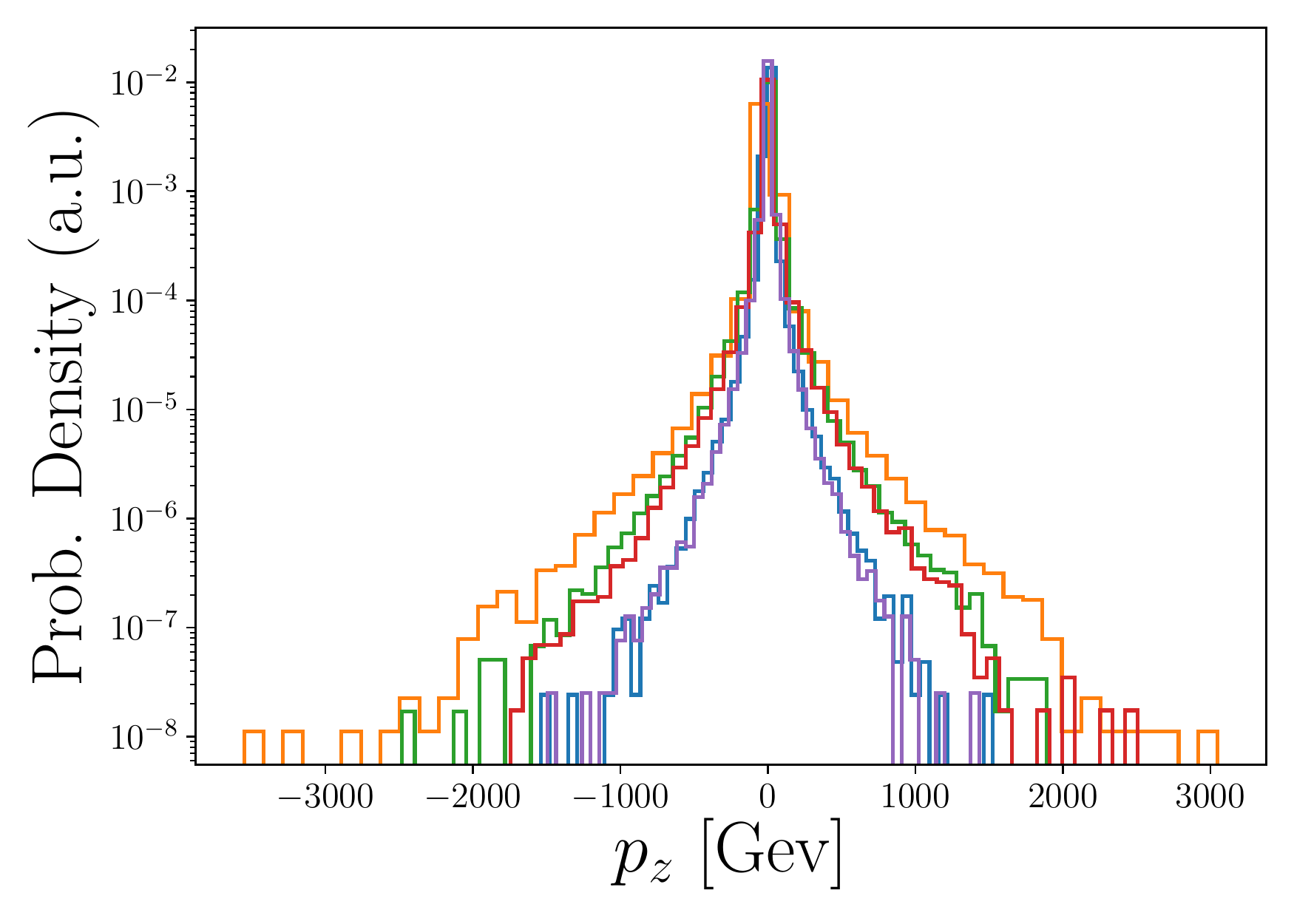} 
\\
\includegraphics[width=0.3\textwidth]{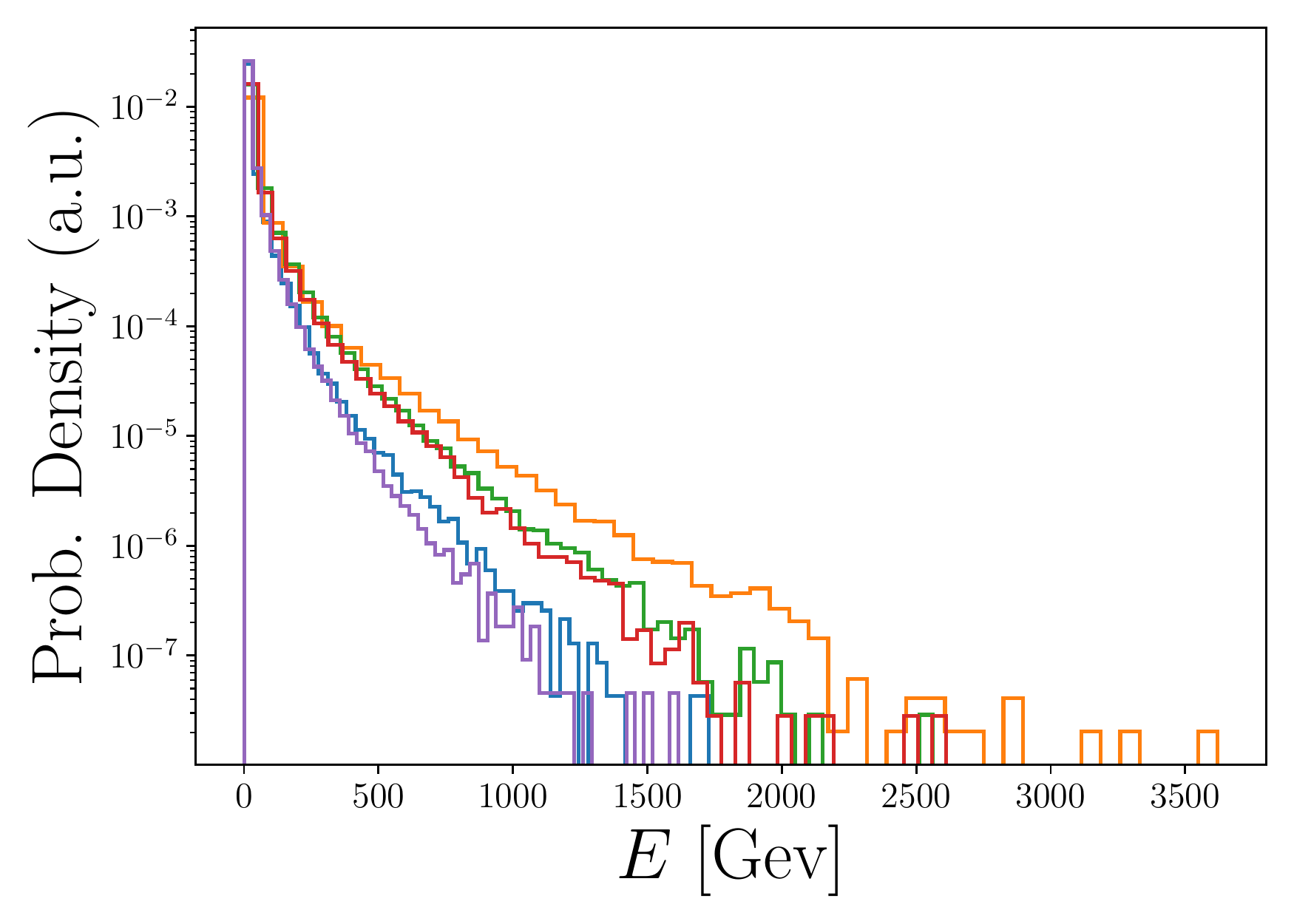} 
\includegraphics[width=0.3\textwidth]{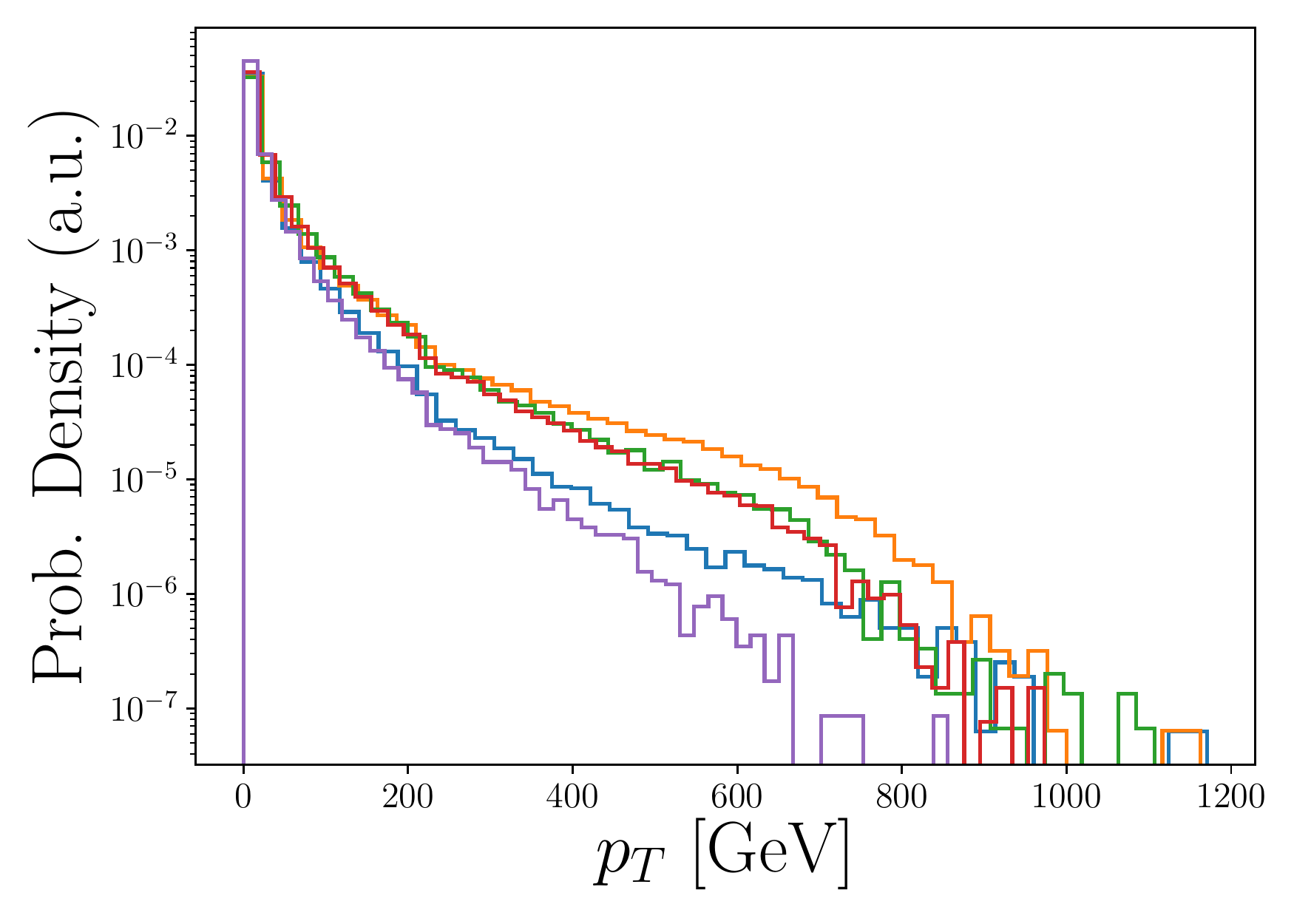} 
\includegraphics[width=0.3\textwidth]{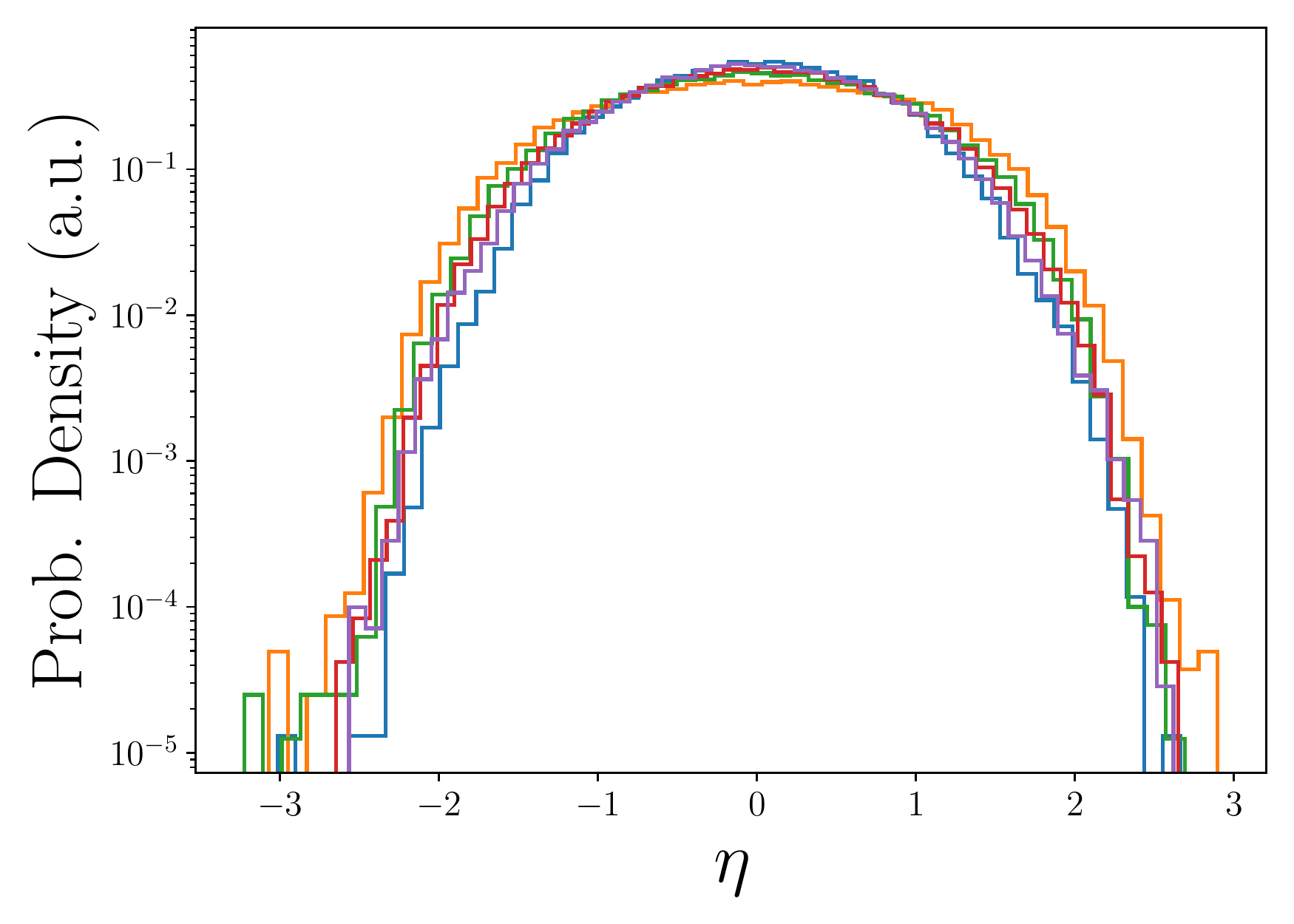} 
\\
\includegraphics[width=0.3\textwidth]{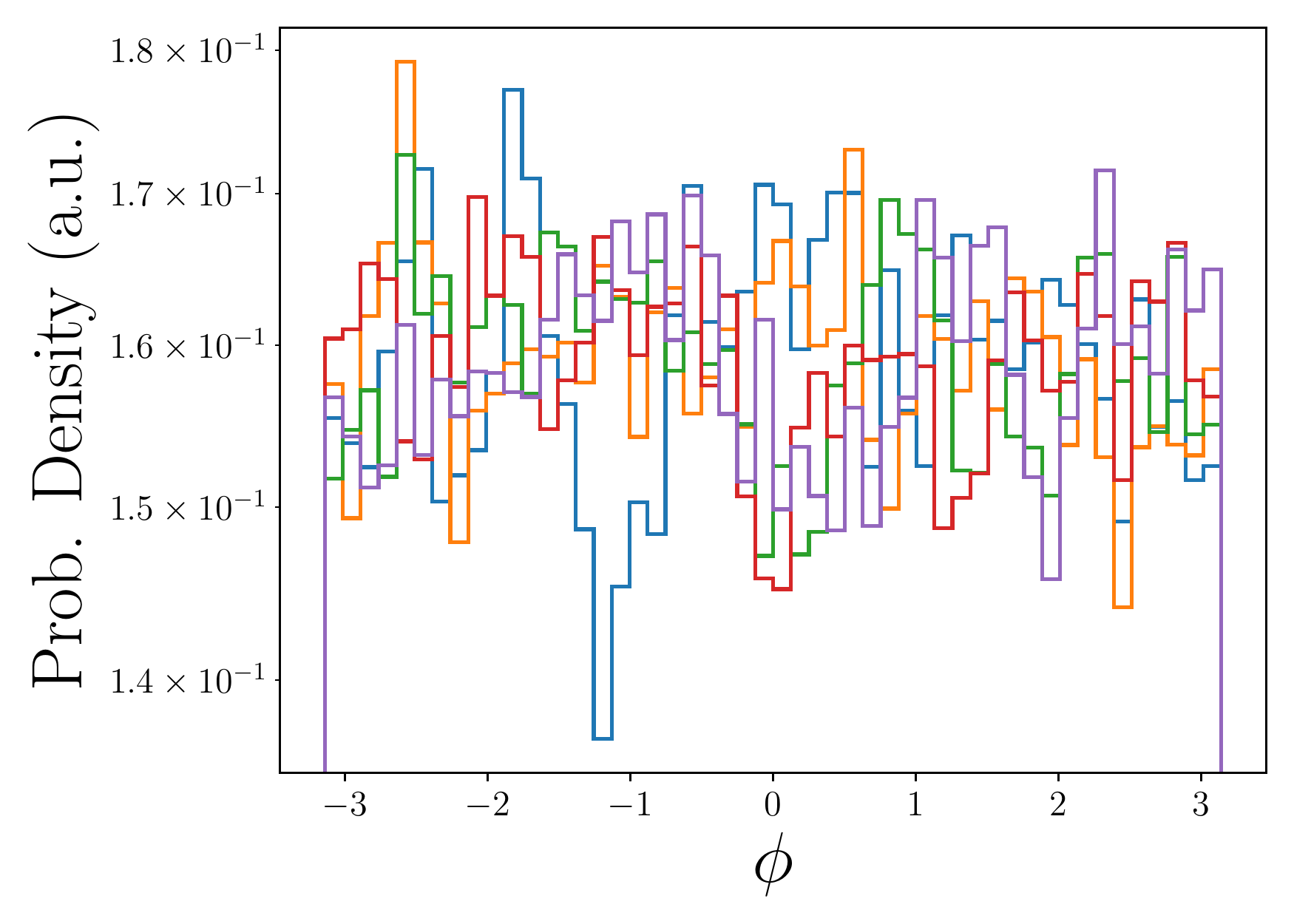} 
\includegraphics[width=0.3\textwidth]{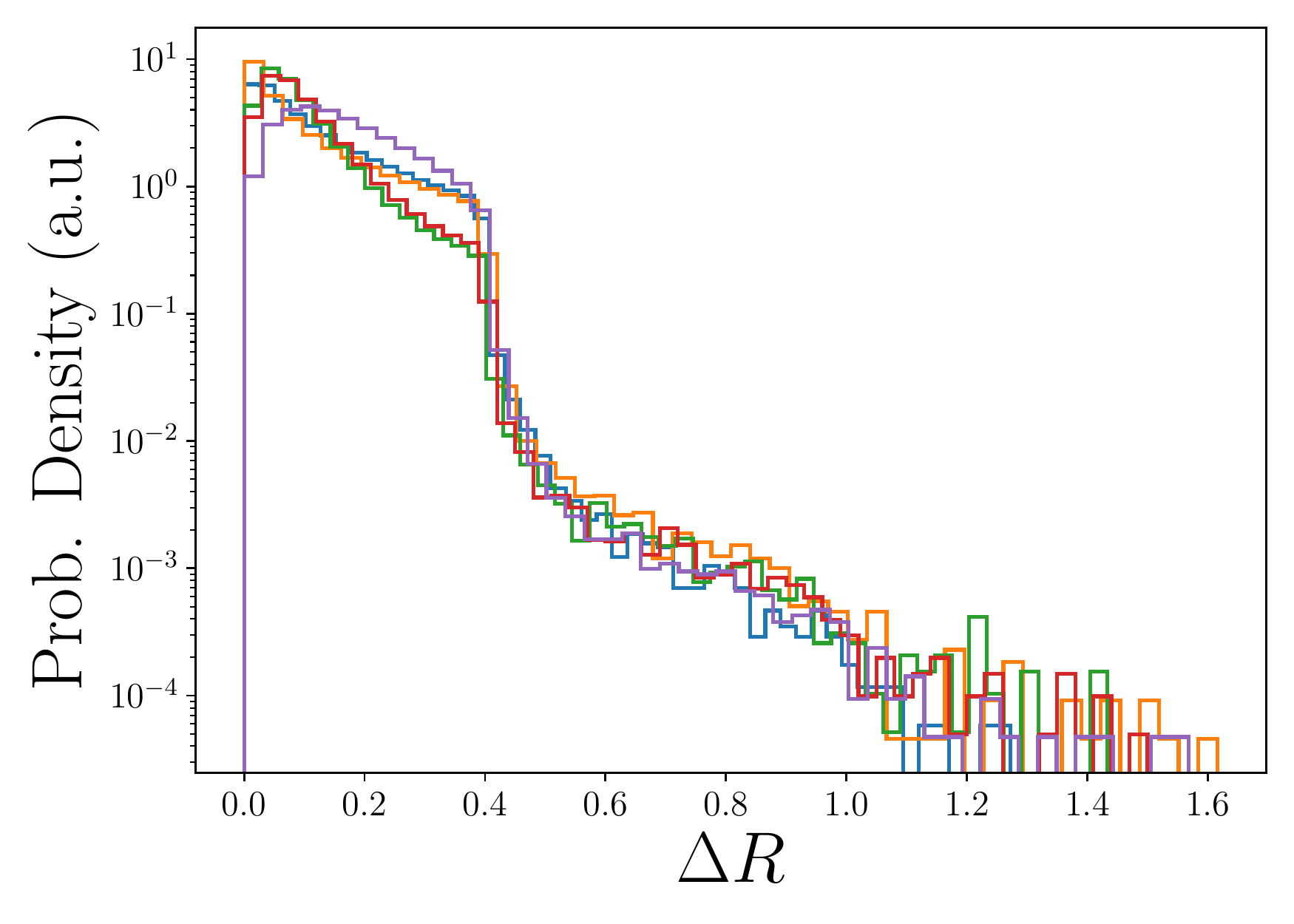}
\includegraphics[width=0.3\textwidth]{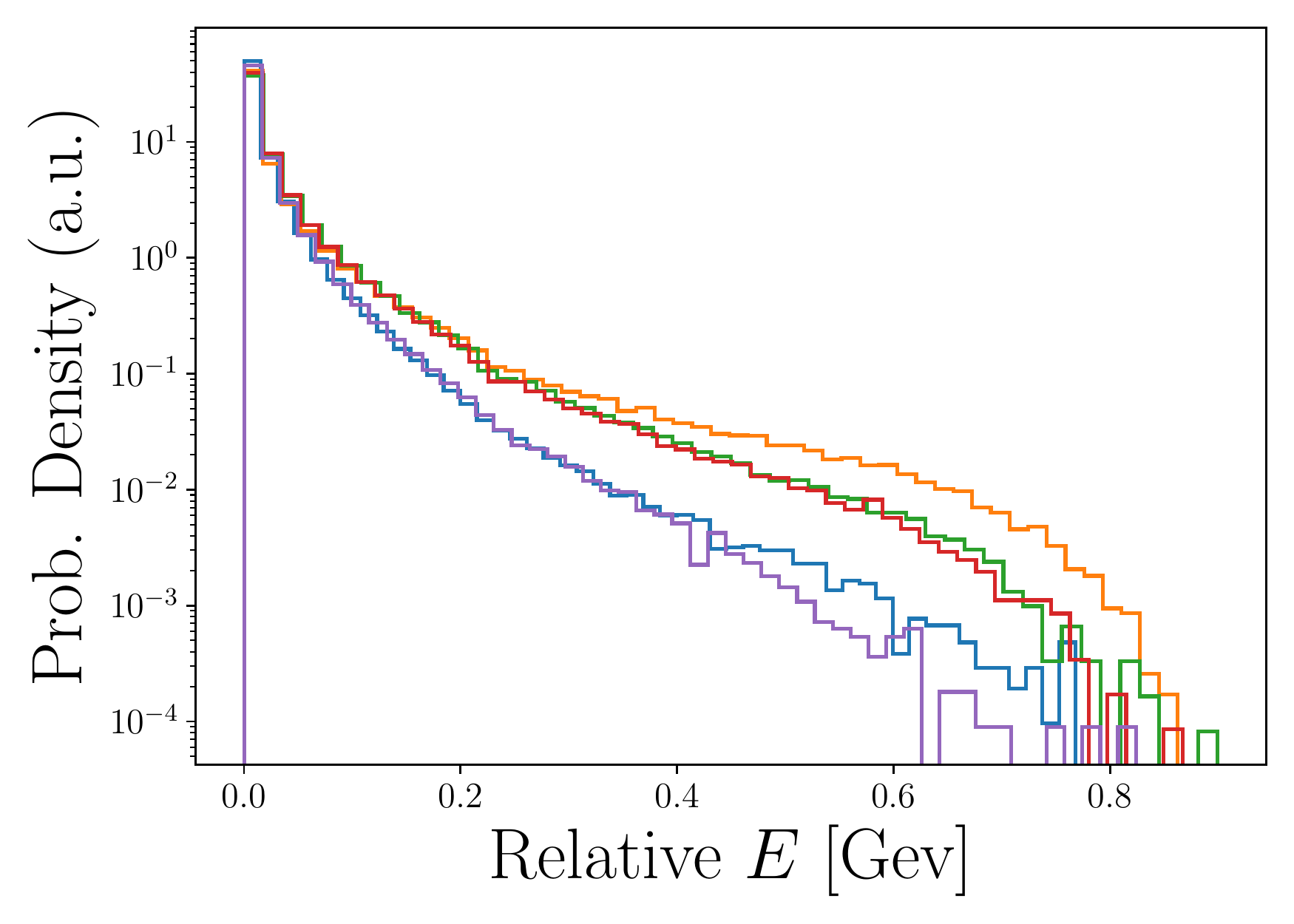} 
\\
\includegraphics[width=0.3\textwidth]{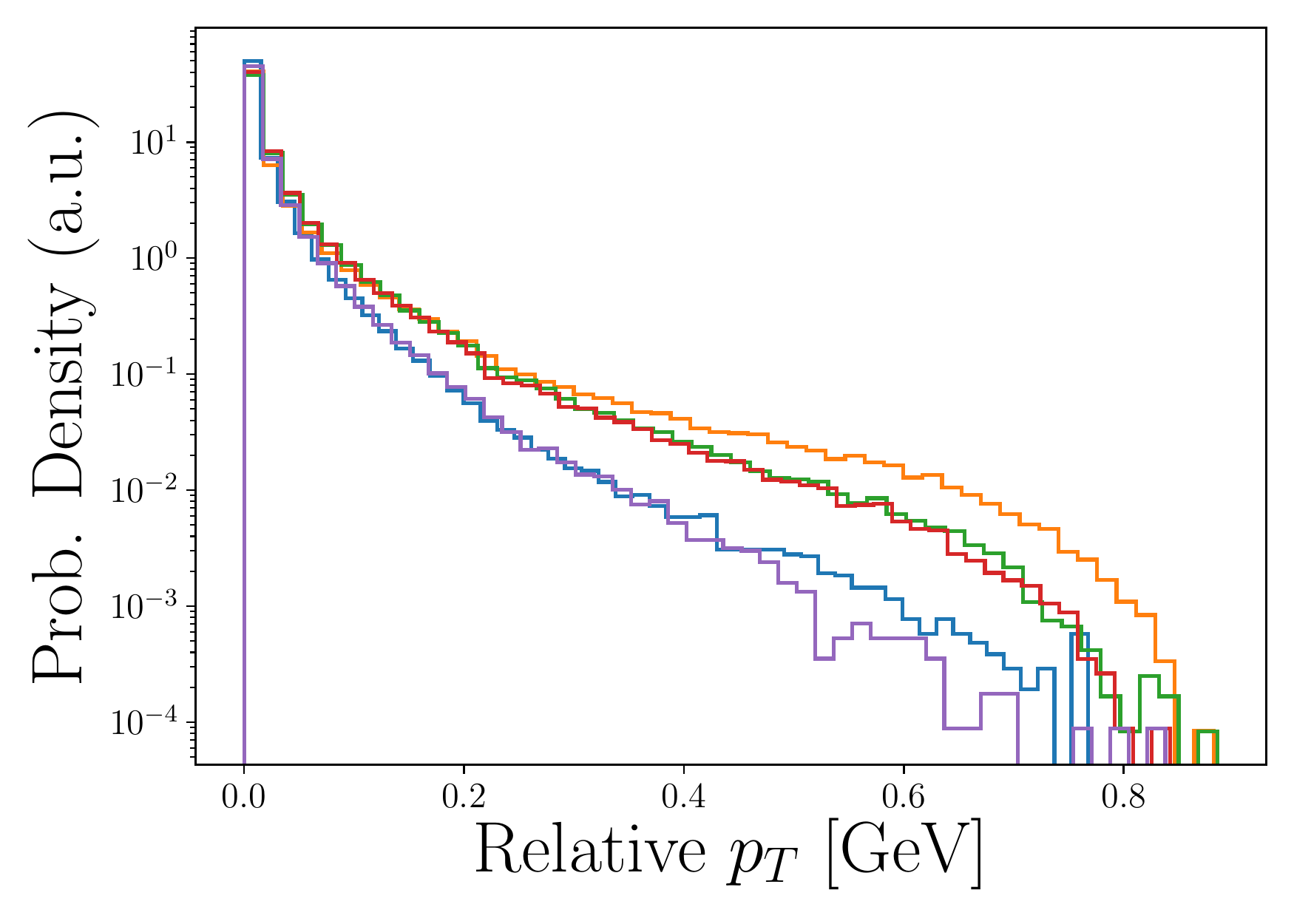} 
\includegraphics[width=0.3\textwidth]{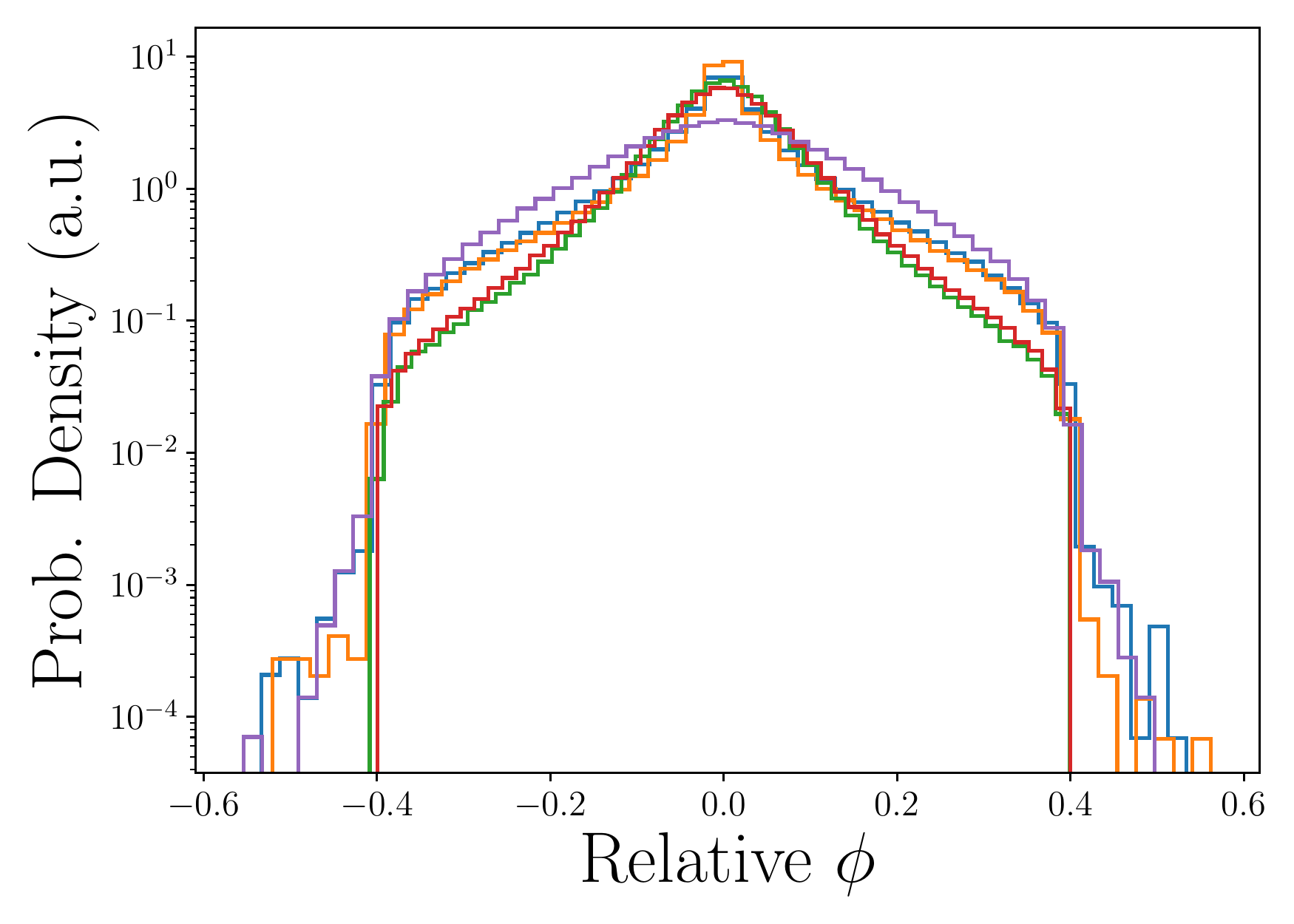} 
\includegraphics[width=0.3\textwidth]{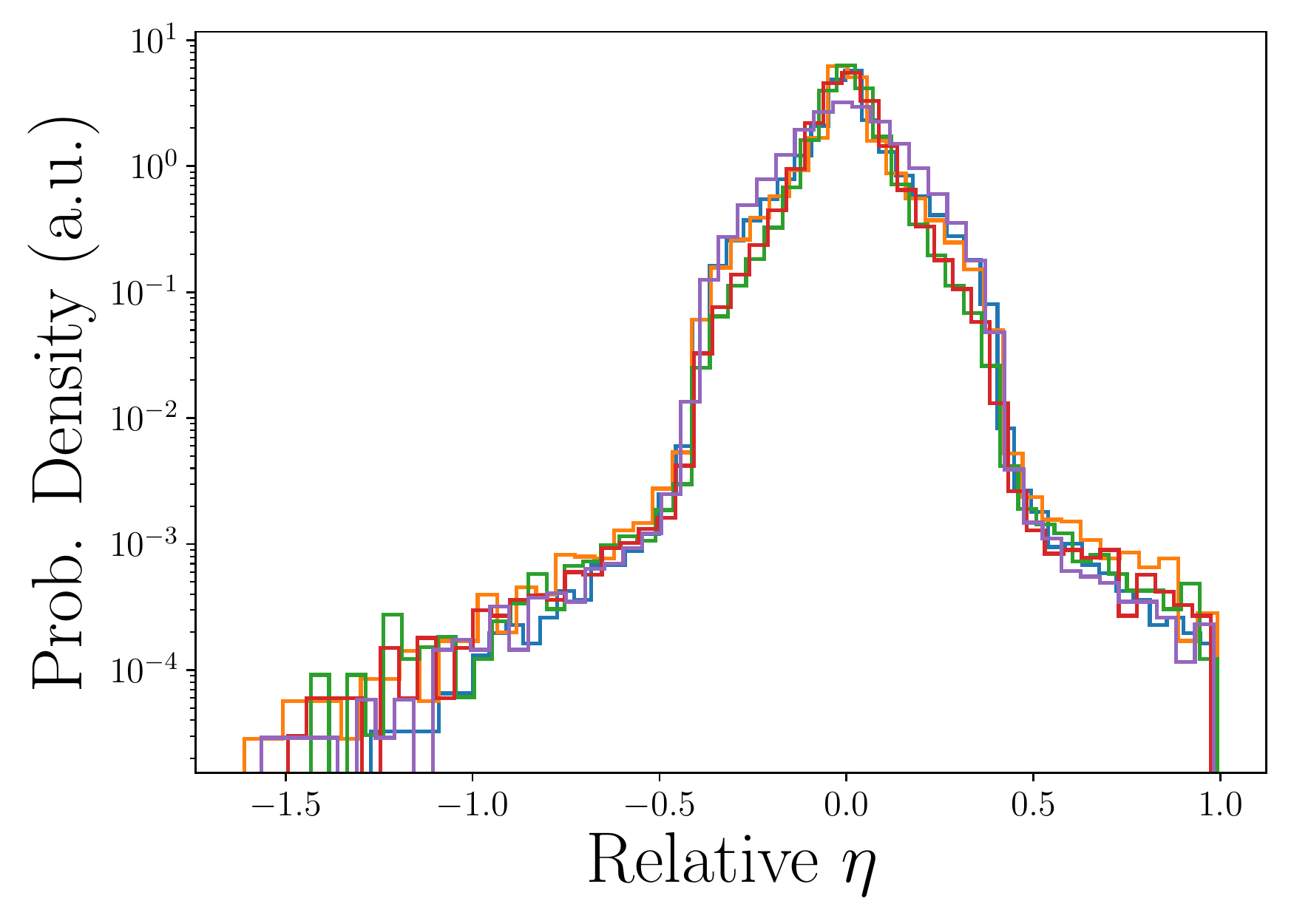} 
\\
\includegraphics[width=0.3\textwidth]{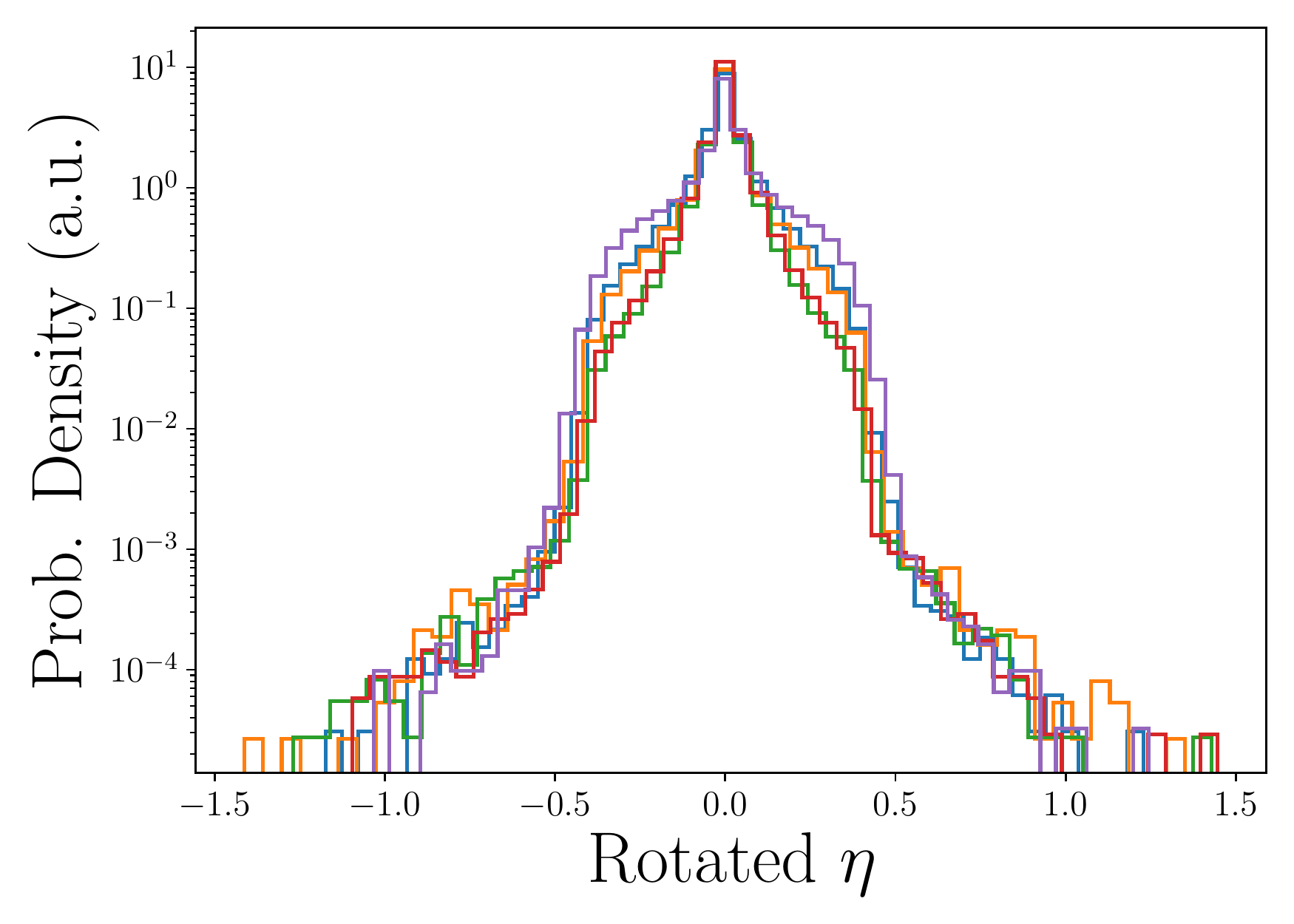} 
\includegraphics[width=0.3\textwidth]{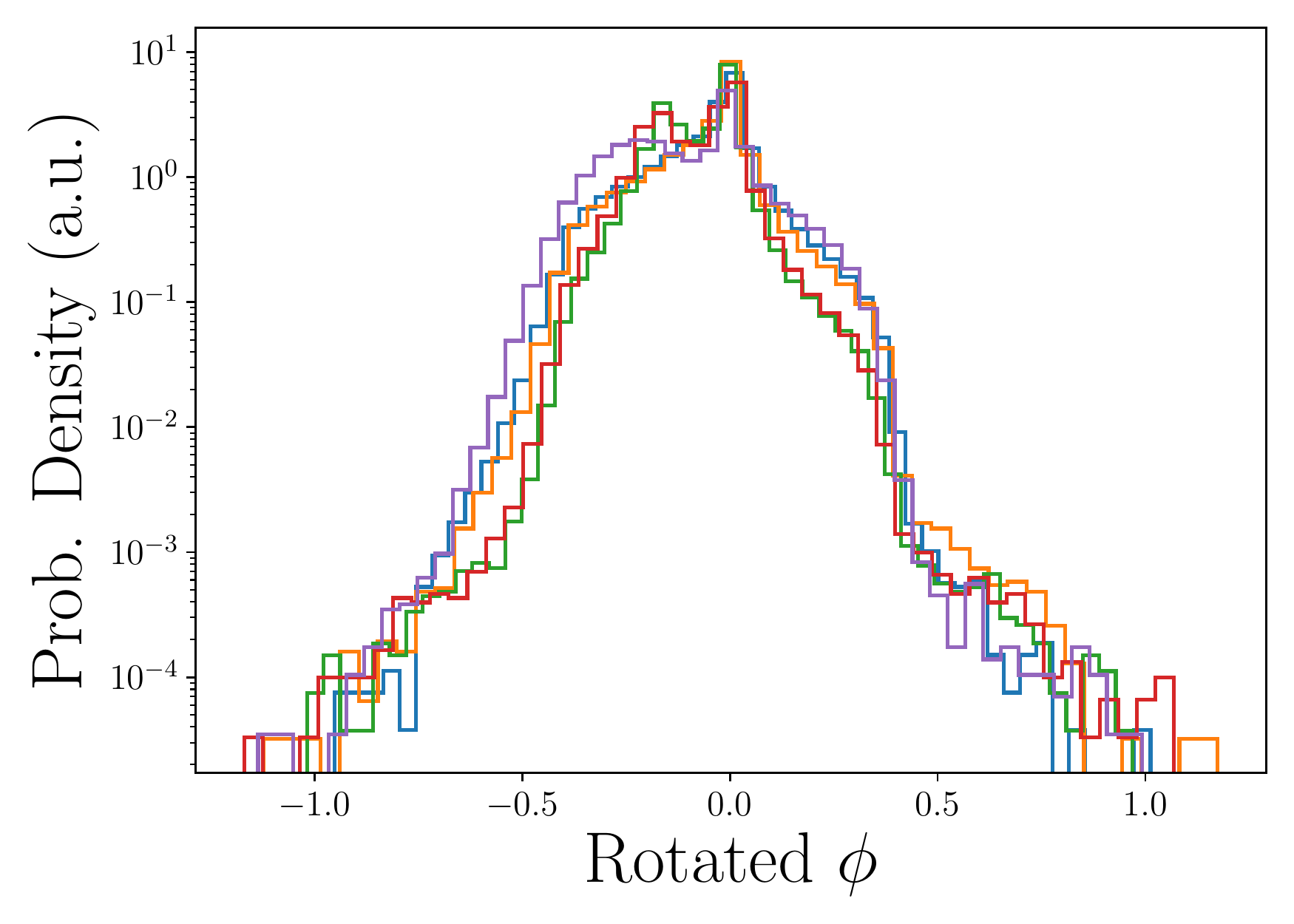} 
\\
\includegraphics[width=0.3\textwidth]{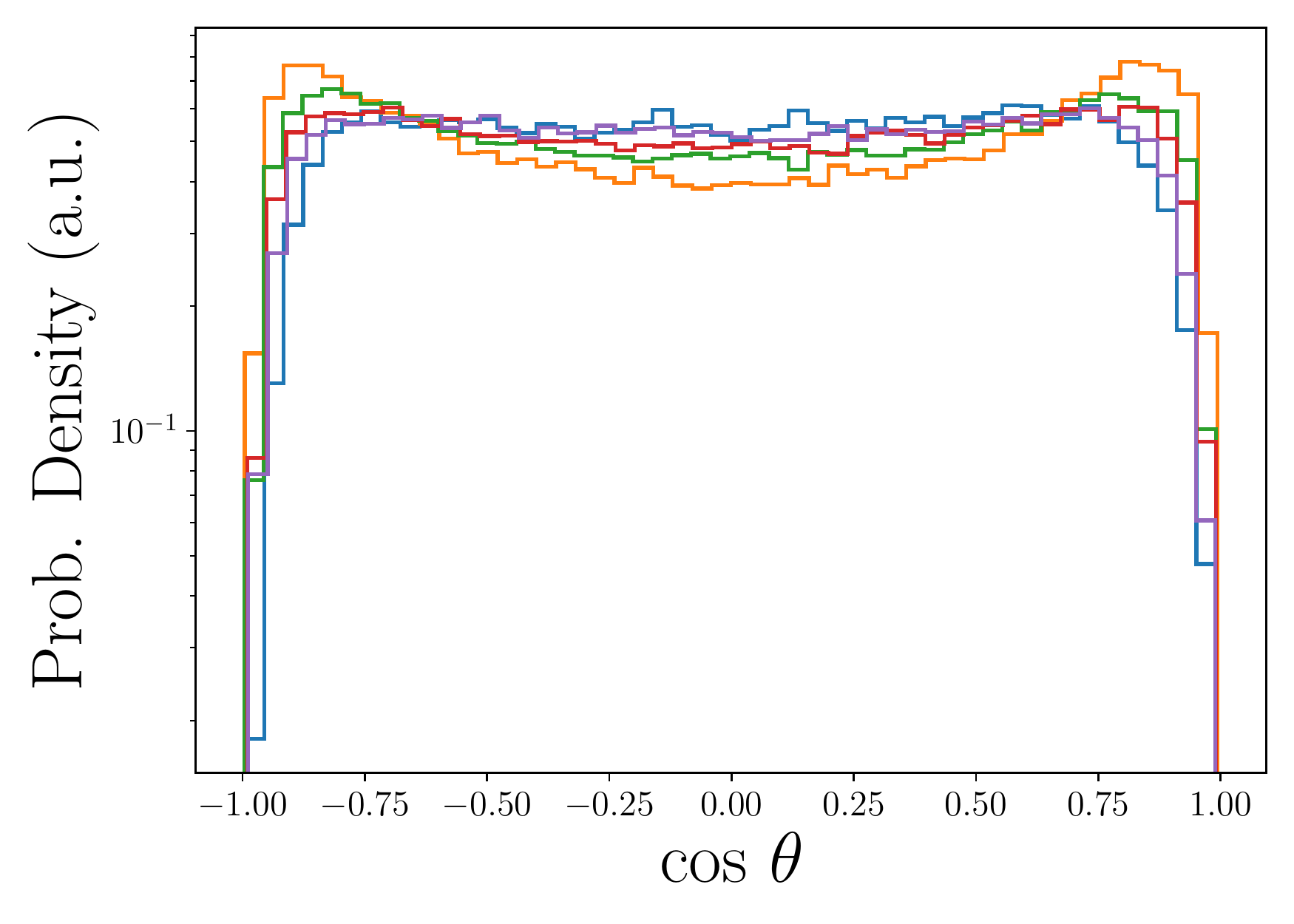} 
\includegraphics[width=0.3\textwidth]{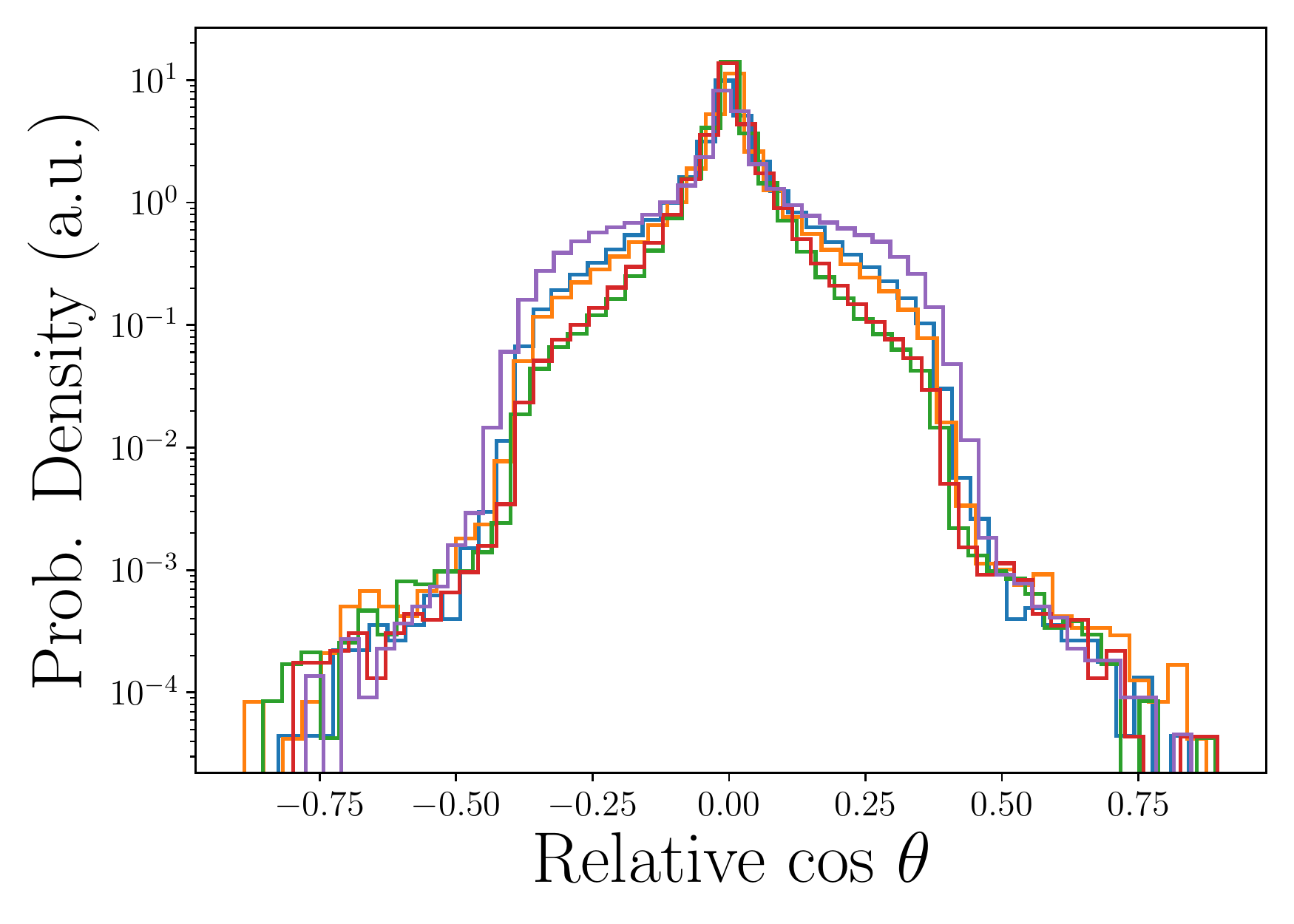} 
\caption{Distributions of kinematic features described in the text for the 150 highest-$\pt$ particles in each jet.\label{fig:jetconstituents}}
\end{figure*}

Jets are clustered from individual reconstructed particles, using the
anti-$\kt$ algorithm~\cite{Cacciari:2008gp,Cacciari:2011ma} with
jet-size parameter $R=0.8$.  Three different jet representations are
considered:
\begin{itemize}
\item A list of 16 HLFs, described in Ref.~\cite{hls4ml}, given as
  input to a DNN.  The 16 distributions are shown in
  Fig.~\ref{fig:HLS} for the five jet classes.
\item An image representation of the jet, derived by considering a
  square with pseudorapidity and azimut distances
  $\Delta\eta=\Delta\phi=2R$, centered along the jet axis.  The image
  is binned into $100\times 100$ pixels.  Such a pixel size is
  comparable to the cell of a typical LHC electromagnetic calorimeter,
  but much coarser than the typical angular resolution of a tracking
  device for the $\pt$ values relevant to this task.  Each pixel is
  filled with the scalar sum of the $\pt$ of the particles in that
  region.  These images are obtained by considering the 150
  highest-$\pt$ constituents for each jet. This jet representation is
  used to train a CNN classifier.  The average jet images for the five
  jet classes are shown in Fig.~\ref{fig:jetImage}. For comparison,
  a randomly chosen set of images is shown in Fig.~\ref{fig:OnejetImage}.
\item A constituent list for up to 150 particles, in which each
  particle is represented by 16 features, computed from the particle
  four-momenta: the three Cartesian coordinates of the momentum
  ($p_x$, $p_y$, and $p_z$), the absolute energy $E$, $\pt$, the
  pseudorapidity $\eta$, the azimuthal angle $\phi$, the distance
  $\Delta R = \sqrt{\Delta \eta^2 + \Delta \phi^2}$ from the jet
  center, the relative energy $E^{\mathrm{rel}} =
  E^\mathrm{particle}/E^\mathrm{jet}$ and relative transverse momentum
  $\pt^{\mathrm{rel}} = \pt^\mathrm{particle}/\pt^\mathrm{jet}$
  defined as the ratio of the particle quantity and the jet quantity,
  the relative coordinates $\eta^{\mathrm{rel}}=\eta^\mathrm{particle}
  - \eta^\mathrm{jet}$ and $\phi^{\mathrm{rel}}=\phi^\mathrm{particle}
  - \phi^\mathrm{jet}$ defined with respect to the jet axis,
  $\cos{\theta}$ and $\cos{\theta^{\mathrm{rel}}}$ where
  $\theta^\mathrm{rel} = \theta^\mathrm{particle} -
  \theta^\mathrm{jet}$ is defined with respect to the jet axis, and
  the relative $\eta$ and $\phi$ coordinates of the particle after
  applying a proper Lorentz transformation (rotation) as described in
  Ref.~\cite{rotation}.  Whenever less than 150 particles are
  reconstructed, the list is filled with zeros. The distributions of
  these features considering the 150 highest-$\pt$ particles in the
  jet are shown in Fig.~\ref{fig:jetconstituents} for the five jet
  categories.  This jet representation is used for a RNN with a GRU
  layer and for JEDI-net.
\end{itemize}

\section{JEDI-net}
\label{sec:models}

\begin{figure}[htb]
  \begin{centering}
  \includegraphics[width=0.4\textwidth]{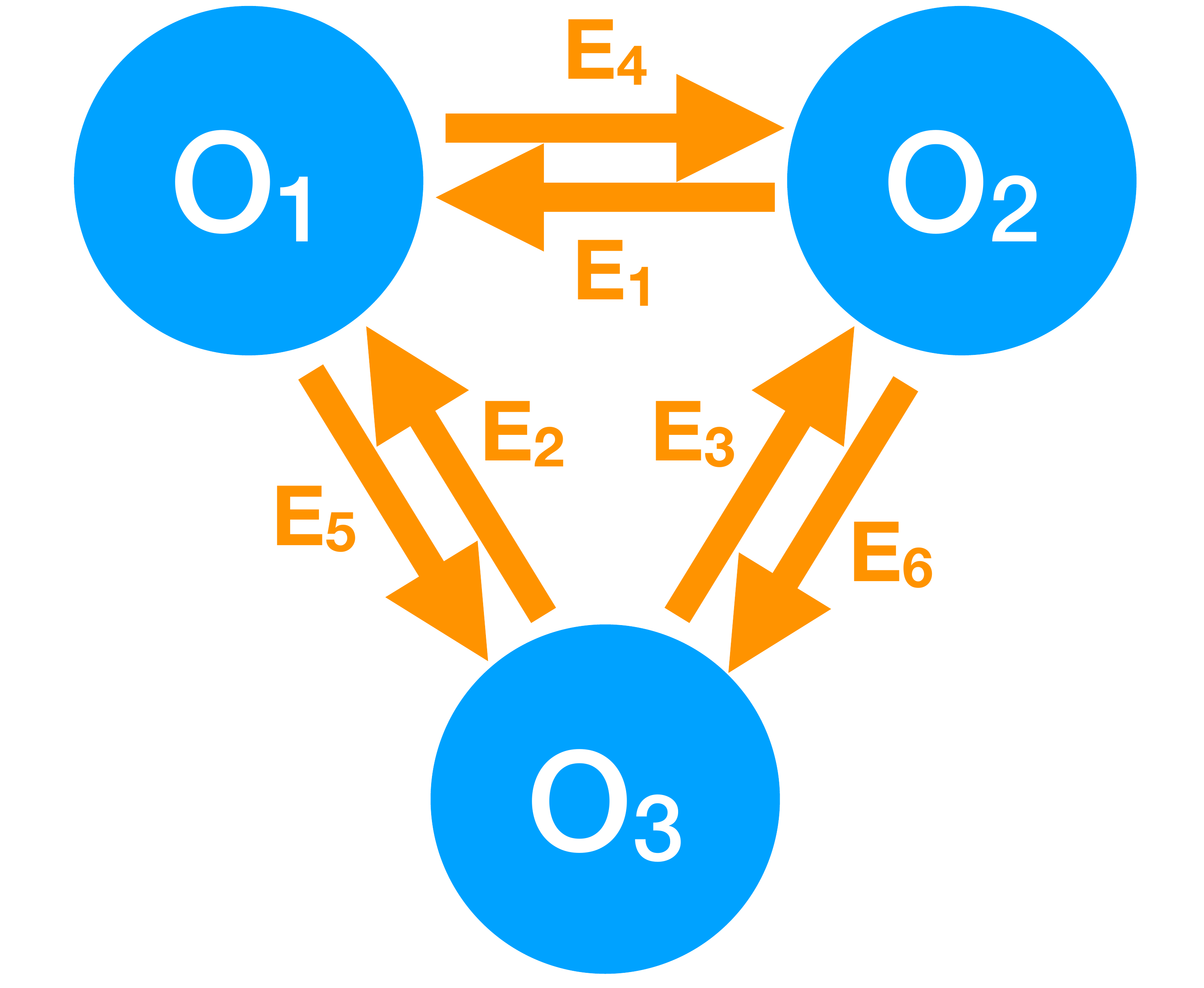}
  \caption{An example graph with three fully connected vertices and the corresponding six edges.}
  \label{fig:3graph}
  \end{centering}
\end{figure}


In this work, we apply an IN~\cite{interactionnetwork} architecture to learn a representation of a given input graph (the set of constituents in a jet) and use it to accomplish a classification task (tagging the jet). One can see the IN architecture as a processing algorithm to learn a new representation of the initial input. This is done replacing a set of input features, describing each individual vertex of the graph, with a set of engineered features, specific of each vertex but whose values depend on the connection between the vertices in the graph.

The starting point consists of building a graph for each input jet. 
The $N_O$ particles in the jet are represented by the vertices of the graph, fully interconnected through directional edges, for a total of $N_E = N_O \times (N_O-1)$ edges. 
An example is shown in Fig.~\ref{fig:3graph} for the case of a three-vertex graph. 
The vertices and edges are labeled for practical reasons, but the network architecture ensures that the labeling convention plays no role in creating the new representation.

Once the graph is built, a receiving matrix ($R_R$) and a sending matrix ($R_S$) are defined. 
Both matrices have dimensions $N_O \times N_E$.
The element $(R_R)_{ij}$ is set to 1 when the $i^{\mathrm{th}}$ vertex receives the $j^{\mathrm{th}}$ edge and is 0 otherwise. 
Similarly, the element $(R_S)_{ij}$ is set to 1 when the $i^{\mathrm{th}}$ vertex sends the $j^{\mathrm{th}}$ edge and is 0 otherwise. 
In the case of the graph of Fig.~\ref{fig:3graph}, the two matrices take the form:
\begin{align}
    R_S &= 
    \bordermatrix{~ & E_1 & E_2 & E_3 & E_4 & E_5 & E_6\cr
                  O_1 & 0 & 0 & 0 & 1 & 1 & 0\cr
                  O_2 & 1 & 0 & 0 & 0 & 0 & 1\cr
                  O_3 & 0 & 1 & 1 & 0 & 0 & 0}\\
     R_R &= 
    \bordermatrix{~ & E_1 & E_2 & E_3 & E_4 & E_5 & E_6\cr
                  O_1 & 1 & 1 & 0 & 0 & 0 & 0\cr
                  O_2 & 0 & 0 & 1 & 1 & 0 & 0\cr
                  O_3 & 0 & 0 & 0 & 0 & 1 & 1}.
\end{align}

The input particle features are represented by an input matrix $I$.
Each column of the matrix corresponds to one of the graph vertices,
while the rows correspond to the $P$ features used to represent each
vertex. In our case, the vertices are the particles inside the jet,
each represented by its array of features (i.e., the 16 features shown
in Fig.~\ref{fig:jetconstituents}).  Therefore, the $I$ matrix has
dimensions $P \times N_O$.

\begin{figure*}[htb]
  \centering
    \includegraphics[width=0.8\textwidth]{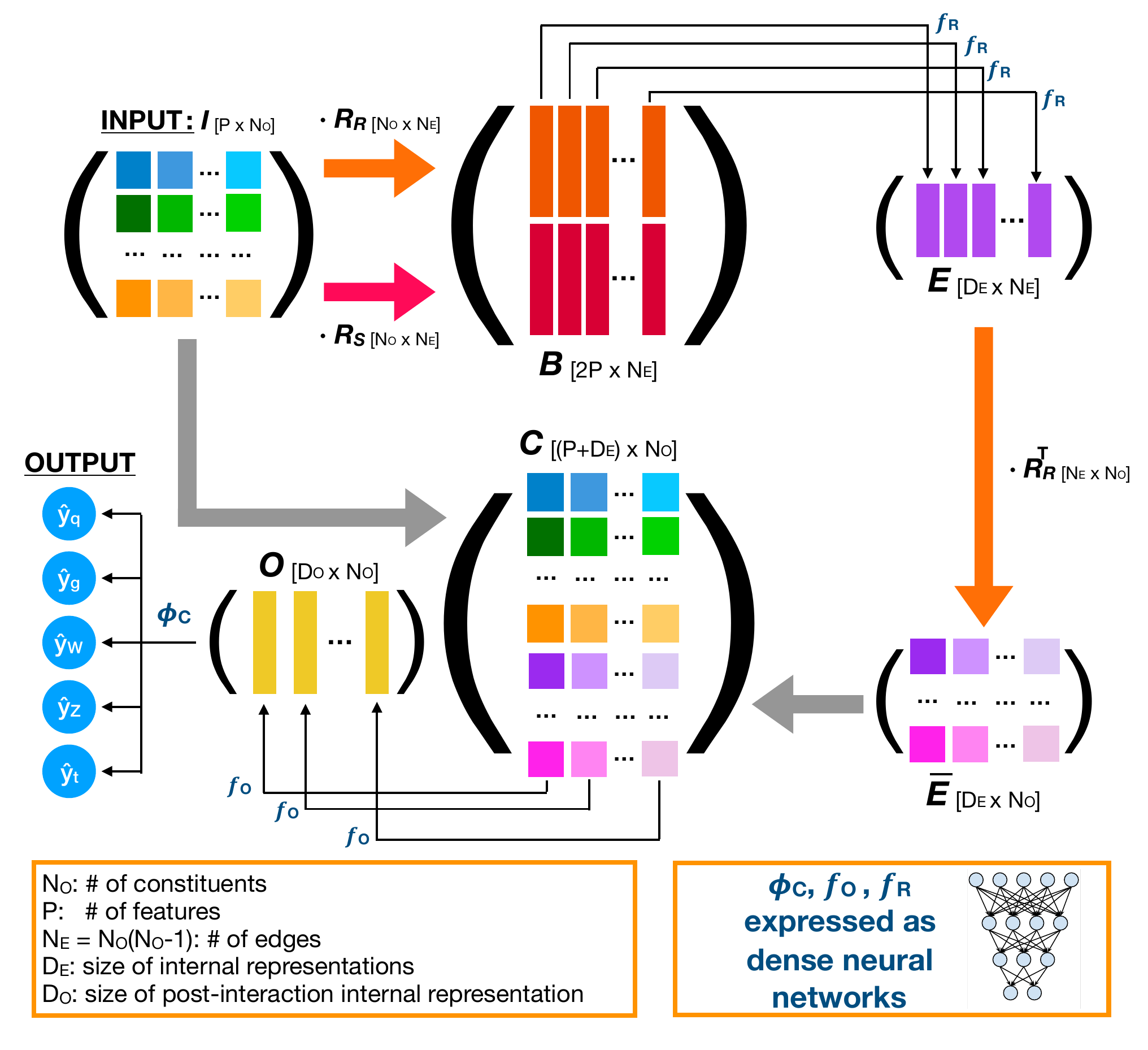}
    \caption{A flowchart illustrating the interaction network scheme.}
    \label{fig:flow}
\end{figure*}

The $I$ matrix is processed by the IN in a series of steps,
represented in Fig.~\ref{fig:flow}.  The $I$ matrix is multiplied by
the $R_R$ and $R_S$ matrices and the two resulting matrices are then
concatenated to form the $B$ matrix, having dimension $2P \times N_E$:
\begin{align}
  B &= \begin{pmatrix}I \times R_R \\ I \times R_S\end{pmatrix}.
\end{align}
Each column of the $B$ matrix represents an edge, i.e. a particle-to-particle interaction. 
The $2P$ elements of each column are the features of the sending and receiving vertices for that edge. 
Using this information, a $D_E$-dimensional hidden representation of the interaction edge is created through a trainable function $f_R: \mathbb{R}^{2P} \mapsto \mathbb{R}^{D_E}$.
This gives a matrix $E$ with dimensions $D_E \times N_E$.
The cumulative effects of the interactions received by a given vertex are gathered by summing the $D_E$ hidden features over the edges arriving to it. 
This is done by computing $\overline{E} = E R_R^\top$ with dimensions $D_E \times N_O$, which is then appended to the initial input matrix $I$:
\begin{align}
C &= \begin{pmatrix}I\\\overline{E}\end{pmatrix}.
\end{align}

At this stage, each column of the $C$ matrix represents a constituent in the jet, expressed as a $(P+D_E)$-dimensional feature vector, containing the $P$ input features and the $D_E$ hidden features representing the combined effect of the interactions with all the connected particles. 
A trainable function $f_O: \mathbb{R}^{P + D_E} \mapsto \mathbb{R}^{D_O}$ is used to build a post-interaction representation of each jet constituent.
The function $f_O$ is applied to each column of $C$ to build the post-interaction matrix $O$ with dimensions $D_O \times N_O$.

A final classifier $\phi_C$ takes as input the elements of the $O$ matrix and returns the probability for that jet to belong to each of the five categories. This is done in two ways: (i) in one case, we define the quantities $\overline{O}_i = \sum_j O_{ij}$, where $j$ is the index of the vertex in the graph (the particle, in our case), and the $i\in [0, D_E]$ index runs across the $D_E$ outputs of the $f_O$ function. The $\overline{O}$ quantities are used as input to $\phi_C: \mathbb{R}^{D_O} \mapsto \mathbb{R}^{N}$. This choice allows to preserve the independence of the architecture on the labeling convention adopted to build the $I$, $R_R$, and $R_S$ matrices, at the cost of losing some discriminating information in the summation. (ii) Alternatively, the $\phi_C$ matrix is defined directly from the $D_O \times N_O$ elements of the $O$ matrix, flattened into a one-dimensional array. The full information from $O$ is preserved, but $\phi_C$ assumes an ordering of the $N_O$ input objects. In our case, we rank the input particles in descending order by $\pt$. 

The trainable functions $f_O$, $f_R$, and $\phi_C$ consist of three DNNs. Each of them has two hidden layers, the first (second) having $N_n^1$ ($N_n^2 = \lfloor N_n^1/2 \rfloor$) neurons.
The model is implemented in \textsc{PyTorch}~\cite{pytorch} and trained using an NVIDIA GTX1080 GPU. 
The training (validation) data set consists of 630,000 (240,000) examples, while 10,000 events are used for testing purposes.

The architecture of the three trainable functions is determined by minimizing the loss function through a Bayesian optimization, using the \textsc{GpyOpt} library~\cite{gpyopt2016}, based on \textsc{Gpy}~\cite{gpy2014}. 
We consider the following hyperparameters: 
\begin{itemize}
\item The number of output neurons of the $f_R$ network, $D_E$ (between 4 and 14).
\item The number of output neurons of the $f_O$ network, $D_O$  (between 4 and 14).
\item The number of neurons $N_n^1$ in the first hidden layer of the $f_O$, $f_R$, and $\phi_C$ network (between 5 and 50).
\item The activation function for the hidden and output layers of the $f_R$ network: ReLU~\cite{RELU}, ELU~\cite{ELU}, or SELU~\cite{SELU} functions.
\item The activation function for the hidden and output layers of the $f_O$ network: ReLU, ELU, or SELU.
\item The activation function for the hidden layers of the $\phi_C$ network: ReLU, ELU, or SELU.
\item The optimizer algorithm: Adam~\cite{adam} or AdaDelta~\cite{adadelta}.
\end{itemize}
In addition, the output neurons of the $\phi_C$ network are activated by a softmax function. 
A learning rate of $10^{-4}$ is used. 
For a given network architecture, the network parameters are optimized by minimizing the categorical cross entropy. 
The Bayesian optimization is repeated four times. 
In each case, the input particles are ordered by descending $\pt$ value and the first 30, 50, 100, or 150 particles are considered. 
The parameter optimization is performed on the training data set, while the loss for the Bayesian optimization is estimated on the validation data set.

Tables~\ref{tab:INopt}~and~\ref{tab:INoptWithSum} summarize the result of the Bayesian optimization for the JEDI-net architecture with and without the sum over the columns of the $O$ matrix, respectively. The best result of each case, highlighted in bold, is used as a reference for the rest of the paper. 

\begin{table}[htb]
\centering
\begin{tabular}{c|cccc}
\multirow{2}{*}{Hyperparameter}     &  \multicolumn{4}{c}{Number of jet constituents}\\
                    & 30 & 50 & 100 & {\bf 150} \\
\hline
$N_n^1$             &   6    & 50   & 30   &  {\bf 50}  \\
$D_E$               &   8    & 12  & 4   &  {\bf 14}  \\
$D_O$               &   6     & 14   &  4  & {\bf 10}   \\
$f_R$ activation    &   ReLU  & ReLU   & SELU   &  {\bf SELU}  \\
$f_O$ activation    &   ELU  & ReLU   & ReLU   &  {\bf SELU} \\
$\phi_C$ activation &   ELU  & SELU   & SELU   &   {\bf SELU}  \\
Optimizer           &   Adam  & Adam   & Adam   &  {\bf Adam}  \\
\hline
Optimized loss & 0.84 & 0.58 & 0.62 & {\bf 0.55} \\
\end{tabular}
\caption{Optimal JEDI-net hyperparameter setting for different input data sets, when the summed $\overline{O}_i$ quantities are given as input to the $\phi_C$ network. The best result, obtained when considering up to 150 particles per jet, is highlighted in bold.\label{tab:INoptWithSum}}
\end{table}

\begin{table}[htb]
\centering
\begin{tabular}{c|cccc}
\multirow{2}{*}{Hyperparameter}     &  \multicolumn{4}{c}{Number of jet constituents}\\
                    & 30 & 50 & {\bf 100} & 150 \\
\hline
$N_n^1$             &   50    & 50   & {\bf 30}   &  10  \\
$D_E$               &   12    & 12  & {\bf 10}   &  4  \\
$D_O$               &   6     & 14   &  {\bf 10}  & 14   \\
$f_R$ activation    &   ReLU  & ELU   & {\bf ELU}   &  SELU  \\
$f_O$ activation    &   SELU  & SELU   & {\bf ELU}   &  SELU \\
$\phi_C$ activation &   SELU  & ELU   & {\bf ELU}   &   SELU  \\
Optimizer           &   Adam  & Adam   & {\bf Adam}   &  Adam  \\
\hline
Optimized loss & 0.63 & 0.57 & {\bf 0.56} & 0.62 \\
\end{tabular}
\caption{Optimal JEDI-net hyperparameter setting for different input data sets, when all the $O_{ij}$ elements are given as input to the $\phi_C$ network. The best result, obtained when considering up to 100 particles per jet, is highlighted in bold.\label{tab:INopt}}
\end{table}

For comparison, three alternative models are trained on the three different representations of the same data set described in Sec.~\ref{sec:dataset}: a DNN model taking as input a list of HLFs, a CNN model processing jet images, and a recurrent model applying GRUs on the same input list used for JEDI-net. 
The three benchmark models are optimized through a Bayesian optimization procedure, as done for the INs. Details of these optimizations and the resulting best models are discussed in Appendix~\ref{appendix:otherModelOpt}.

\section{Results}
\label{sec:results}

\begin{figure*}
\centering
\includegraphics[width=0.4\textwidth]{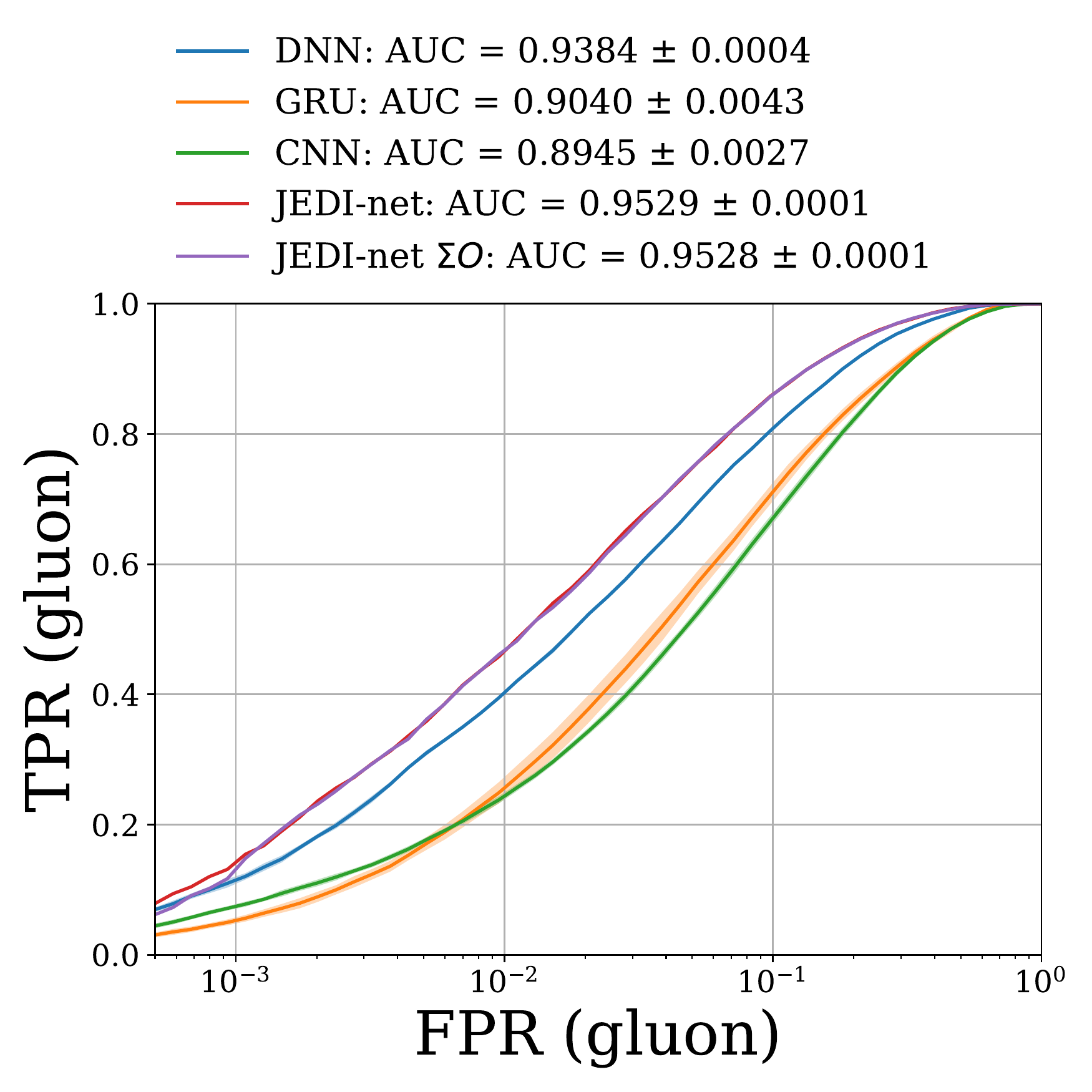}
\includegraphics[width=0.4\textwidth]{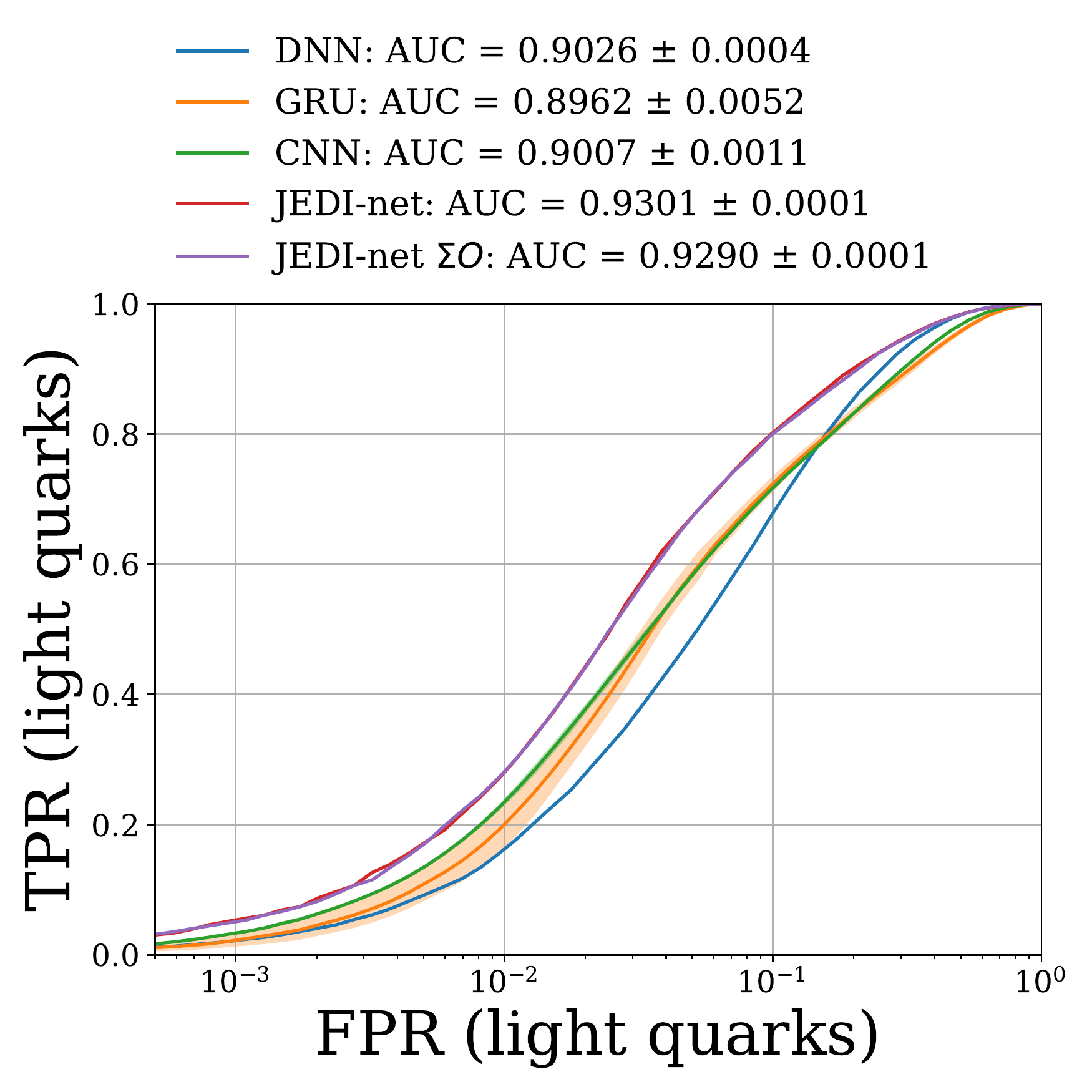} \\
\includegraphics[width=0.4\textwidth]{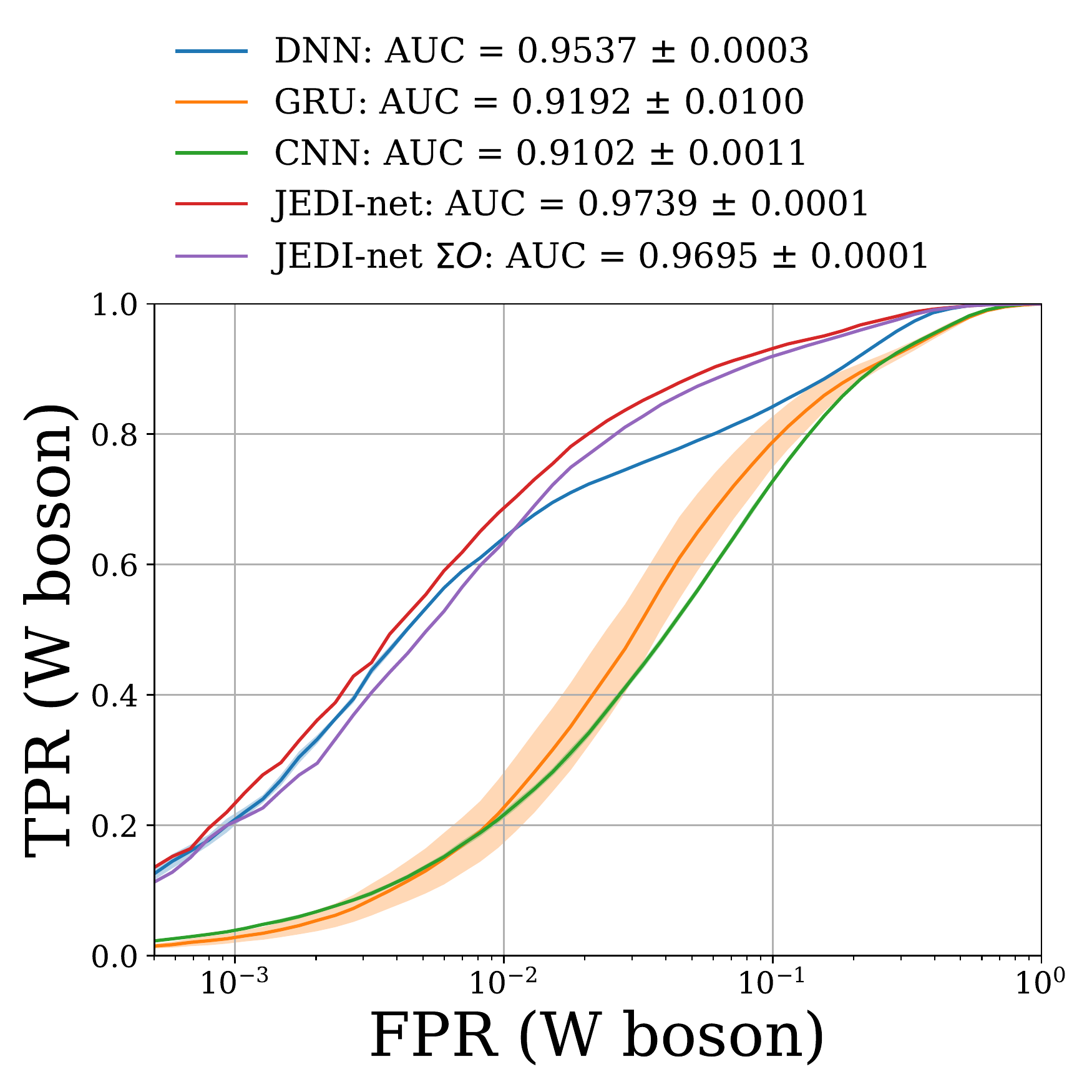}
\includegraphics[width=0.4\textwidth]{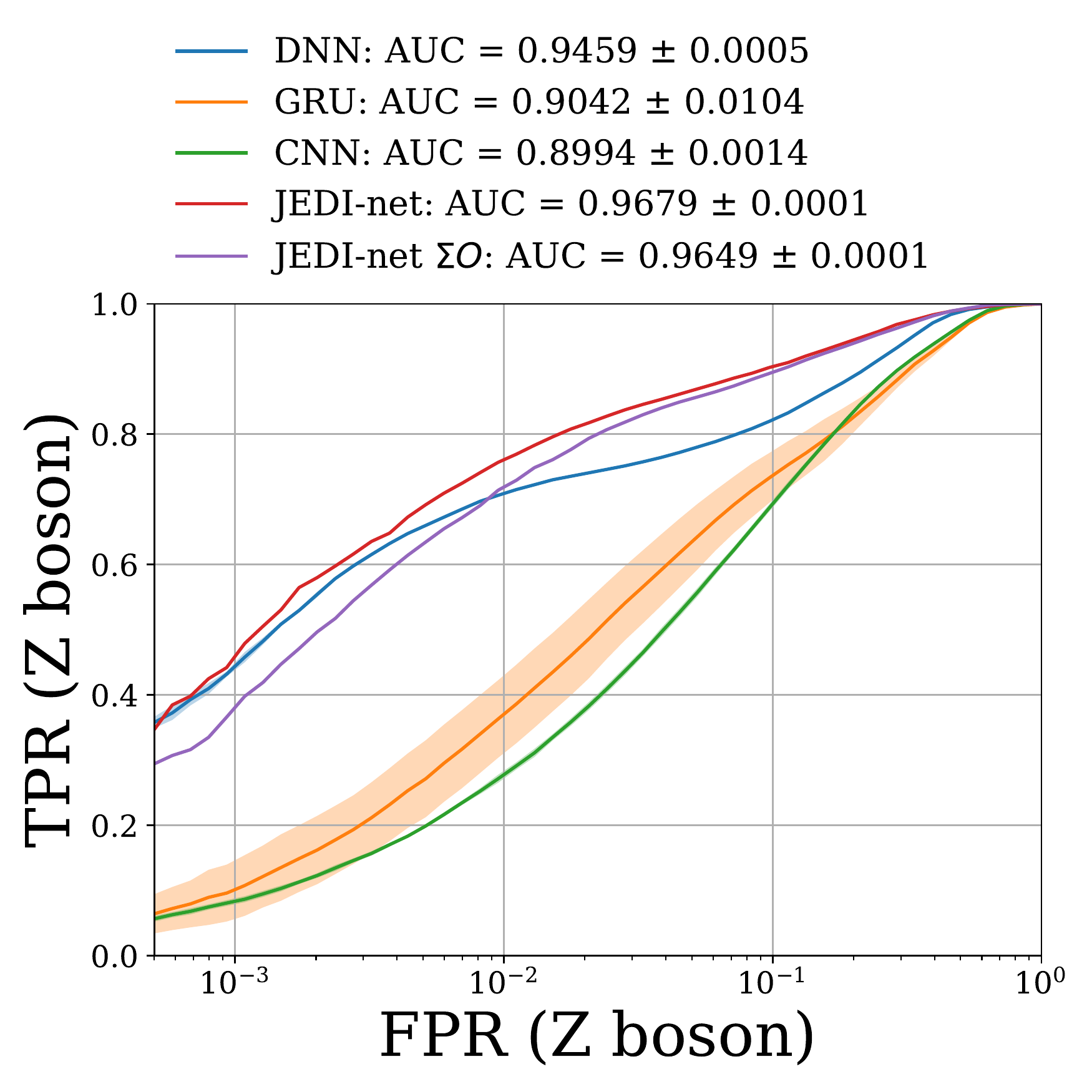} \\
\includegraphics[width=0.4\textwidth]{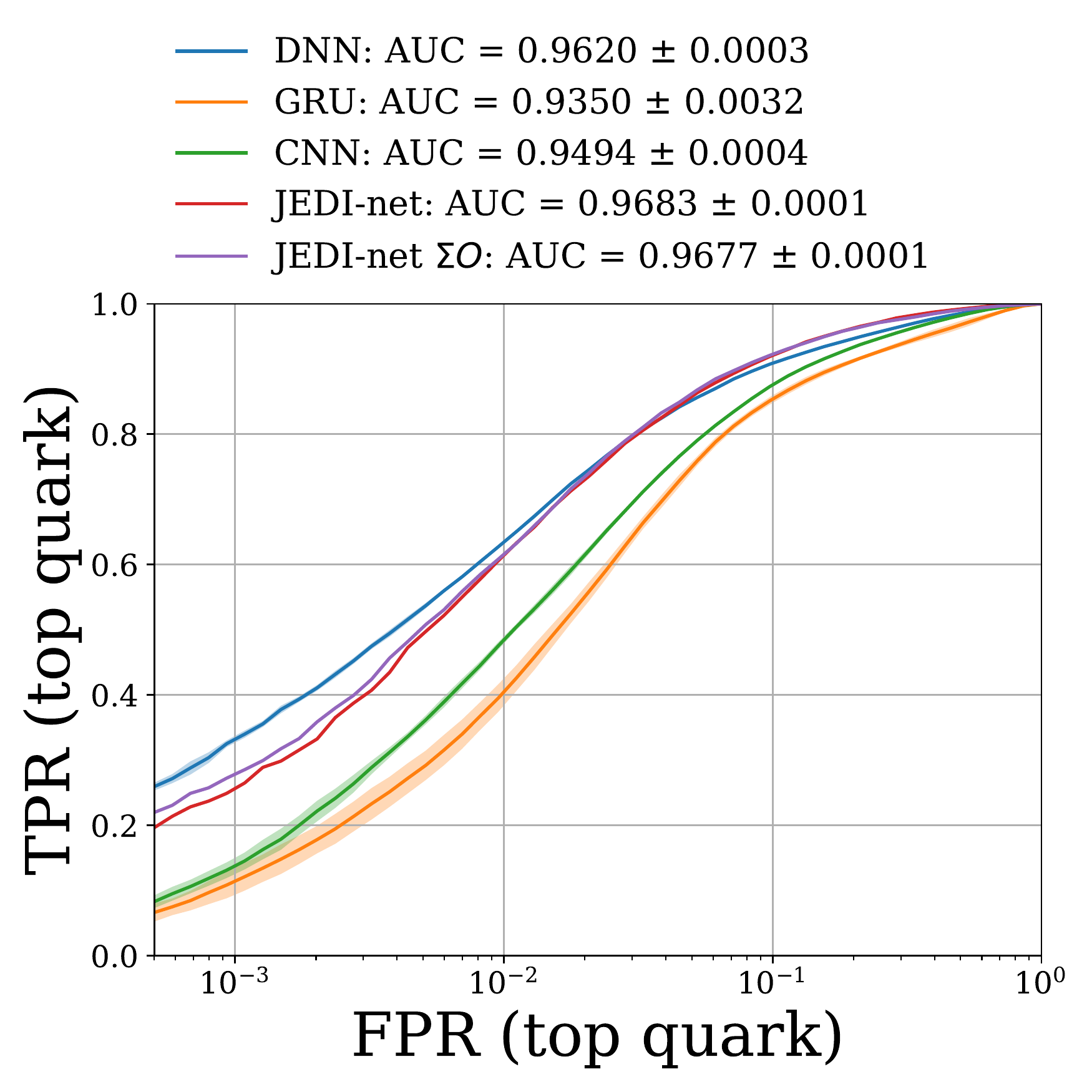}
\caption{ROC curves for JEDI-net and the three alternative models, computed for gluons (top-left), light quarks (top-right), $\PW$ (center-left) and $\PZ$ (center-right) bosons, and top quarks (bottom). The solid lines represent the average ROC curves derived from 10 $k$-fold trainings of each model. The shaded bands around the average lines are represent one standard deviation, computed with the same 10 $k$-fold trainings.\label{fig:ROC}}
\end{figure*}

Figure~\ref{fig:ROC} shows the receiver operating characteristic (ROC) curves obtained for the optimized JEDI-net tagger in each of the five jet categories, compared to the corresponding curves for the DNN, CNN, and GRU alternative models. 
The curves are derived by fixing the network architectures to the optimal values based on Table~\ref{tab:INopt} and App.~\ref{appendix:otherModelOpt} and performing a $k$-fold cross-validation training, with $k=10$.
The solid lines represent the average ROC curve, while the shaded bands quantify the $\pm 1$ RMS dispersion. 
The area under the curve (AUC) values, reported in the figure, allow for a comparison of the performance of the different taggers.

\begin{table*}[htb]
\small
\centering
\begin{tabular}{c|ccccc}
\multirow{2}{*}{Jet category}   & \multirow{2}{*}{DNN} & \multirow{2}{*}{GRU} & \multirow{2}{*}{CNN} & \multirow{2}{*}{JEDI-net} & JEDI-net \\
               &  & &        &  & with $\sum O$  \\
\hline
\multicolumn{6}{c}{TPR for FPR=$10\%$}\\
\hline
gluon        &  $0.830 \pm 0.002$ & $0.740 \pm 0.014$ & $0.700 \pm 0.008$ & $0.878 \pm 0.001$ & ${\bf 0.879 \pm 0.001}$ \\
light quarks & $0.715 \pm 0.002$ & $0.746 \pm 0.011$ & $0.740 \pm 0.003$ & ${\bf 0.822 \pm 0.001}$ & $0.818 \pm 0.001$ \\
$\PW$ boson    & $0.855 \pm 0.001$ & $0.812 \pm 0.035$ & $0.760 \pm 0.005$ & ${\bf 0.938 \pm 0.001}$ & $0.927 \pm 0.001$ \\
$\PZ$ boson    & $0.833 \pm 0.002$ & $0.753 \pm 0.036$ & $0.721 \pm 0.006$ & ${\bf 0.910 \pm 0.001}$ & $0.903 \pm 0.001$ \\
top quark    & $0.917 \pm 0.001$ & $0.867 \pm 0.006$ & $0.889 \pm 0.001$ & $0.930 \pm 0.001$ & ${\bf 0.931 \pm 0.001}$ \\
\hline
\multicolumn{6}{c}{TPR for FPR=$1\%$}\\
\hline
gluon        & $0.420 \pm 0.002$ & $0.273 \pm 0.018$ & $0.257 \pm 0.005$ & ${\bf 0.485 \pm 0.001}$ & $0.482 \pm 0.001$ \\
light quarks & $0.178 \pm 0.002$ & $0.220 \pm 0.037$ & $0.254 \pm 0.007$ & ${\bf 0.302 \pm 0.001}$ & $0.301 \pm 0.001$ \\
$\PW$ boson    & $0.656 \pm 0.002$ & $0.249 \pm 0.057$ & $0.232 \pm 0.006$ & ${\bf 0.704 \pm 0.001}$ & $0.658 \pm 0.001$ \\
$\PZ$ boson    & $0.715 \pm 0.001$ & $0.386 \pm 0.060$ & $0.291 \pm 0.005$ & ${\bf 0.769 \pm 0.001}$ & $0.729 \pm 0.001$ \\
top quark    & ${\bf 0.651 \pm 0.003}$ & $0.426 \pm 0.020$ & $0.504 \pm 0.005$ & $0.633 \pm 0.001$ & $0.632 \pm 0.001$ \\
\end{tabular}
\caption{True positive rates (TPR) for the optimized JEDI-net taggers and the three alternative models (DNN, CNN, and GRU), corresponding to a false positive rate (FPR) of 10\% (top) and 1\% (bottom). The largest TPR value for each case is highlighted in bold.\label{tab:TPR}}
\end{table*}

The algorithm's tagging performance is quantified computing the true positive rate (TPR) values for two given reference false positive rate (FPR) values (10\% and 1\%). 
The comparison of the TPR values gives an assessment of the tagging performance in a realistic use case, typical of an LHC analysis. 
Tables~\ref{tab:TPR} shows the corresponding FPR values for the optimized JEDI-net taggers, compared to the corresponding values for the benchmark models. 
The largest TPR value for each class is highlighted in bold. 
As shown in Fig.~\ref{fig:ROC} and Table~\ref{tab:TPR}, the two JEDI-net models outperform the other architectures in almost all cases. The only notable exception is the tight working point of the top-jet tagger,  for which the DNN model gives a TPR higher by about $2\%$, while the CNN and GRU models give much worse performance. 

The TPR values for the two JEDI-net models are within $1\%$. The only exception is observed for the tight working points of the $\PW$ and $\PZ$ taggers, for which the model using the $\overline{O}$ sums shows a drop in TPR of $\sim 4\%$. In this respect, the model using summed $\overline{O}$ features is preferable (despite this small TPR loss), given the reduced model complexity (see Section~\ref{sec:resources}) and its independence on the labeling convention for the particles embedded in the graph and for the edges connecting them.

\begin{figure*}[htbp!]
\centering
\includegraphics[width=0.19\textwidth]{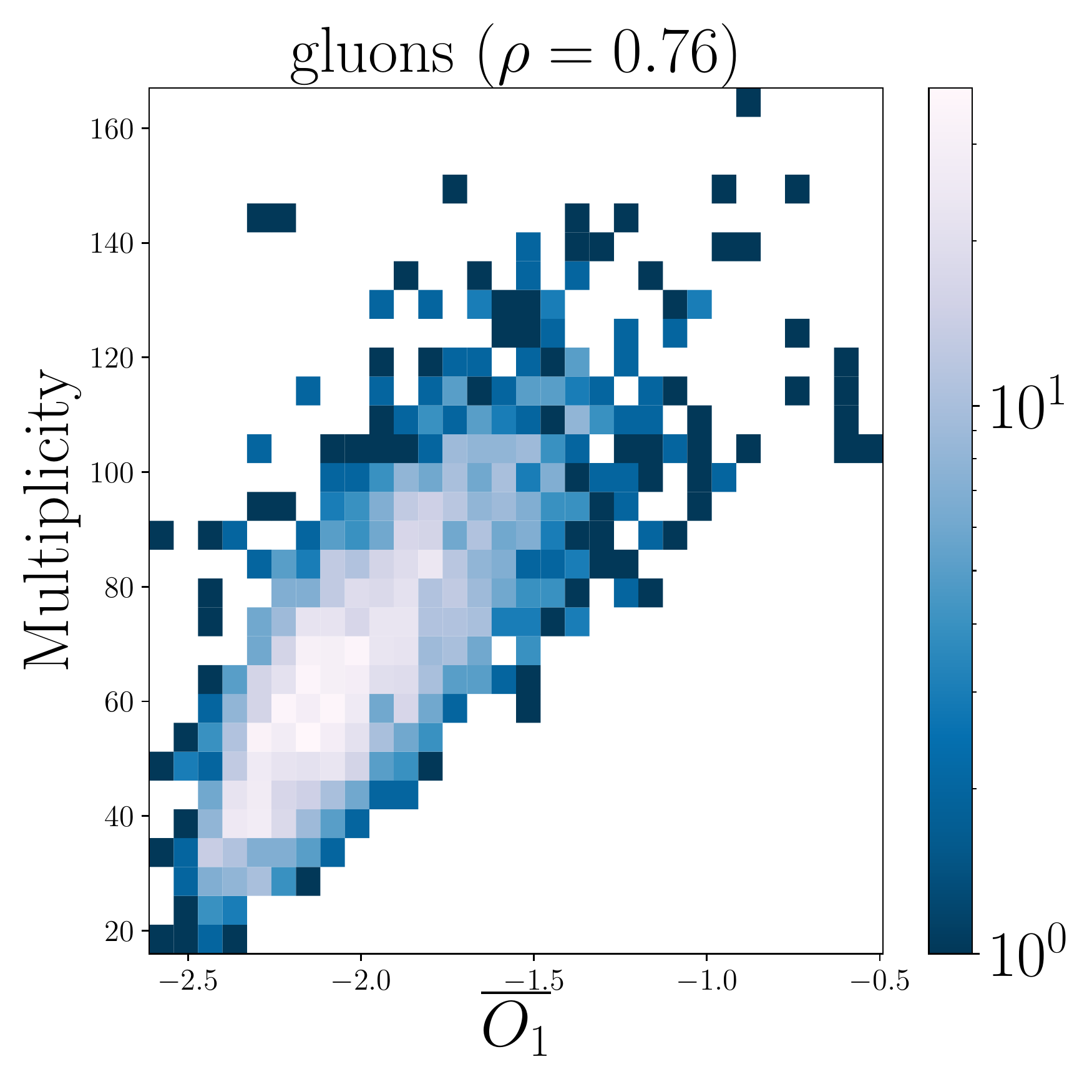}
\includegraphics[width=0.19\textwidth]{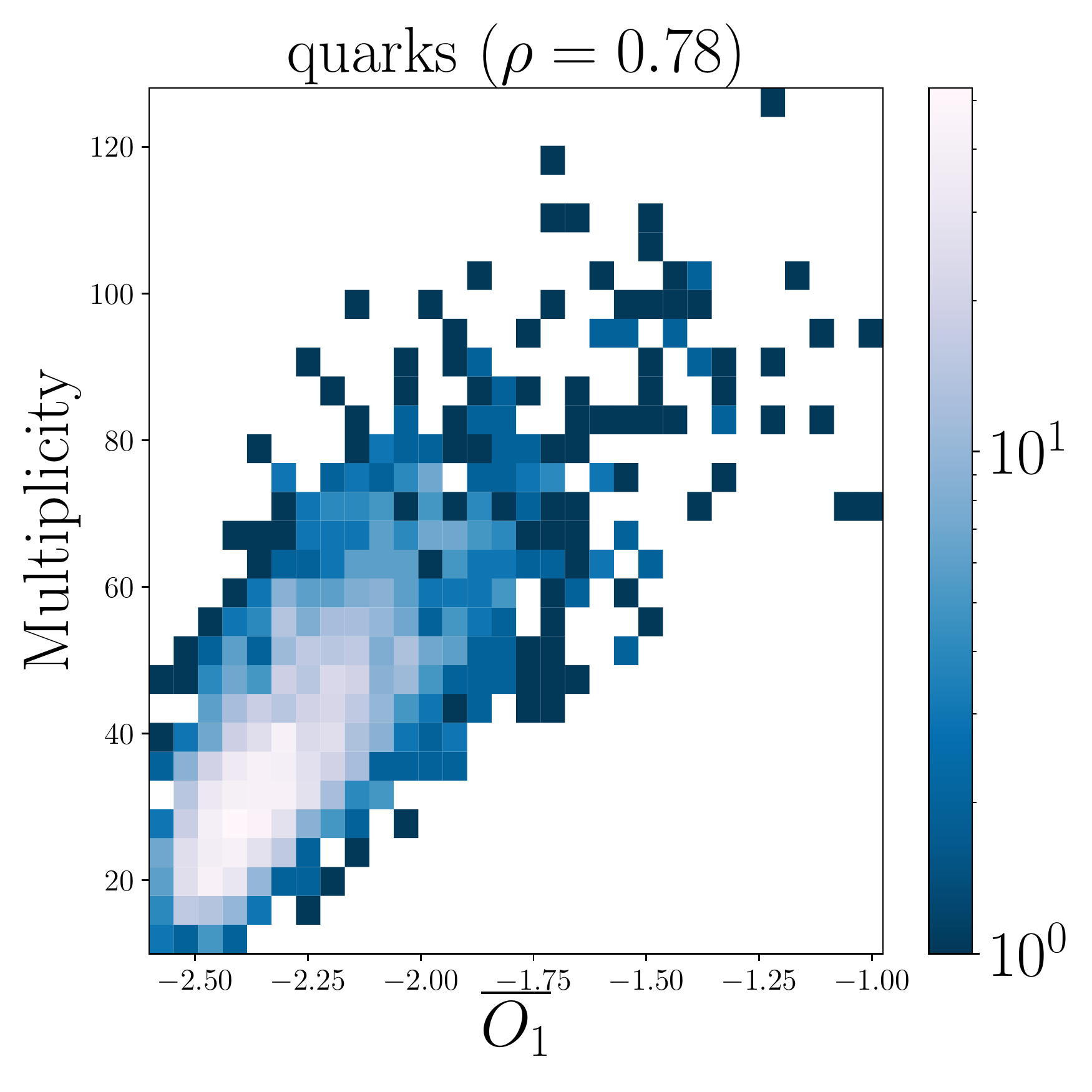}
\includegraphics[width=0.19\textwidth]{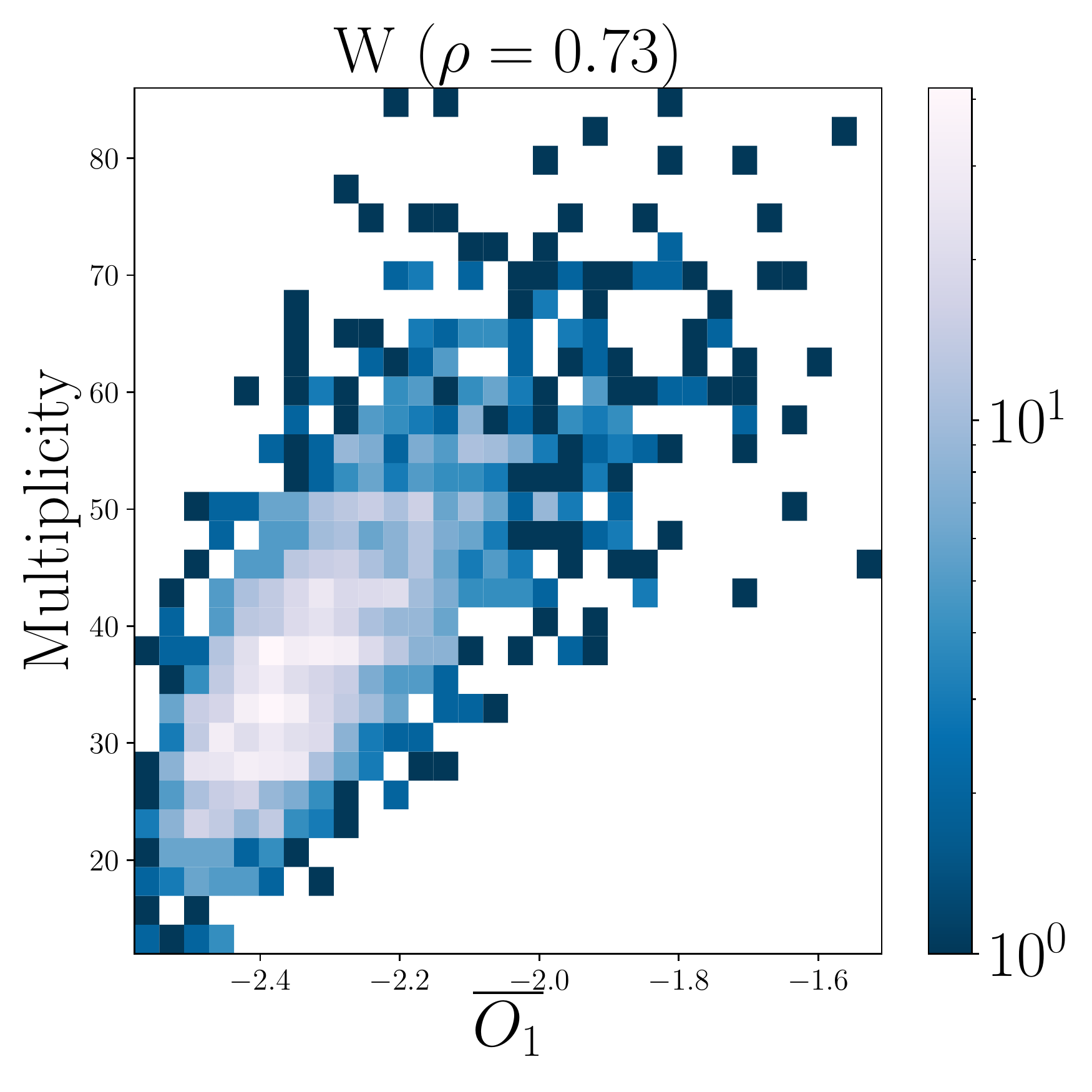} 
\includegraphics[width=0.19\textwidth]{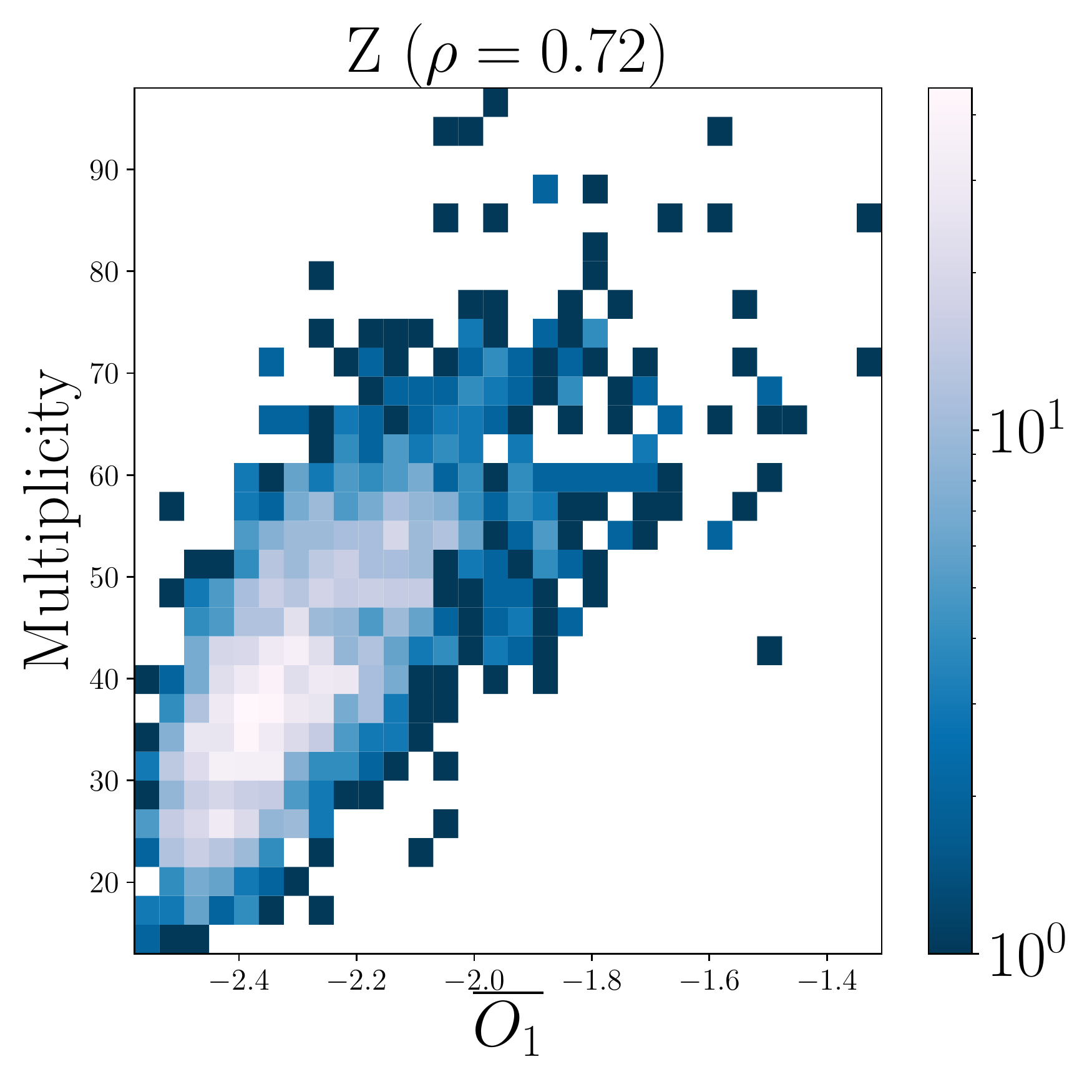}
\includegraphics[width=0.19\textwidth]{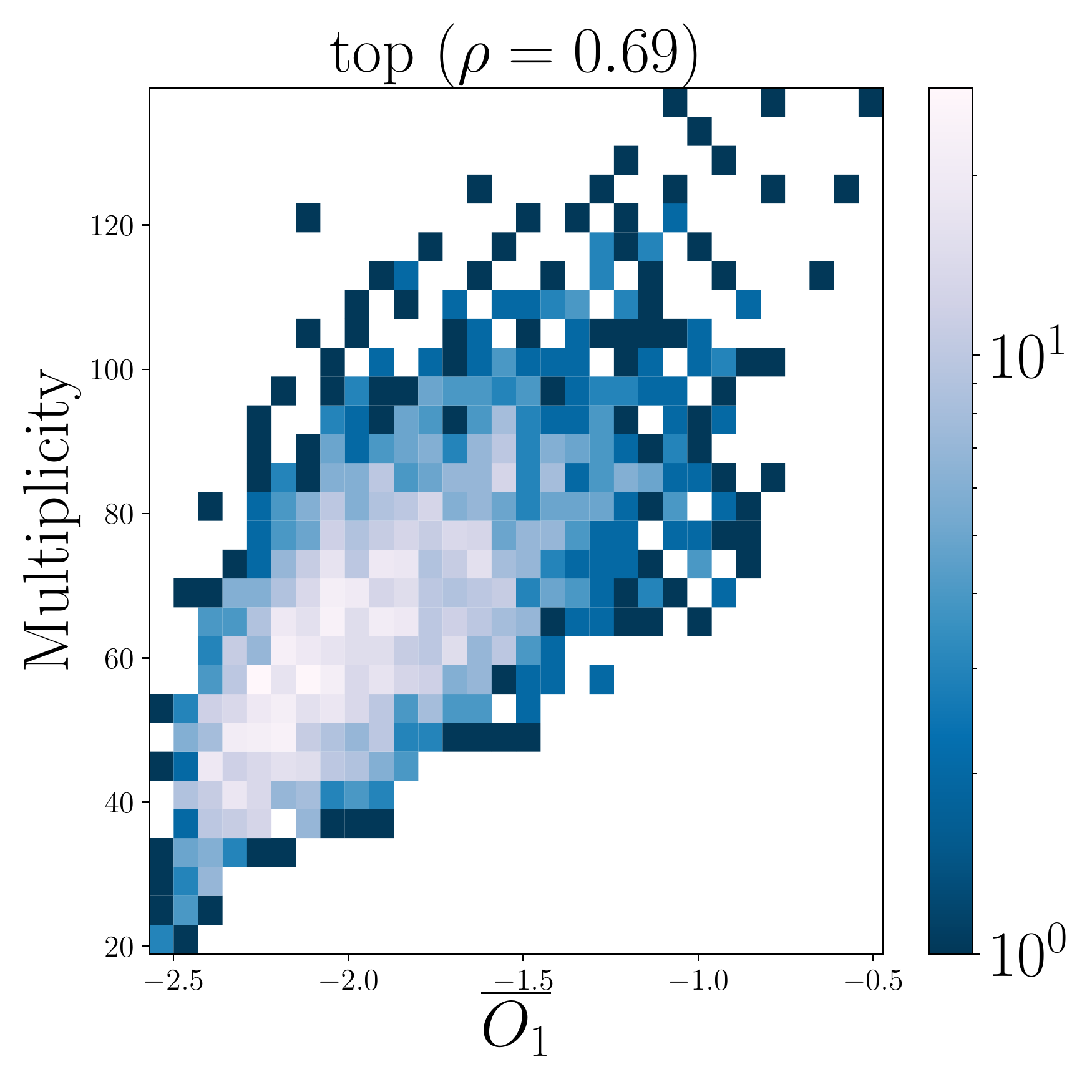} \\
\includegraphics[width=0.19\textwidth]{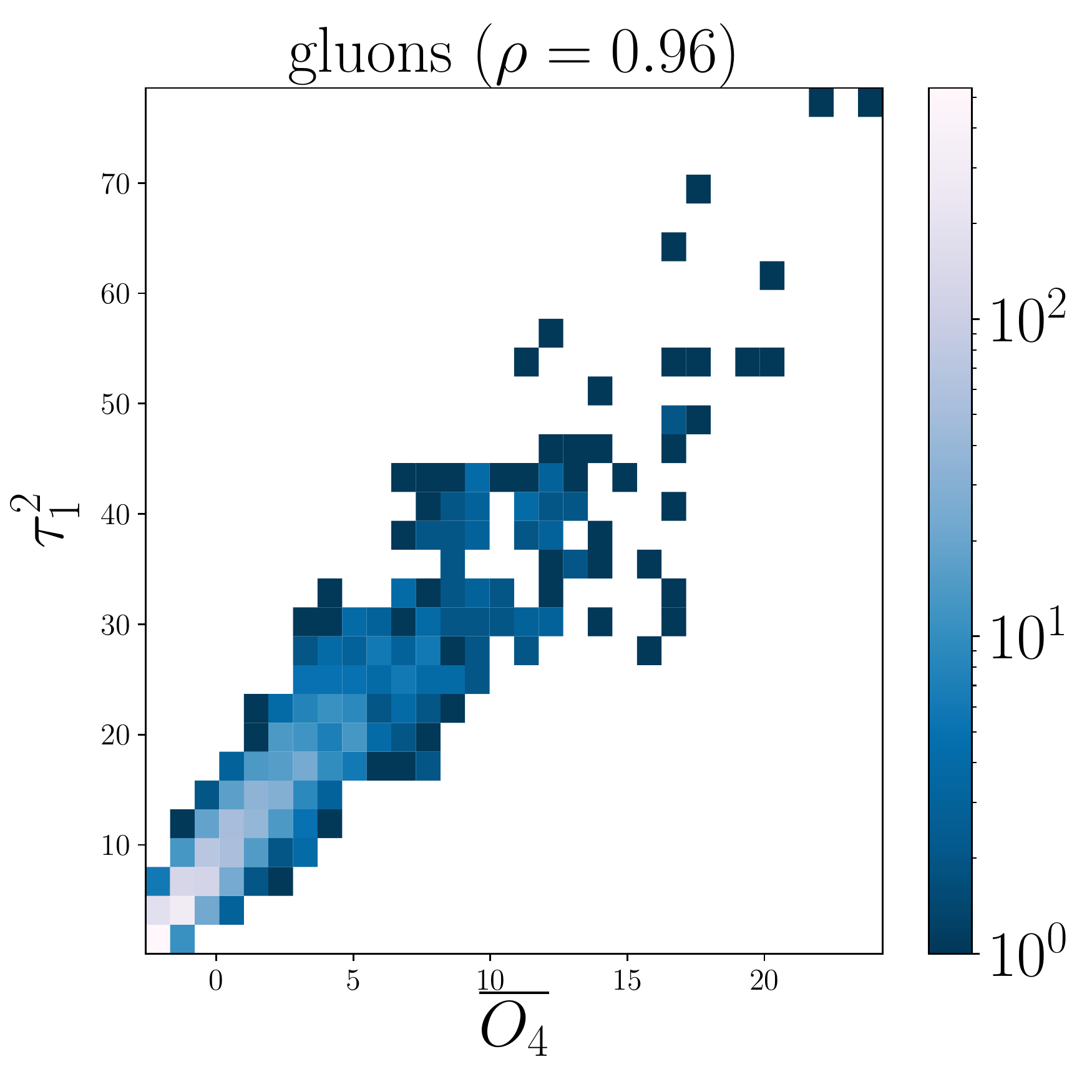}
\includegraphics[width=0.19\textwidth]{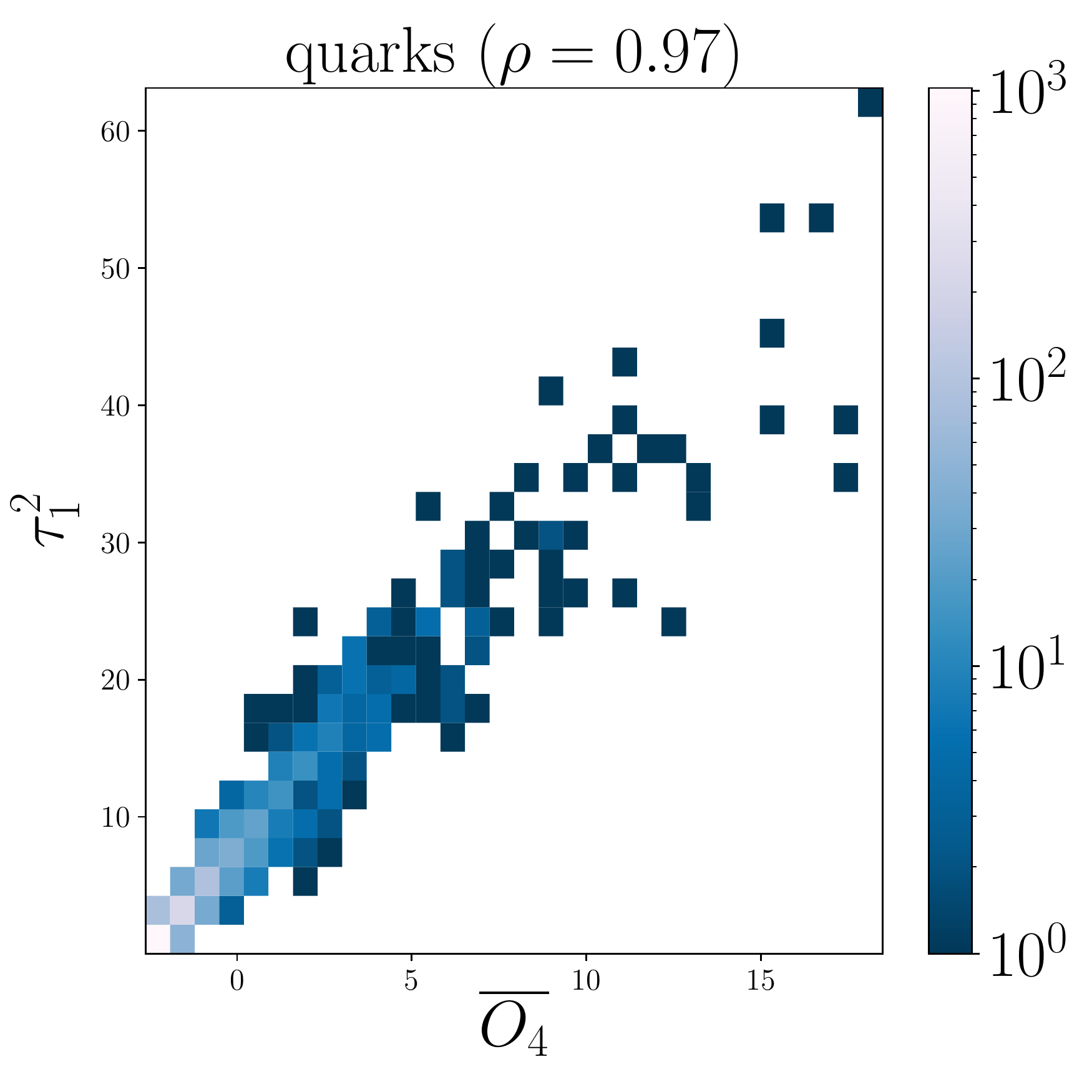}
\includegraphics[width=0.19\textwidth]{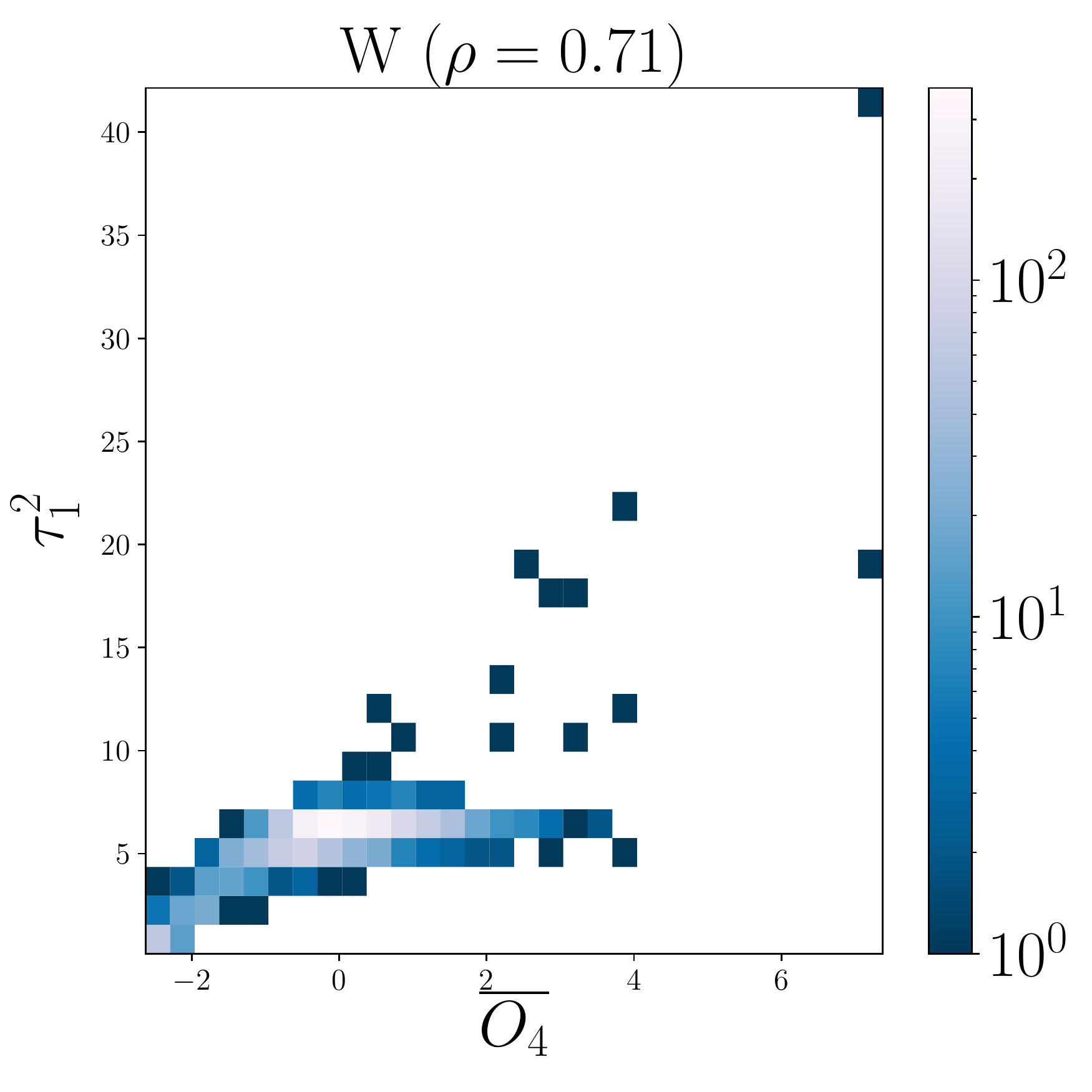} 
\includegraphics[width=0.19\textwidth]{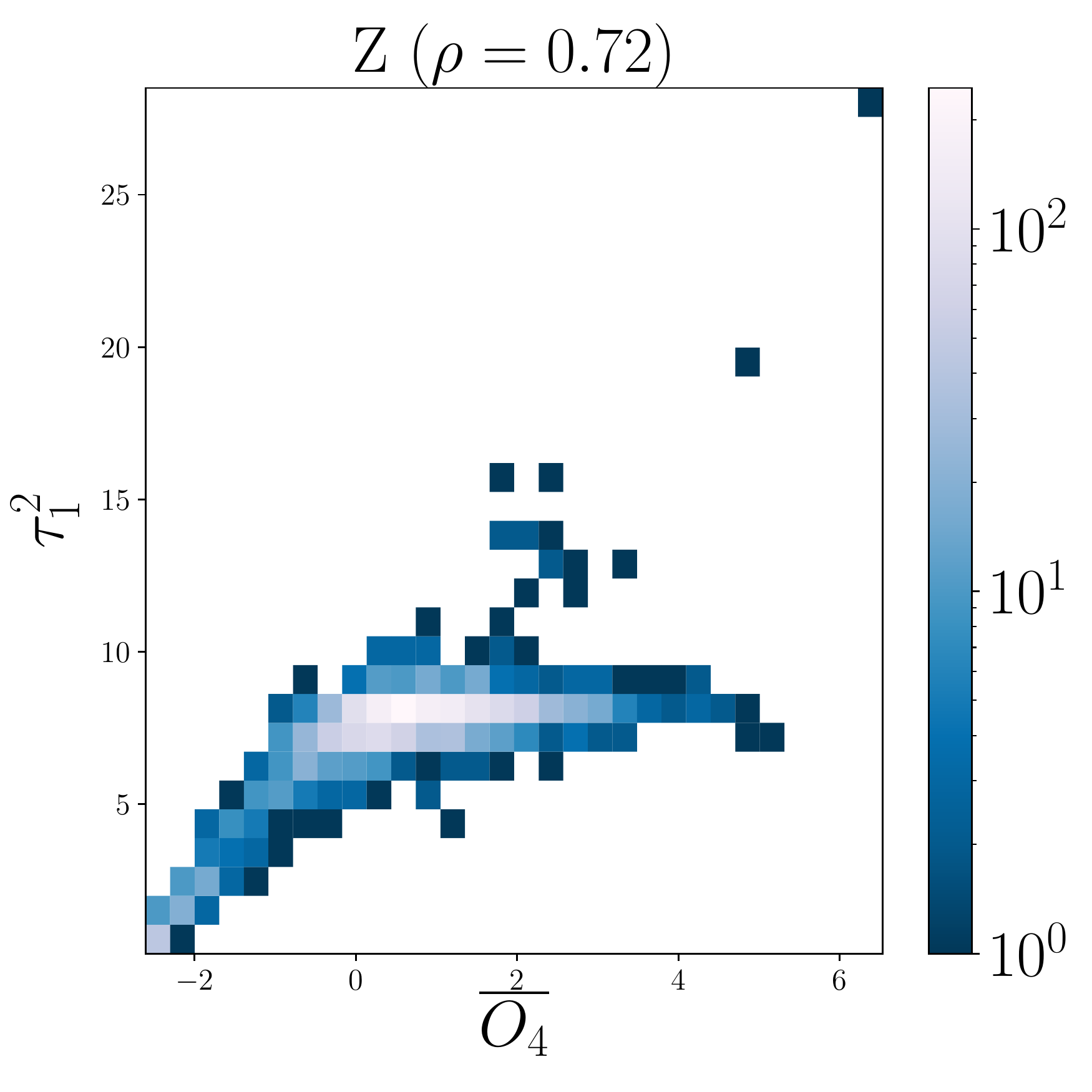}
\includegraphics[width=0.19\textwidth]{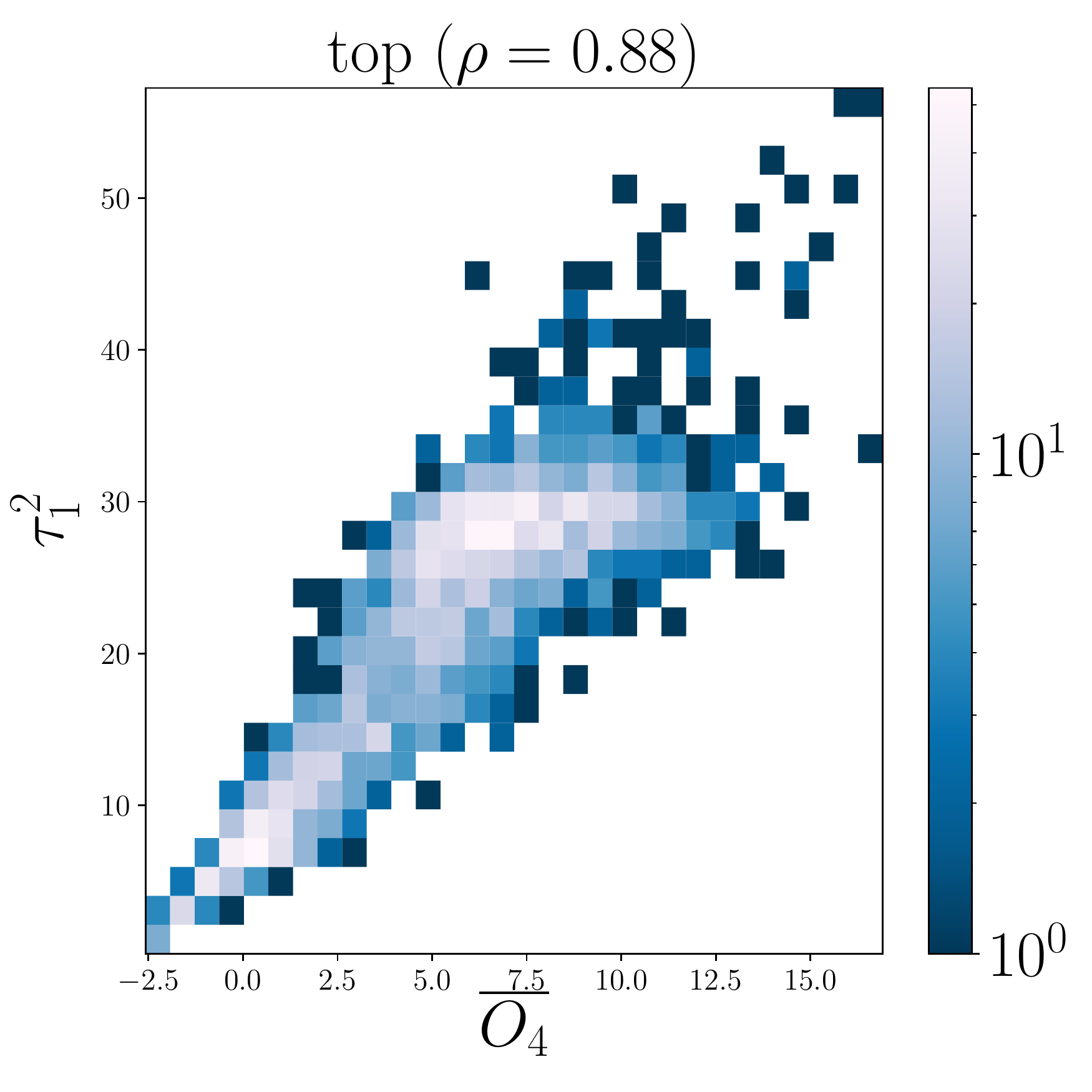} \\
\includegraphics[width=0.19\textwidth]{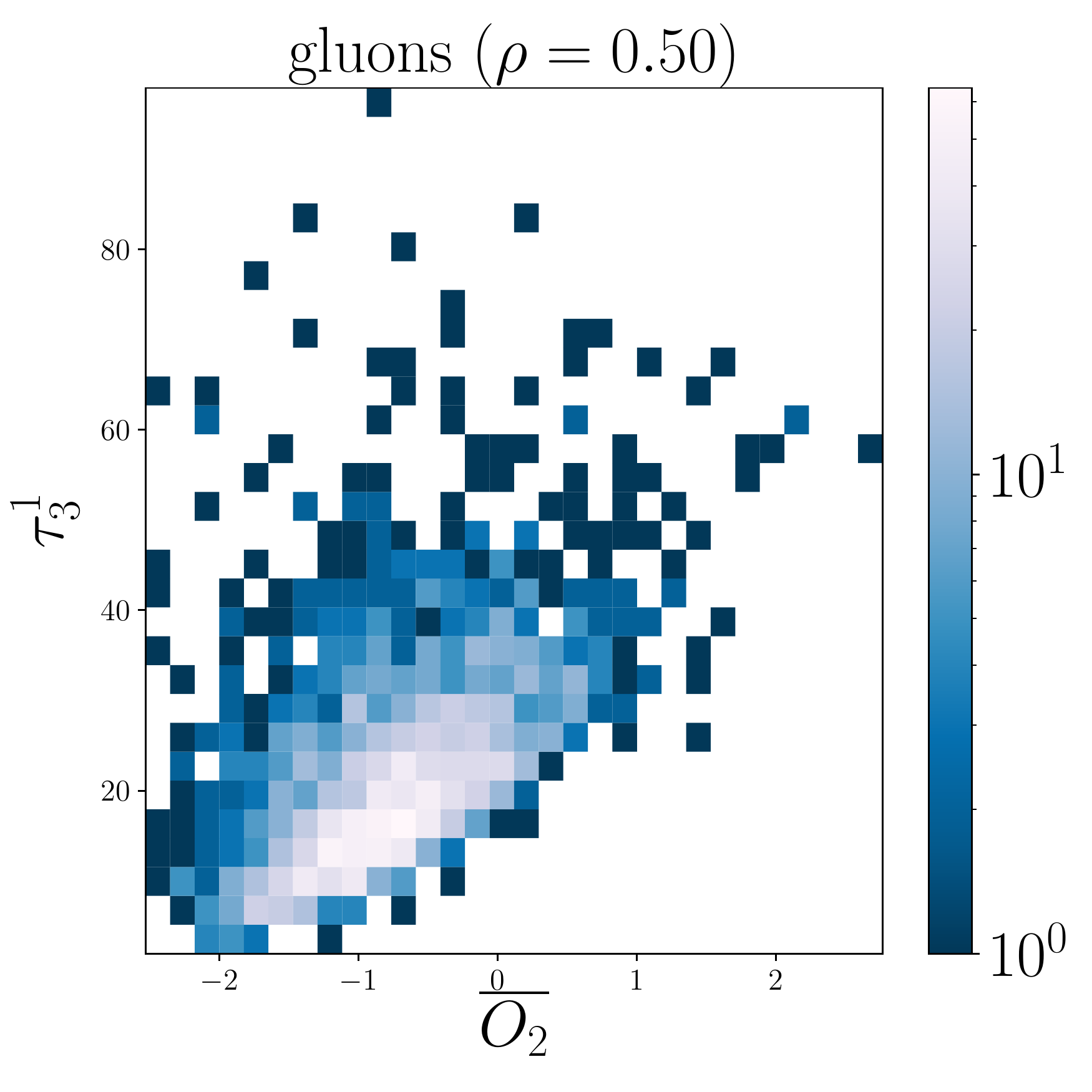}
\includegraphics[width=0.19\textwidth]{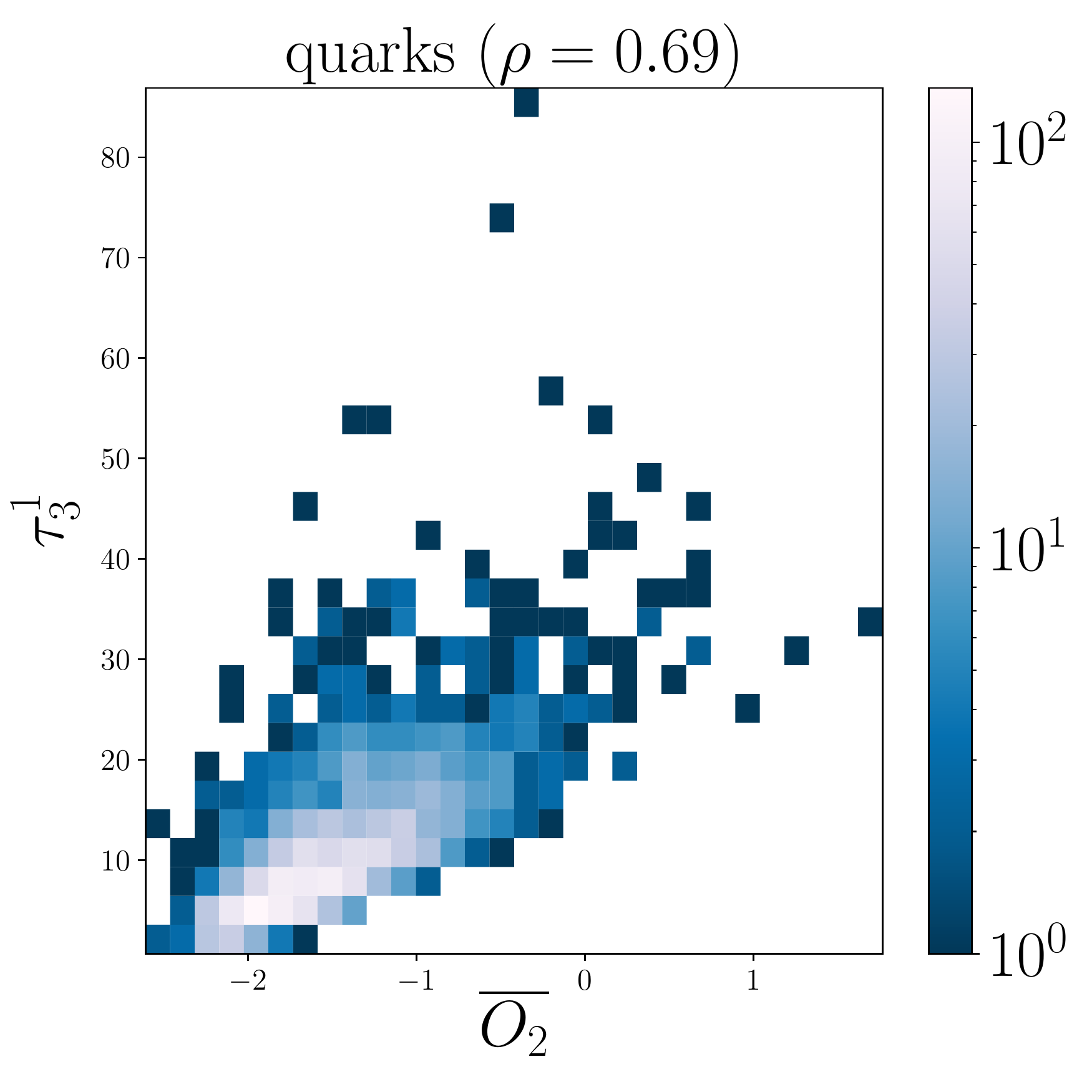}
\includegraphics[width=0.19\textwidth]{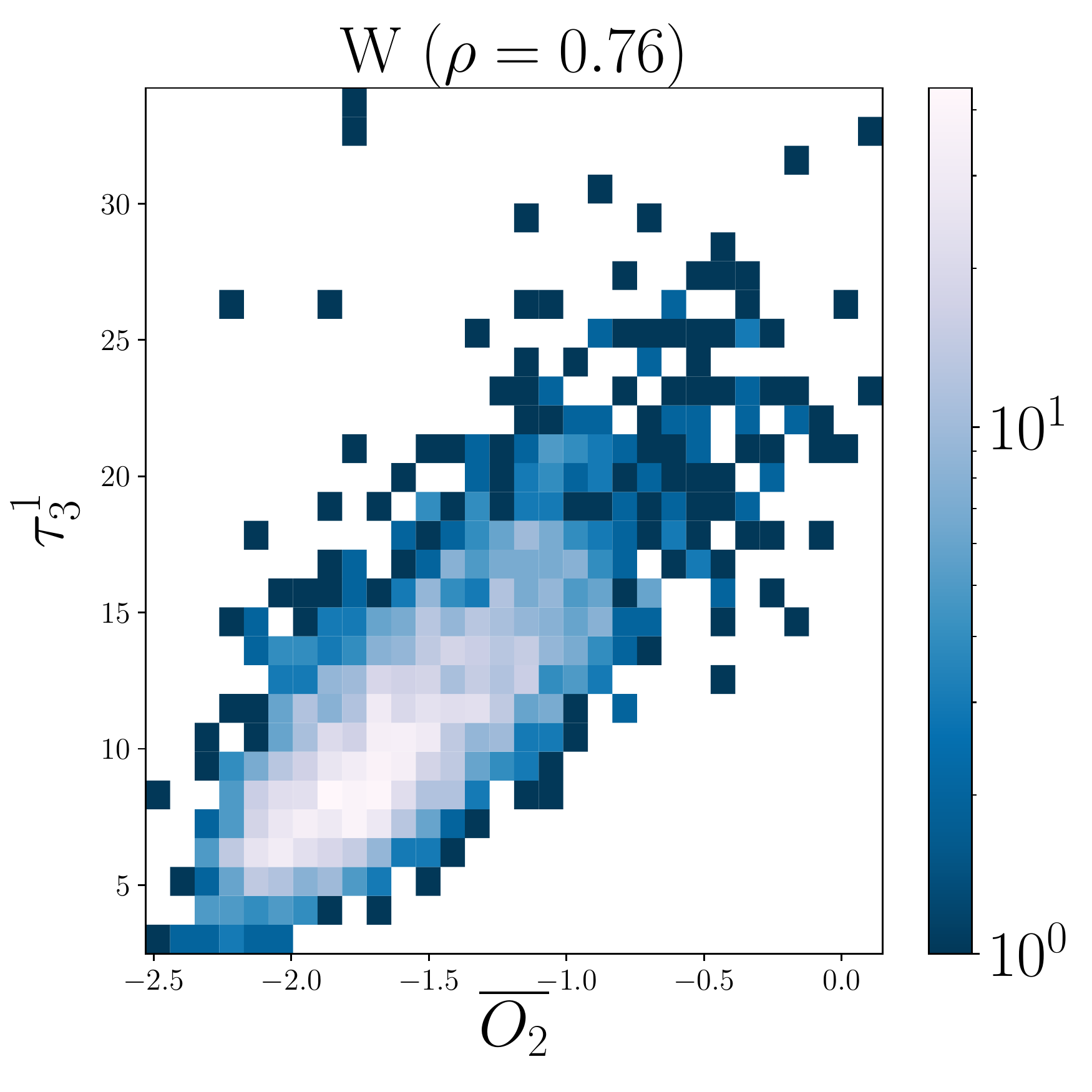} 
\includegraphics[width=0.19\textwidth]{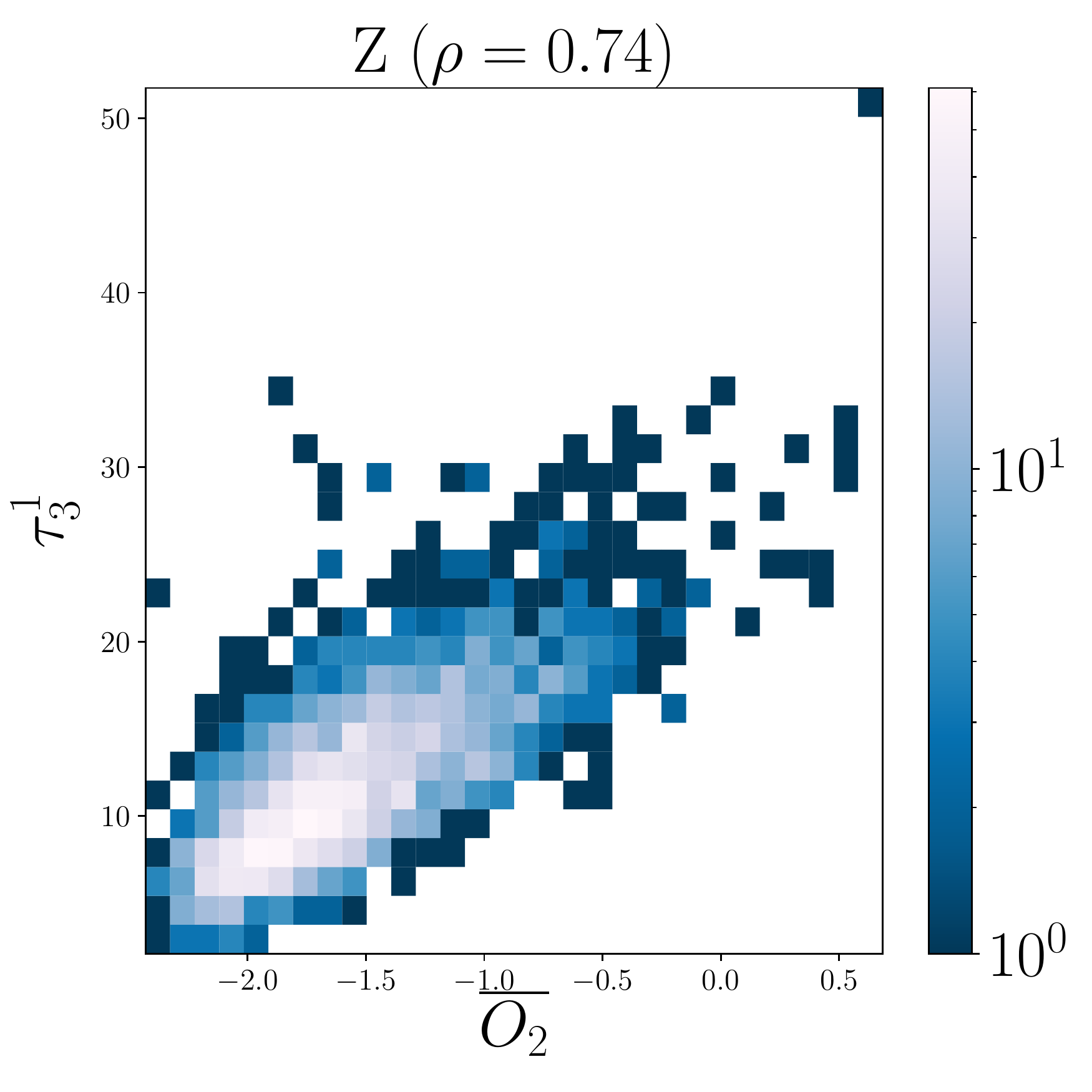}
\includegraphics[width=0.19\textwidth]{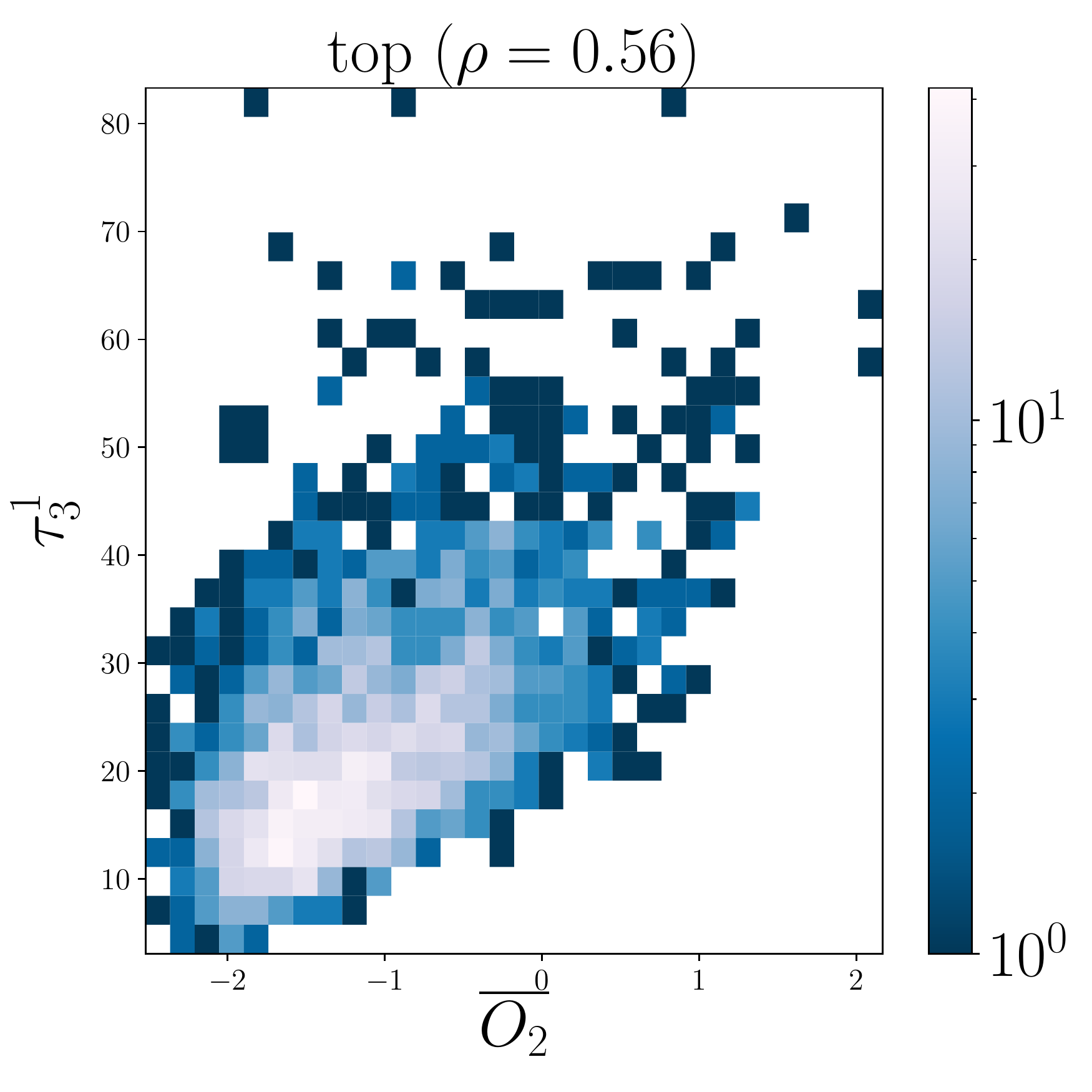} \\
\includegraphics[width=0.19\textwidth]{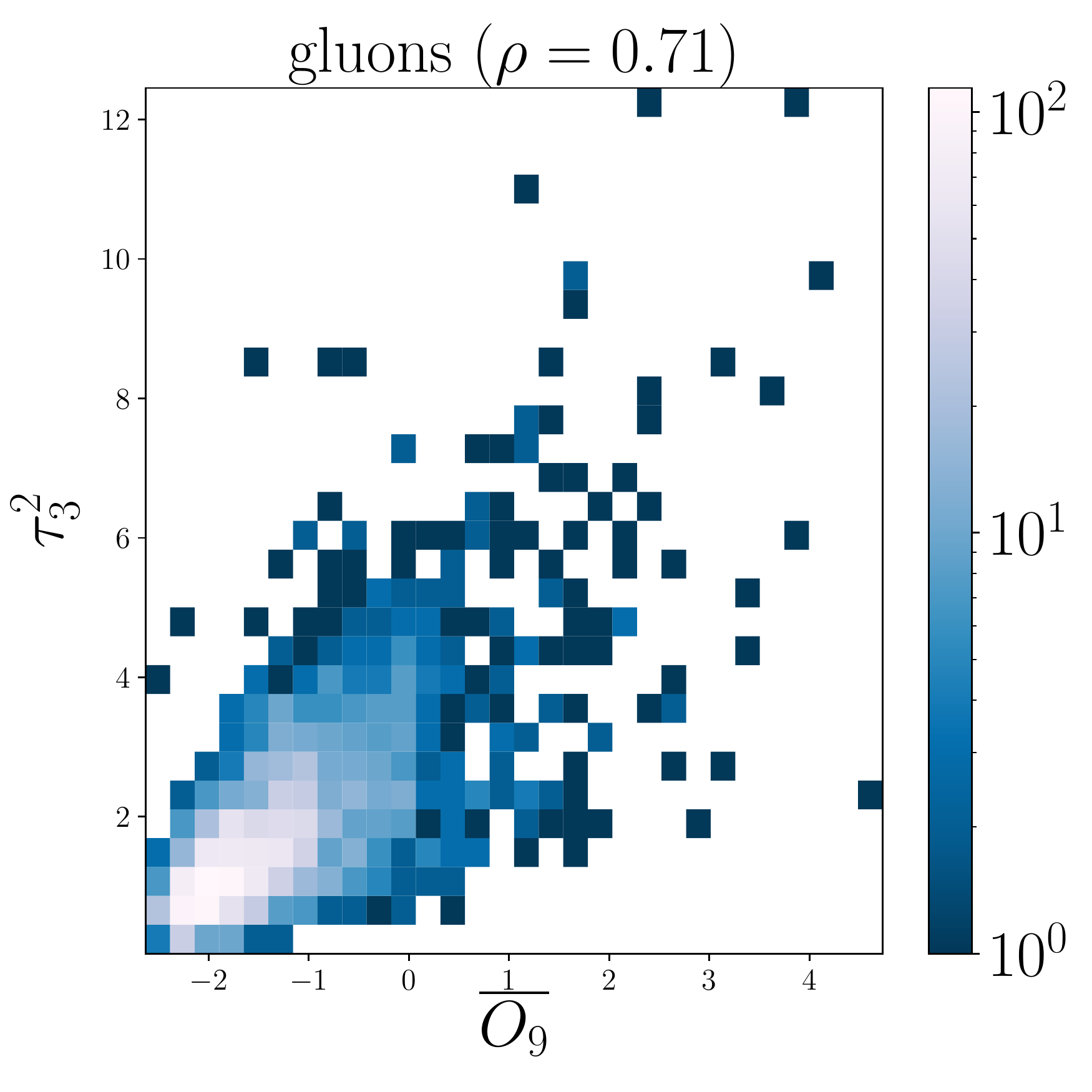}
\includegraphics[width=0.19\textwidth]{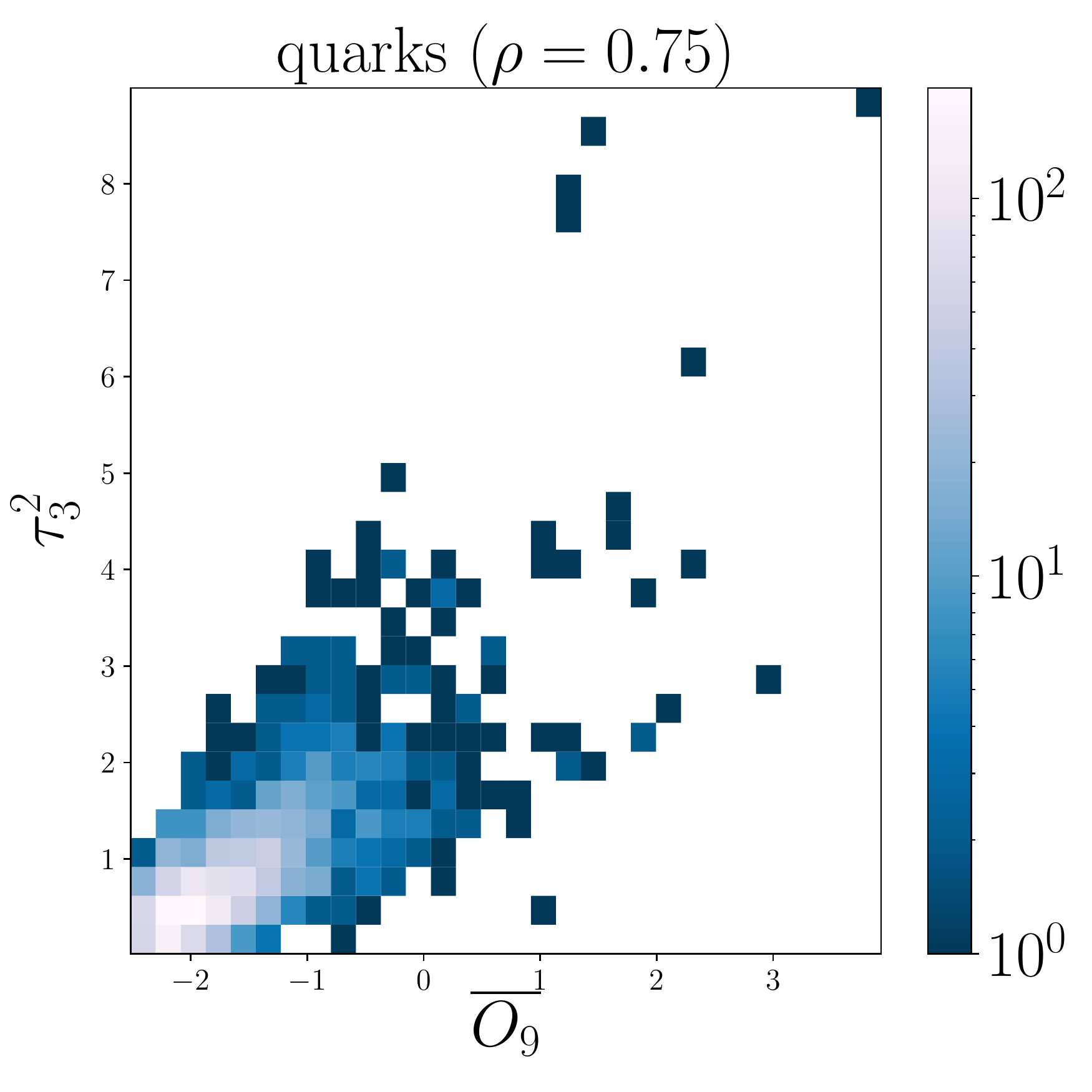}
\includegraphics[width=0.19\textwidth]{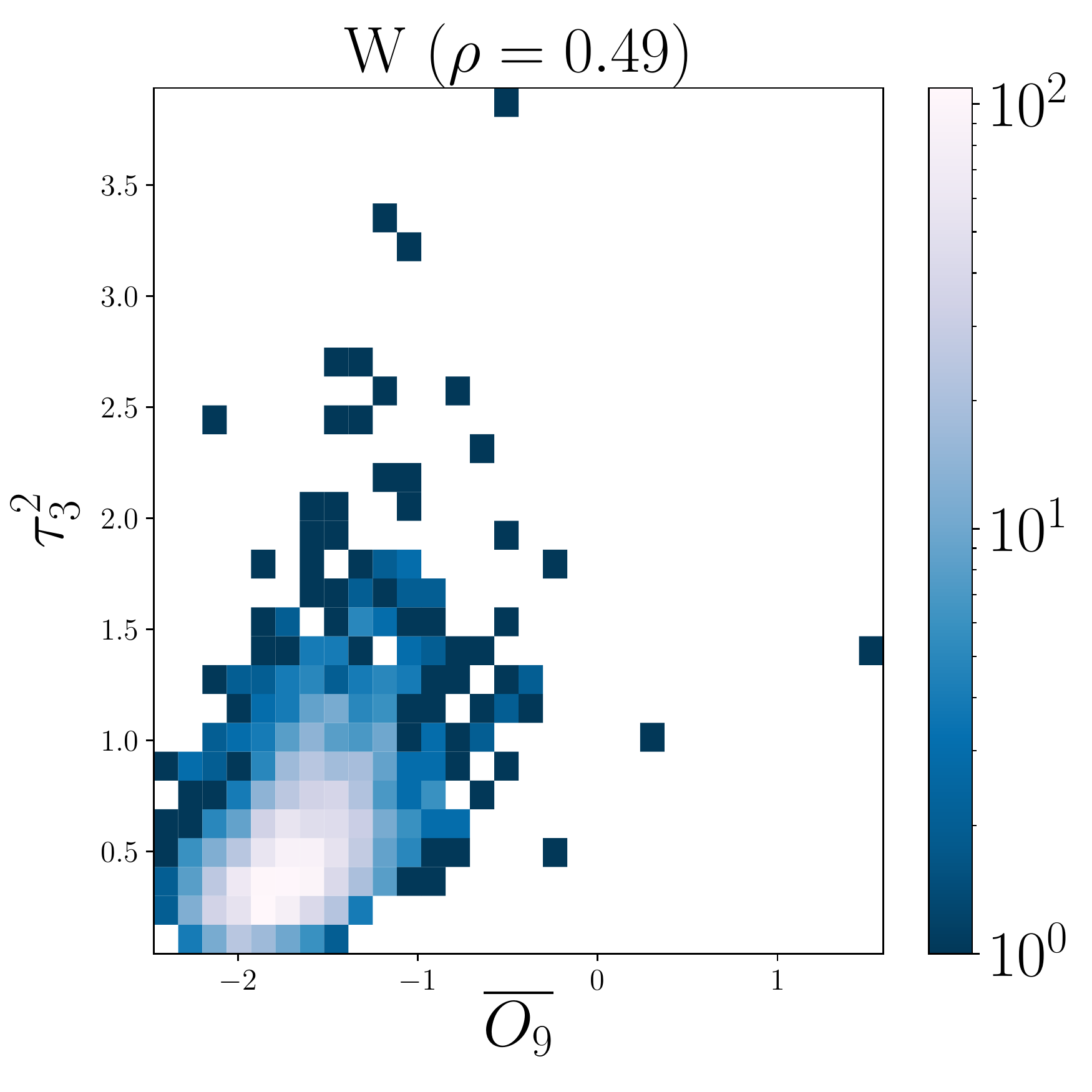} 
\includegraphics[width=0.19\textwidth]{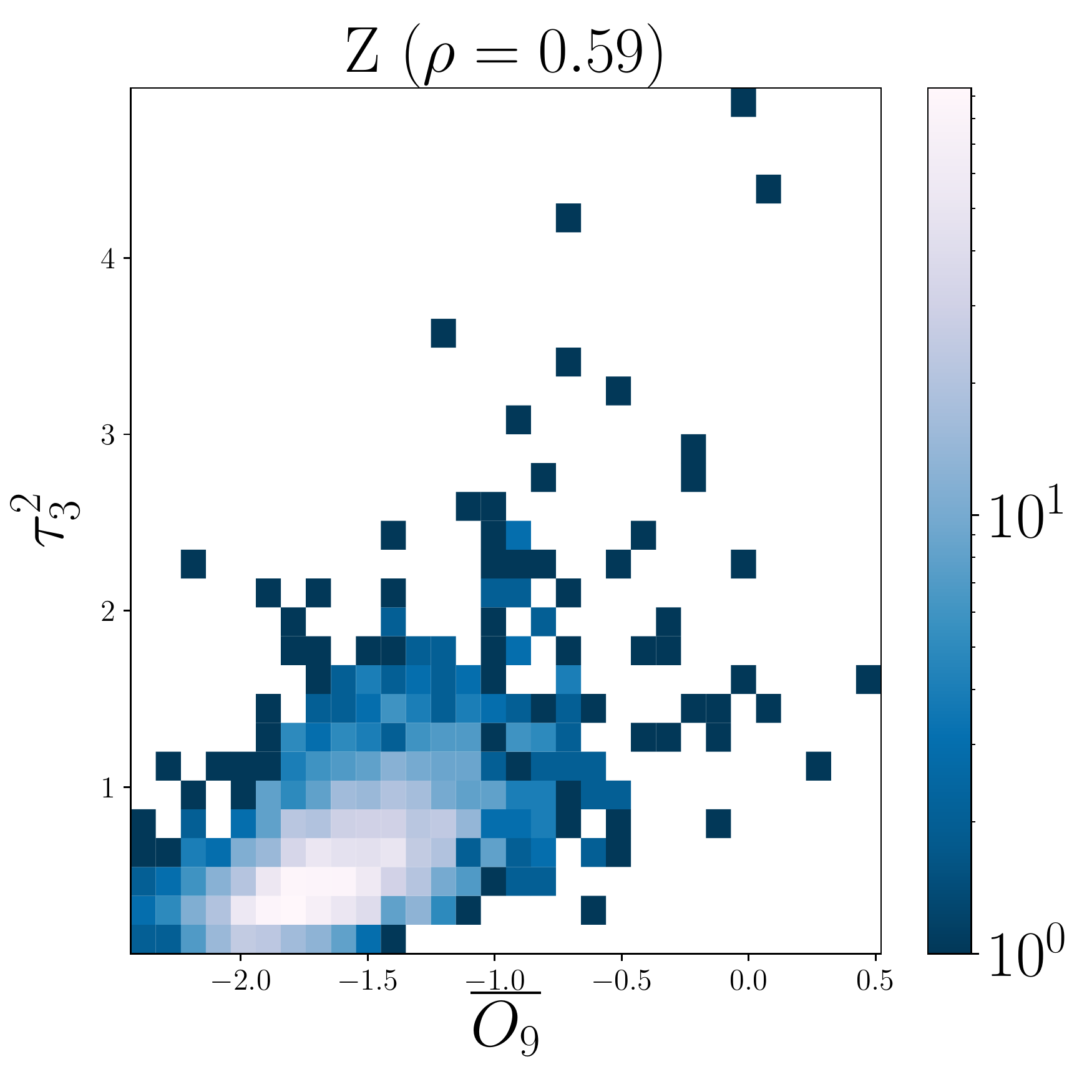}
\includegraphics[width=0.19\textwidth]{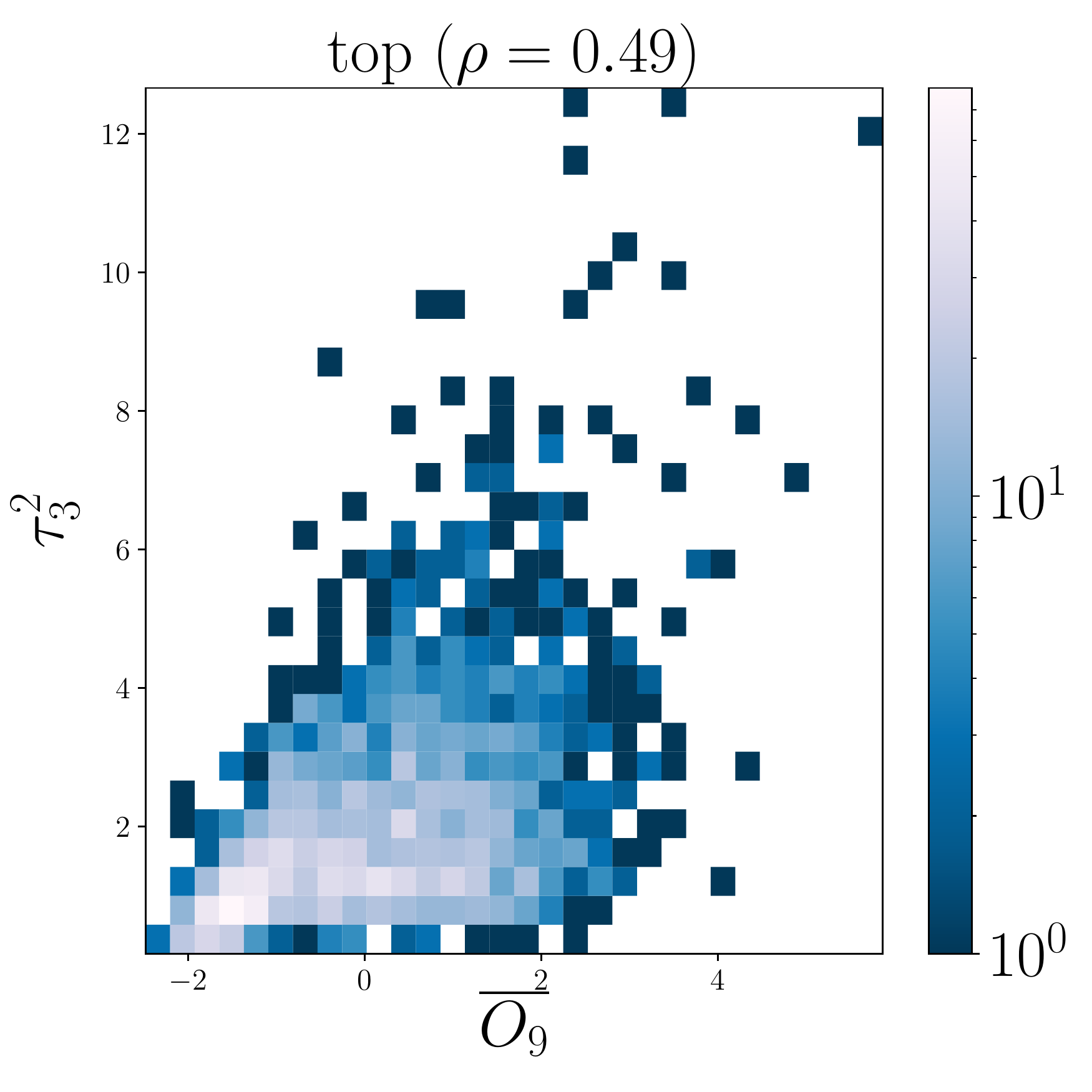} \\
\caption{Two-dimensional distributions between (top to bottom) $\overline{O}_1$ and constituents multiplicty, $\overline{O}_4$ and $\tau_1^{(\beta=2)}$, $\overline{O}_2$ and $\tau_3^{(\beta=1)}$, $\overline{O}_9$ and $\tau_3^{(\beta=2)}$ , for jets originating from (right to left) gluons, light flavor quarks,  $\PW$ bosons, $\PZ$ bosons, and top quarks. For each distribution, the linear correlation coefficient $\rho$ is reported.\label{fig:Ocorr}}
\end{figure*}

\section{What did JEDI-net learn?}
\label{sec:learned}

In order to characterize the information learned by JEDI-net, we consider the $\overline{O}$ sums across the $N_O$ vertices of the graph (see Section~\ref{sec:models}) and we study their correlations to physics motivated quantities, typically used when exploiting jet substructure in a search. We consider the HLF quantities used for the DNN model and the $N$-subjettiness variables $\tau_N^{(\beta)}$~\cite{Thaler:2010tr}, computed with angular exponent $\beta=1,2$. 

Not all the $\overline{O}$ sums exhibit an obvious correlation with the considered quantities, i.e.,  the network engineers high-level features that encode other information than what is used, for instance, in the DNN model. 

Nevertheless, some interesting correlation pattern between the physics motivated quantities and the $\overline{O}_i$ sums is observed. The most relevant examples are given in Fig.~\ref{fig:Ocorr}, where the 2D histograms and the corresponding linear correlation coefficient ($\rho$) are shown. The correlation between $\overline{O}_1$ and the particle multiplicity in the jet is not completely unexpected. As long as the $O$ quantities aggregated across the graph have the same order of magnitude, the corresponding sum $\overline{O}$ would be proportional to jet-constituent multiplicity.

The strong correlation between the $\overline{O}_4$ and $\tau^{(\beta=2)}_{1}$ (with $\rho$ values between 0.69 and 0.97, depending on the jet class) is much less expected. The $\tau_1^\beta$ quantities assume small values when the jet constituents can be arranged into a single {\it sub-jet} inside the jet. Aggregating information from the constituent momenta across the jet, the JEDI-net model based on the $\overline{O}$ quantities learns to build a quantity very close to $\tau^{(\beta=2)}_1$. The last two rows of Fig.~\ref{fig:Ocorr} show two intermediate cases: the correlation between $\overline{O}_2$ and $\tau^{(\beta=1)}_3$ and between $\overline{O}_9$ and $\tau^{(\beta=2)}_3$. The two $\overline{O}$ sums considered are correlated to the corresponding substructure quantities, but with smaller (within 0.48 and 0.77) correlation coefficients. 

\section{Resource comparison}
\label{sec:resources}

Table~\ref{tab:Resources} shows a comparison of the computational resources needed by the different models discussed in this paper. The best-performing JEDI-net model has more than twice the number of trainable parameters than the DNN and GRU model, but approximately a factor of $6$ less parameters than the CNN model. The JEDI-net model based on the summed $\overline{O}$ features achieves comparable performance with about a factor of $4$ less parameters, less than the DNN and GRU models. While being far from expensive in terms of number of parameters, the JEDI-net models are expensive in terms of the number of floating point operations (FLOP). The simple model based on $\overline{O}$ sums, using as input a sequence of 150 particles, uses 458~MFLOP. The increase is mainly due to the scaling with the number of vertices in the graph. Many of these operations are the $\times 0$ and $\times 1$ products involving the elements of the $R_R$ and $R_S$ matrices. The cost of these operations could be reduced with an IN implementation optimized for inference, e.g., through an efficient sparse-matrix representation.

\begin{table}[htb]
\small
\centering
\begin{tabular}{c|ccc}
Model   & Number of  & Number of  & Inference \\
        & parameters & FLOP & time$/$batch [ms] \\
\hline
DNN & 14725 & 27~k & $ 1.0 \pm  0.2$ \\
CNN & 205525 & 400~k & $57.1 \pm    0.5$ \\
GRU & 15575 & 46~k & $  23.2 \pm    0.6$ \\
JEDI-net & 33625 & 116~M &  $  121.2 \pm    0.4$\\
JEDI-net & \multirow{2}{*}{8767} & \multirow{2}{*}{458~M} &  \multirow{2}{*}{$402 \pm 1$}\\
with $\sum O$ & & \\
\end{tabular}
\caption{Resource comparison across models. The quoted number of parameters refers only to the trainable parameters for each model. The inference time is measured by applying the model to batches of 1000 events 100 times: the 50\% median quantile is quoted as central value and the 10\%-90\% semi-distance is quoted as the uncertainty. The GPU used is an NVIDIA GTX 1080 with 8 GB memory, mounted on a commercial desktop with an Intel Xeon CPU, operating at a frequency of 2.60GHz. The tests were executed in \textsc{Python} 3.7 with no other concurrent process running on the machine.\label{tab:Resources}}
\end{table}

In addition, we quote in Table~\ref{tab:Resources} the average inference time on a GPU.
The inference time is measured applying the model to 1000 events, as part of a \textsc{Python} application based on \textsc{TensorFlow}~\cite{tensorflow}. 
To this end, the JEDI-net models, implemented and trained in \textsc{PyTorch}, are exported to \textsc{ONNX}~\cite{onnx} and then loaded as \textsc{TensorFlow} graph. 
The quoted time includes loading the data, which occurs for the first inference and is different for different event representations, that is smaller for the JEDI-net models than for the CNN models. 
The GPU used is an NVIDIA GTX 1080 with 8~GB memory, mounted on a commercial desktop with an Intel Xeon CPU, operating at a frequency of 2.60~GHz. 
The tests were executed in \textsc{Python} 3.7, with no other concurrent process running on the machine. 
Given the larger number of operations, the GPU inference time for the two IN models is much larger than for the other models.

The current IN algorithm is costly to deploy in the online selection environment of a typical LHC experiment.
A dedicated R\&D effort is needed to reduce the resource consumption in a realistic environment in order to benefit from the improved accuracy that INs can achieve. 
For example, one could trade model accuracy for reduced resource needs by applying neural network pruning~\cite{NIPS1989_250,DBLP:journals/corr/HanMD15}, reducing the numerical precision~\cite{DBLP:journals/corr/abs-1710-09282,DBLP:journals/corr/GuptaAGN15}, and limiting the maximum number of particles in each jet representation.

\section{Conclusions}
\label{sec:conclusions}

This paper presents JEDI-net, a jet tagging algorithm based on interaction networks.  Applied to a data set of jets from light-flavor quarks, gluons, vector bosons, and top quarks, this algorithm achieves better performance than models based on dense, convolutional, and recurrent neural networks, trained and optimized with the same procedure on the same data set.
As other graph networks, JEDI-net offers several practical advantages that make it particularly suitable for deployment in the data-processing workflows of LHC experiments: it can directly process the list of jet constituent features (e.g. particle four-momenta), it does not assume specific properties of the underlying detector geometry, and it is insensitive to any ordering principle applied to the input jet constituents. 
For these reasons, the implementation of this and other graph networks is an interesting prospect for future runs of the LHC. On the other hand, the current implementation of this model demands large computational resources and a large inference time, which make the use of these models problematic for real-time selection and calls for a dedicated program to optimize the model deployment on typical L1 and HLT environments.

The quantities engineered by one of the trained IN models exhibit interesting correlation patterns with some of the jet substructure quantities proposed in literature, showing that the model is capable of learning some of the relevant physics in the problem. 
On the other hand, some of the engineered quantities do not exhibit striking correlation patterns, implying the possibility of a non trivial insight to be gained by studying these quantities.  

\section*{Acknowledgments}
We are grateful to Caltech and the Kavli Foundation for their support of undergraduate student research in cross-cutting areas of machine learning and domain sciences. We would also like to thank the Taylor W. Lawrence Research Fellowship and Mellon Mays Fellowship for supporting E.~A.~M. and making this research effort possible. 
This work was conducted at ``\textit{iBanks},'' the AI GPU cluster at Caltech. 
We acknowledge NVIDIA, SuperMicro  and the Kavli Foundation for their support of ``\textit{iBanks}.''
This project has received funding from the European Research Council (ERC) under the European Union's Horizon 2020 research and innovation program (grant agreement n$^o$ 772369) and is partially supported by the U.S. Department of Energy, Office of High Energy Physics Research under Caltech Contract No. DE-SC0011925.
J.~M.~D. is supported by Fermi Research Alliance, LLC under Contract No. DE-AC02-07CH11359 with the U.S. Department of Energy, Office of Science, Office of High Energy Physics.

\appendix

\section*{Appendix}

\section{Alternative models}
\label{appendix:otherModelOpt}

The three benchmark models considered in this work are derived through a Bayesian optimization of their hyperparameters, performed using the \textsc{GpyOpt} library~\cite{gpyopt2016}, based on \textsc{Gpy}~\cite{gpy2014}. 
For each iteration, the training is performed using early stopping to prevent over-fitting and to allow a fair comparison between different configurations. 
The data set for training (validation) consists of 630,000 (240,000) jets, with 10,000 jets used for testing purposes. 
The loss for the Bayesian optimization is estimated on the validation data set.
The CNN and GRU networks are trained on four different input data sets, obtained considering the first 30, 50, 100, or 150 highest-$\pt$ jet constituents. 
The DNN model is trained on quantities computed from the full list of particles. 

The DNN model consists on a multilayer perceptron, alternating dense layers to dropout layers. 
The optimal architecture is determined optimizing the following hyperparameters: 
\begin{itemize}
    \item Number of dense layers ($N_{DL}$) between 1 and 3.
    \item Number of neurons per dense layer ($n_{n}$): $10, 20, \ldots, 100$.
    \item Activation functions for the dense layers: ReLU, ELU, or SELU.
    \item Dropout rate: between 0.1 and 0.4.
    \item Batch size: 50, 100, 200, or 500.
    \item Optimization algorithm: Adam, Nadam~\cite{nadam}, or AdaDelta.
\end{itemize}  
The optimization process gives as output an optimal architecture with three hidden layers of 80 neurons each, activated by ELU functions. 
The best dropout rate is found to be 0.11, when a batch size of 50 and the Adam optimizer are used. 
This optimized network gives a loss of 0.66 and an accuracy of 0.76.

The CNN model consists of two-dimensional convolutional layers with batch normalization, followed by a set of dense layers. 
A $2\times2$ max pooling layer is applied after the fist convolutional layer. 
The optimal architecture is derived optimizing the following hyperparameters:
\begin{itemize}
    \item Number of convolutional layers $N_{CL}$ between 1 and 3.
    \item Number of convolutional filters $n_{f}$ in each layer (10, 15, 20, 25, or 30).
    \item Convolutional filter size: $3\times 3$, $5\times 5$, $7\times 7$, or $9\times 9$.
    \item Max pooling filter size: $2\times 2$, $3\times 3$, or $5\times 5$.
    \item Activation functions for the convolutional layers (ReLU, ELU, or SELU).
    \item Number of dense layers $N_{DL}$ between 1 and 3.
    \item Number of neurons $n_{n}$ per dense layer: $10, 20, \ldots, 60$.
    \item Activation functions for the dense layers: ReLU, ELU, or SELU.
    \item Dropout rate: between 0.1 and 0.4.
    \item Batch size: 50, 100, 200, or 500.
    \item Optimization algorithm: Adam, Nadam, or AdaDelta.
\end{itemize}  
The stride of the convolutional filters is fixed to 1 and ``same'' padding is used. 
Table~\ref{tab:CNNopt} shows the optimal sets of hyperparameter values, obtained for the four different data set representations. 
While the optimal networks are equivalent in performance, we select the network obtained for $\leq 50$ constituents, because it has the smallest number of parameters. 

\begin{table}[htb]
\centering
\begin{tabular}{c|cccc}
\multirow{2}{*}{Hyperparameter}   &  \multicolumn{4}{c}{Number of jet constituents}\\
                 & 30 & {\bf 50} & 100 & 150 \\
\hline
$N_{CL}$         &  3  & {\bf 1}   &  1     &  3  \\
$n_{f}$          &  20  & {\bf 10}   & 30     &  30  \\
Filter size      &  $3\times 3$  &  $\mathbf{3\times 3}$  & $3\times 3$    &  $3\times 3$  \\
Max pooling size     &  $2\times 2$  & $\mathbf{5\times 5}$   & $5\times 5$    &  $2\times 2$  \\
Conv. activation &  ReLU  &  {\bf ELU}  & ELU   &   ReLU \\
$N_{DL}$         &  2  &  {\bf 3 } & 3  &  3  \\
$n_{n}$          &  60  & {\bf 50}   &  60    &  60  \\
Dense activation &  SELU  & {\bf ELU}   &  ELU   &  ELU  \\
Dropout          &  0.11  & {\bf 0.1}   & 0.4    &  0.1  \\
Batch size       &  200  &  {\bf 500}  & 100     &  50  \\
Optimizer        &  Adam  & {\bf Adam}   & Adam   &  Adam  \\
\hline
Optimized loss     & 0.88 & {\bf 0.73} & 0.74 & 0.74 \\
Optimized accuracy & 0.67 & {\bf 0.74} & 0.74 & 0.74 \\
\end{tabular}
\caption{Optimal CNN hyperparameter setting for different input data sets. The best configuration, used as a benchmark for comparison, is highlighted in bold.\label{tab:CNNopt}}
\end{table}

The recurrent model consists of a GRU layer feeding a set of dense layers. The following hyperparameters are considered:
\begin{itemize}
    \item Number of GRU units: 50, 100, 200, 300, 400, or 500.
    \item Activation functions for the GRU layers: ReLU, ELU, or SELU.
    \item Number of dense layers: between 1 and 4.
    \item Number of neurons per dense layer: 10, 20, $\ldots$, 100.
    \item Activation functions for the dense layers: ReLU, ELU, or SELU.
    \item Dropout rate: between 0.1 and 0.4.
    \item Batch size: 50, 100, 200, or 500.
    \item Optimization algorithm: Adam, Nadam, or AdaDelta.
\end{itemize}  
The best hyperparameter values are listed in Table~\ref{tab:GRUopt}.
As for the CNN model, the best performance is obtained when the list of input particles is truncated at 50 elements.

\begin{table}[htb]
\centering
\begin{tabular}{c|cccc}
\multirow{2}{*}{Hyperparameter}   &  \multicolumn{4}{c}{Number of jet constituents}\\
                 & 30 & {\bf 50} & 100 & 150 \\
\hline
$n_{u}$          & 100    & {\bf 50}   &  200 &  50  \\
$N_{DL}$         &  3      & {\bf 1}     & 3 &  4 \\
$n_{n}$          &  70     & {\bf 40}    & 40 &  100  \\
Dense activation &  SELU   & {\bf SELU} &  ReLU  & ELU   \\
Dropout          &  0.40   & {\bf 0.10}  &  0.22 &  0.10  \\
Batch size       &  500    & {\bf 500}   & 500 &  500  \\
Optimizer        & Adam   & {\bf Adam}  &   Adam & AdaDelta  \\
\hline
Optimized loss & 0.78 & {\bf 0.71} & 0.78 & 0.85 \\
Optimized accuracy & 0.72 & {\bf 0.75} & 0.73 & 0.68 \\
\end{tabular}
\caption{Optimal GRU hyperparameter settings for different input data sets. The best configuration, used as a benchmark for comparison, is highlighted in bold.\label{tab:GRUopt}}
\end{table}

\section{Performance on public top tagging data set}
\label{sec:toptagging}
\newcommand{\bkgrej}{1/\epsilon_B (\epsilon_S=30\%)}

In this appendix, we retrain and evaluate the performance of JEDI-net on a public top tagging data set~\cite{Butter:2017cot,Kasieczka:2019dbj} used to benchmark many neural networks architectures for the task of differentiating top quark jets from light quark jets.
To select the hyperparameters of the model (with and without the sum over particles), we performed a Bayesian optimization.
We scan $N_n^1$ from $16$ to $256$, $D_E$ from $4$ to $64$, $D_O$ from  $4$ to $64$, ReLU, ELU, or SELU activation functions for $f_R$, $f_O$, and $\phi_C$, and either the Adam or Adadelta optimizers with an initial learning rate of $10^{-3}$.
We report three metrics for the performance of the network on the top tagging data set: model accuracy, area under the ROC curve (AUC), and background rejection power at a fixed signal efficiency of 30\%, $\bkgrej$.  
In Table~\ref{tab:topperf}, the accuracy, AUC, and $\bkgrej$ values are listed for each model considered.  
The performance of JEDI-net compared to other models developed for this data set is approaching state-of-the-art~\cite{Kasieczka:2019dbj}.


\begin{table}
\centering
\begin{tabular}{c|cccc}
  Model & JEDI-net & JEDI-net with $\sum O$\\
  \hline
  Number of constituents & 150 & 150 \\
   $N_n^1$ & 64 & 256 \\
  $D_E$  & 64 & 64 \\
  $D_O$ & 16 & 32 \\
  $f_R$ activation & ReLU & SELU \\
  $f_O$ activation & SELU & ReLU \\
  $\phi_C$ activation & ReLU & SELU\\
  Optimizer & Adam & Adam\\
  Number of parameters & 169906 & 148962\\
  \hline
  Accuracy & 0.9263 & 0.9300 \\
  AUC & 0.9786 & 0.9807 \\
  $\bkgrej$ & 590.4 & 774.6
\end{tabular}
\caption{\label{tab:topperf} The optimized hyperparameters, number of
  trainable parameters, and performance metrics of the JEDI-net models on the top tagging data set. Performance metrics are evaluated on the test sample. We
  quote the area under the ROC curve (AUC), the accuracy, and the
  background rejection at a signal efficiency of 30\%.}
\end{table}

\bibliographystyle{lucas_unsrt_epjc}
\bibliography{bib}
\end{document}